\documentstyle[epsf]{thesis}
\newcommand{\resetcounters}{
	    \setcounter{equation}{0}
	    \setcounter{figure}{0}
	    \setcounter{table}{0}   }
\newcommand{\bold}[1]{\mbox{\boldmath $#1$}}
\newcommand{\bphi}{\bold{\vec{\phi}}}
\newcommand{\bPi}{\bold{\vec{\Pi}}}
\newcommand{\bvec}[1]{\mbox{ {\boldmath $\vec{#1}$}}}
\newcommand{\bpsi}{\bold{\vec{\psi}}}
\begin{document}
\thesistitle{\bf Theory of Nuclear Magnetic 
Relaxation in Haldane Gap Materials: 
	  An Illustration of the Use of (1+1)-Dimensional
	  Field Theory Techniques}
\author{Jacob. S. Sagi}
\previousdegrees{B.\ Sc. (Physics) 
                 University of Toronto, 1991}
\degreetitle{Doctor of Philosophy}
\department{Department of Physics}
\signaturelines{4}
\copyrightyear{August 1995}
\titlepage
\authorizepage
\begin{center} 
   {\Huge \bf Abstract}
\end{center}

A comprehensive theory of nuclear magnetic relaxation in $S=1$ Haldane
gap materials is developed using nonlinear-$\sigma$, boson and fermion 
models. We find that at temperatures much smaller than the lowest gap
the dominant contribution to the relaxation rate comes from two magnon
processes with $T_1^{-1} \sim e^{-\Delta_m/T}$, where $\Delta_m$ is the
smallest gap corresponding to a polarization direction
perpendicular to the field direction. As the gap closes, we
find that the dominant contribution comes from one magnon processes, and
the result depends on the symmetry of the Hamiltonian. Overall the models
agree qualitatively, except near the critical regime, where the fermion
model is shown to be the best description.
We include a thorough discussion of the effects of interchain couplings,
nearest neighbour hyperfine interactions and crystal structure,
and introduce a new theory of impurities corresponding to broken
chain ends weakly coupled to bulk magnons. The work is then applied 
to recent measurements on NENP. We find overall fair agreement between
available $T^{-1}_1$ data and our calculations. We finish by suggesting
further experimental tests of our conclusions. 

\addcontentsline{toc}{chapter}{Abstract} \vspace{2ex}
\newpage
\addcontentsline{toc}{chapter}{Table of Contents}
\tableofcontents
\listoffigures

\chapter{Introduction and Background}
\resetcounters

\section{Introduction}

In 1983, Haldane derived his famous result stating
that integer spin one dimensional
Heisenberg antiferromagnets featured a gap in their low energy excitation
spectrum \cite{hal1}. Since then, much effort has been devoted
to further exploration of such systems, both experimentally and theoretically.
The purpose of this work is to develop a theoretical framework
for the understanding of low energy experiments on one dimensional
Haldane gap materials.
In particular, we focus on the nuclear magnetic
relaxation rate, $T_1^{-1}$, although the
work has relevance to many other techniques. By studying this thesis,
it is hoped that the reader can become familiar with the tools used
to understand integer spin Heisenberg antiferromagnetic chains with 
anisotropies, and can apply these tools to the analysis of real systems.

There are, essentially, three important models that have so far been used
to describe the system. In the later sections of this
chapter we review the competing descriptions of $S=1$
antiferromagnetic spin-chains, paying some attention to their
advantages and shortcomings. We start by outlining 
the traditional spin-wave theory\footnote{see \cite{kittel} for
a comprehensive discussion of this topic} used to model
antiferromagnetism in higher dimensions.
After illustrating the deficiencies in this approach, we describe the
Nonlinear $\sigma$ (NL$\sigma$) model in some detail. This is
followed by an analysis of a simplified yet closely related
Landau-Ginsburg boson model.
Last, we discuss a free fermion model used recently
to successfully treat the case of anisotropic spin-chains.
We end the chapter with background on the nuclear magnetic relaxation
rate, $1/T_1$, for nuclear spins coupled to the spin-chain through
hyperfine interactions.

Chapter Two focuses on the details of the models,
building the tools necessary for a detailed
analysis. We discuss the temperature
and magnetic
field dependence of the NL$\sigma$ model and its possible relevance to the
spectrum of the boson model, as well as cite
some exact results available in cases
of high symmetry. We also diagonalize the free boson
and fermion field theories, including on-site anisotropy effects. We derive
matrix elements of the uniform spin operator between one particle states
of magnetic excitations. These are used to compare the different models.

Chapter Three explicitly
describes the calculation of NMR $T_1^{-1}$, considering various symmetries
of the Hamiltonian. We identify the leading mechanisms for low temperature
relaxation in the presence of a magnetic field. We discuss
three regimes corresponding to different magnitudes of
the applied external magnetic field, giving expressions for the rate in
each case. We discover that at temperatures much lower than the smallest
gap the uniform part of the spin operator contributes most to the
relaxation rate; in
the absence of interactions, this
corresponds to two magnon processes. The rate is found
to be $T_1^{-1} \sim e^{-\Delta_m/T}$, where $\Delta_m$ is the smallest gap
corresponding to a polarization direction
perpendicular to the magnetic field. As the externally applied magnetic
field approaches a critical value, one of the gaps closes and we find the 
dominant process to be one magnon, corresponding to contributions from 
the staggered part of the spin. In this regime, we show that the fermion model
is the best description and that the expression for $T_1^{-1}$ depends
on the symmetry of the Hamiltonian.

Chapter Four deals with intrinsic effects which must be taken into account when
analyzing experimental data. We discuss nearest neighbour
hyperfine interactions; we show that these will contribute to order
$A_{nn}/A$, the ratio of the nearest neighbour coupling to the local
coupling. We also consider interchain couplings and show that they 
introduce a natural infrared cutoff to the diverging density of states
at the gap; for sufficiently long chains, they also densely fill the energy
intervals between states along a finite chain. Finally, we
introduce a new impurity theory to explain the effects of nearly free 
spin-$\frac{1}{2}$ chain end degrees of freedom. We find that the states
formed by such end spins in the gap, can give rise to non-trivial relaxation
when coupled to the bulk excitations.

Chapter Five applies the theory to recent
experiments on the well studied material, $Ni(C_2H_8N_2)_2NO_2(ClO_4)$ 
(NENP). We take 
a close look at the crystal structure of NENP and identify possible terms
which may be present in the Hamiltonian. We also note the fact, hitherto 
neglected, that NENP possesses two inequivalent chains in each unit cell.
The results of Chapters Three and Four are then used to analyze
experimental data. We find reasonable agreement for a magnetic 
field placed along
the crystal $c$-axis of NENP, and an unexpected discrepancy for a magnetic
field placed along the chain axis. The impurity theory is used to model
low field data with qustionable results.

The final chapter proposes 
further experimental tests of the theoretical predictions of this work.
We suggest elastic neutron, electron spin resonance and further NMR
studies to verify our own.

\section{Spin-wave Theory }

The Heisenberg Hamiltonian describing the isotropic
antiferromagnetic spin-chain is 
\begin{eqnarray}
H = J\sum_{i} \bvec{S}_i \cdot \bvec{S}_{i+1} \mbox{\hspace{.5in}}
J>0
\label{hamiltonian}
\end{eqnarray}

This arises naturally from the Hubbard model for insulators at
`half' filling \cite{anderson}. To understand where this might come from,
we follow the case where there is a triplet of possible spin states per site.
On each site there are a number of valence electrons (eight valence 
electrons in the $d$ shell of
$Ni^{+2}$, for example);
the degenerate electronic levels are split in a way
determined by Hund's rules\footnote{Hund's rules maximize the total electronic
spin and the total angular momentum of the electrons in the valence shell.} 
and the symmetry of the crystal fields
surrounding the ion.
In some special
cases (as with $Ni^{+2}$ in a field with octahedral symmetry), a
degenerate triplet of
states lies lowest. The ensuing low energy physics can be
essentially described
using effective spin 1 operators \cite{abragam}. 
By `half' filling, we mean that there is an effective $S=1$
triplet of
states for every site in the chain (ie. there are two singly
occupied orbitals on each site. Other orbitals are either empty
or doubly occupied. Spins in singly occupied orbitals are aligned by
Hund's rules.) Antiferromagnetism comes from
allowing a small 
amplitude for nearest neighbour hopping which is highly
suppressed by coulomb repulsion
from the electrons already occupying the site.

In the quantum case, the spin operators have the commutation
relations:
\begin{eqnarray}
[S^a_i,S^b_j] = i \delta_{ij} \sum_c \epsilon^{abc} S^c_i,
\mbox{\hspace{.2in}} 
\bvec{S}_i \cdot \bvec{S}_i = s(s+1)
\end{eqnarray}
where $\delta_{ij}$ is the Kronecker Delta Function and
$\epsilon^{abc}$ is the
completely antisymmetric Levi-Civita symbol. 

It is easy to see that the classical N\'{e}el ground state with
alternating
spins is not the quantum ground state. To this end we write the
Hamiltonian in
terms of raising and lowering spin operators:
\[ S^{\pm} \equiv S^x \pm iS^y  \]
\begin{eqnarray}
H = J\sum_{i} \left [ S_i^z S_{i+1}^z + \frac{1}{2} (S^+_i
S^-_{i+1} +
S^-_i S^+_{i+1}) \right ]
\end{eqnarray}
The N\'{e}el ground state is composed of spins alternating in
quantum numbers
$s^z$ between sites.
\begin{eqnarray}
|\mbox{N\'{e}el}> = |s^z_1=+1,s^z_2=-1,s^z_3=+1,...,s^z_N=-1>
\end{eqnarray}
This state is clearly not an eigenstate of the above Hamiltonian
since upon acting on it, the $S^+_i S^-_{i+1}$
terms in
the Hamiltonian generate states with $m_i=0$.
To proceed in understanding the low energy properties one usually
assumes that the ground state is {\em approximately} N\'{e}el with
quantum
fluctuations. The picture is that of zero point motion about the
positions of the
classical N\'{e}el spins. What we will shortly see is that the
assumption of
small fluctuations breaks down in one dimension. 

The conventional approach
makes use of the Holstein-Primakov transformation. One begins by
dividing the chain into 
two sublattices, ``A'' and ``B'', with adjacent sites on separate
sublattices.
On sublattice ``A'' one defines
\begin{eqnarray} \label{prim1}
S^z_i = s-a^{\dagger}_i a_i, \mbox{\hspace{.2in}} 
S^-_i = a^{\dagger}_i \sqrt{2s-a^{\dagger}_i a_i}
\end{eqnarray}
On sublattice ``B'' we have
\begin{eqnarray} \label{prim2}
S^z_{i+1} = -s+b^{\dagger}_i b_i, \mbox{\hspace{.2in}}
S^+_{i+1} = b^{\dagger}_i \sqrt{2s-b^{\dagger}_i b_i}
\end{eqnarray}
$a$ and $b$ are usual bosonic operators with commutation
relations:
\begin{eqnarray}
[a,a^{\dagger}] = [b,b^{\dagger}] = 1, \mbox{\hspace{.2in}} [a,a]
= [b,b] = 0
\end{eqnarray}
It can be checked that (\ref{prim1}) and (\ref{prim2}) preserve
the correct spin
commutation relations and the constraint $\bvec{S}_i \cdot
\bvec{S}_i = s(s+1)$.
The N\'{e}el ground state is one without bosons. So far no
approximations have entered
into the picture. However, to make progress, we assume that $s$
is large. This is 
equivalent to a semi-classical approximation since for $s
\rightarrow \infty$ the
commutator of the spins will have much smaller eigenvalues than
the square of the 
spin variables
\begin{eqnarray}
[S^a,S^b] = i\sum_c \epsilon^{abc}S^c = O(s) \ll O(s^2)
\end{eqnarray}
We expand the spin operators to give
\begin{eqnarray}
S^-_i = a^{\dagger}_i \sqrt{2s}, \mbox{\hspace{.2in}} S^+_{i+1} =
b^{\dagger}_i \sqrt{2s}
\end{eqnarray}
To leading order, the Hamiltonian reduces to 
\begin{eqnarray} \label{hal1}
H = J\sum_i \left [ -s^2 + s(2a^{\dagger}_i a_i + 2b^{\dagger}_i
b_i + 
a_i b_i + b_{i-1} a_i + b^{\dagger}_i a^{\dagger}_i +
a^{\dagger}_i b^{\dagger}_{i-1})
+ O(1)  \right ]
\end{eqnarray}
Fourier transforming:
\begin{eqnarray}
a_j = \frac{1}{\sqrt{N}} \sum_{n=-\frac{N}{2}}^{\frac{N}{2}}
e^{i2\pi jn/N} a_k
\end{eqnarray}
with a similar expression for $b$; $N$ is the number of sites
on each
sublattice. Ignoring the constant term, we rewrite (\ref{hal1}) as 
\begin{eqnarray}
H = 2Js\sum_k \left [ a^{\dagger}_k a_k + b^{\dagger}_k b_k +
(1+ e^{2ika})(a_k b_{-k} + b_{k}^{\dagger} a^{\dagger}_{-k})
\right ]
\end{eqnarray}
$2a$ is the sublattice spacing and $k=\pi n/Na$ for $n \in
[-\frac{N}{2}, \frac{N}{2}]$.
We now make the Bogoliubov transformation,
\begin{eqnarray}
c_k = u_k a_k - v_k b_{-k}^{\dagger} \nonumber \\
d_k = u_k b_{-k} - v_k a^{\dagger}_k
\end{eqnarray}
where, 
\begin{eqnarray}
u_k = \frac{ie^{ika/2}}{\sqrt{2}} \left ( 1 + \csc(ka) \right )^{1/2}
\nonumber \\
v_k = \frac{-ie^{-ika/2}}{\sqrt{2}} \left (-1 + \csc(ka) \right )^{1/2}
\end{eqnarray}
The $d$'s and $c$'s are spin wave operators corresponding to
magnetic excitations (magnons) with $S^z = \pm1$ respectively.
This transformation preserves the commutation relations and the
Hamiltonian can
now be written as
\begin{eqnarray}
H = 2Js\sum_k \sin(|ka|) \left ( c^{\dagger}_k c_k +
d^{\dagger}_k d_k \right )
\end{eqnarray}

We see that this low energy description implies the spins are
in some 
coherent state of $a$'s and $b$'s built on the
N\'{e}el state, but there are long wavelength Goldstone modes
with dispersion relation
$\omega = 2Js|ka|$ which allow each of the sublattice
magnetization vectors to make
long wavelength rotations. To ensure consistency, we now look at
the expectation value of the
magnetization (say, on the ``A'' sublattice), hoping to see that
we get the semi-classical result
$<S^z_i>=s - O(1)$. We invert the Bogoliubov transformation to
get
\begin{eqnarray} \label{bob}
a_k = u_k^{\ast} c_k + v_k d^{\dagger}_k
\end{eqnarray}
and compute
\begin{eqnarray}
<S^z_i>=s - <a^{\dagger}_i a_i> = s - \frac{1}{N}\sum_k
<a^{\dagger}_k a_k> \nonumber \\
= s - a\int \frac{dk}{2\pi} |v_k|^2 = s - a\int \frac{dk}{4\pi}
\left ( \csc(ka) - 1 \right )
\end{eqnarray}
The last line follows from (\ref{bob}) and the fact that
the true ground state has zero spin wave occupation number.
The problem is now apparent: low wavelength modes cause $<s-S^z>$
to diverge
logarithmically. The semi classical picture of a N\'{e}el-like
ground state is
completely off. This is special to 1 dimension and actually
arises as a general 
consequence of Coleman's theorem \cite{coleman}; it states that
in $(1+1)$ dimensions,
infrared divergences
associated with Goldstone bosons will always wash out the
classical value of the
order parameter rendering spontaneous symmetry breaking of
continuous symmetries impossible.
One is therefore forced to look elsewhere in order to describe
the low energy physics of
the Heisenberg model.

\section{Non-Linear $\sigma$ (NL$\sigma$) Model}

The most consistent continuum model derivable from Eqn. (\ref{hamiltonian})
is the Nonlinear $\sigma$ model. In addition to being a continuum model
(valid only in the long wave length limit), it is also based on a large
$s$ approximation.
One introduces two
fields: $\bphi$ --- 
corresponding to the N\'{e}el order parameter, and $\bf{\vec{l}}$ 
--- the uniform magnetization. The spin variable, $\bf{\vec{S}}$ is
defined in terms of these fields via
\begin{eqnarray}
{\bf\vec{S}}(x) = (-1)^x s\bphi (x) + {\bf\vec{l}}(x)
\end{eqnarray}
The conventional derivation defines these variables on the
lattice and a continuum
limit is taken in the semi-classical large $s$ approximation
to arrive at the NL$\sigma$ model Hamiltonian 
\cite{leshouches}. There are several problems with this
approach. First, parity is broken in intermediate steps and is
eventually restored in
the continuum limit. Second, and more importantly, the crucial
topological term which
is found in the continuum Hamiltonian is derived without clear
notions of how $1/s$ corrections may be made. 
A much more elegant approach will be reviewed here.
We will make use of
path integral formalism based on independent derivations by
Haldane \cite{hal_p} and
Fradkin and Stone \cite{frad}.\footnote{We note that the derivation in 
reference \cite{frad} is somewhat in error. We correct their mistakes 
using a similar derivation due to Ian Affleck 
(unpublished).} These were motivated by similar
questions about topological
terms in 2-D quantum models thought useful in attempting to describe
high temperature superconductors.

One begins by defining a coherent basis of states for the $S=s$
representation of 
$SU(2)$ \cite{perel}:
\begin{eqnarray}
|\hat{n}> \equiv e^{i(\widehat{ \hat{z} \times \hat{n}}) 
\cdot \vec{S} \theta} |+s>
\end{eqnarray}
where $|+s>$ is the eigenstate of $S^z$ with eigenvalue $s$,
$\hat{z} \cdot \hat{n} = \cos\theta$ and 
$(\widehat{ \hat{z} \times \hat{n}})$ is a unit vector perpendicular
to both $\hat{n}$ and $\hat{z}$.
We see that $|\hat{n}>$ is the state with spin $s$ in the
$\hat{n}$ direction (ie.
$\hat{n} \cdot \vec{S} |\hat{n}> = s|\hat{n}>$.)
This basis is over-complete and not orthogonal. We now make use
of two identities:
\begin{eqnarray}
<\hat{n}_1|\hat{n}_2> =
e^{is\Phi(\hat{n}_1,\hat{n}_2)} \left (\frac{1 + \hat{n}_1 \cdot
\hat{n}_2}{2}  \right )^s \\
1 = \frac{2s+1}{4\pi} \int d^3 \hat{n}  \;
\delta(\hat{n} \cdot \hat{n} -1) |\hat{n}><\hat{n}|
\end{eqnarray}
$\Phi(\hat{n}_1,\hat{n}_2)$ is the area enclosed by 
the geodesic triangle on the sphere
with vertices at $\hat{n}_1,\hat{n}_2$ and the north pole. There
is an ambiguity of 
$4\pi$ in this definition, but this makes little difference when
exponentiated
since $e^{4\pi is}=1$ for $s$ integer or half-integer.
The first identity is most easily proved by using the
$|\frac{1}{2},...,\frac{1}{2}>$
representation of $SU(2)$ while the second follows from the
first. We can now use
these states to write the partition function or path integral of
the system:
\begin{eqnarray}
\mbox{Tr}e^{-\beta H} = \lim_{N \rightarrow \infty} \int \left [
\prod_{l=1}^{N-1}
\frac{2s+1}{4\pi} d^3 \hat{n}(\tau_l) \right ] \prod_{m=2}^{N}
<\hat{n}(\tau_m)|
e^{-\Delta \tau H}|\hat{n}(\tau_{m-1})> 
\end{eqnarray}
\[ N\Delta \tau = \beta,\mbox{\hspace{.2in}} \tau_m - \tau_{m-1} =
\Delta \tau, \mbox{\hspace{.2in}} \tau_1 = \tau_N  \]
With $\Delta \tau \rightarrow 0$, we expand the exponential to
order $O(\Delta \tau)$
to get
\begin{eqnarray}
<\hat{n}(\tau_m)|e^{-\Delta \tau H}|\hat{n}(\tau_{m-1})> \approx
<\hat{n}(\tau_m)|\hat{n}(\tau_{m-1})> -
<\hat{n}(\tau_m)|H|\hat{n}(\tau_m)> \Delta \tau \nonumber \\
= e^{is\Phi(\hat{n}(\tau_m),\hat{n}(\tau_{m-1}))} \left (\frac{1 +
\hat{n}(\tau_m)
\cdot \hat{n}(\tau_{m-1})}{2}  \right )^s -
<\hat{n}(\tau_m)|H|\hat{n}(\tau_m)>
\Delta \tau
\end{eqnarray}
In the limit $N\rightarrow \infty$, the path integral can be
written,
\begin {eqnarray}
\mbox{Tr}e^{-\beta H} & \propto & \int \left [ {\cal D} \hat{n}(\tau)
\right ] e^{-S}  \label{trace} 
\end{eqnarray}
\[ S = \sum_{m=2}^{\infty} \left 
[ <\hat{n}(\tau_m)|H|\hat{n}(\tau_m)> \Delta \tau
- s \left ( i\Phi(\hat{n}(\tau_m),\hat{n}(\tau_{m-1}) 
+ \ln(\frac{1 + \hat{n}(\tau_m)
\cdot \hat{n}(\tau_{m-1})}{2}) \right ) \right ] \]

The last term can be written to second order in $\Delta \tau$ as 
\begin {eqnarray}
-\frac{s \Delta \tau}{4} \; \int_0^{\beta} d\tau \hat{n}(\tau) 
\cdot \partial_{\tau}^2 \hat{n}(\tau) = \frac{s \Delta \tau}{4}
\;
\int_0^{\beta} d\tau \left ( \partial_{\tau} \hat{n}(\tau)
\right )^2
\label{logterm}
\end{eqnarray}
This vanishes in taking the limit $\Delta \tau \rightarrow 0$.
The sum over the phases is just the area enclosed by the vector
$\hat{n}(\tau)$
as it traces its periodic path on the surface of the sphere.
Parametrizing
$\hat{n}$ as
\begin{eqnarray}
\hat{n} = (\sin\theta \cos\phi, \sin\theta \sin\phi, \cos\theta )
\end{eqnarray}
we can write 
\begin{eqnarray}
-is\int d{\cal A} = 
=-is\int d\phi (1- \cos\theta) = -is\int dt \;
(1-\cos\theta)
\; \dot{\phi}
\label{area}
\end{eqnarray}

The Hamiltonian, Eqn. (\ref{hamiltonian}), is a sum over a chain
of spins.
We must therefore extend the path integral to all sites. This is
done by 
indexing each of the coherent states with a position label, $x$,
and making
the substitution
\begin{eqnarray}
|\hat{n}(\tau_m)> \longleftrightarrow \bigotimes_x
|\hat{n}(\tau_m,x)>
\end{eqnarray}
Note that
\begin{eqnarray}
<\hat{n}(\tau,x)|\bvec{S}(x)|\hat{n}(\tau,x)> = s\hat{n}(\tau,x)
\end{eqnarray}
It is useful to write $\hat{n}(\tau,x)$ in terms of a staggered and 
a uniform part which are slowly varying in the limit of large $s$:
\begin{eqnarray}
\hat{n}(\tau,x) = (-1)^x \bphi(\tau,x) + \bvec{l}(\tau,x)/s
\end{eqnarray}
To leading order in $1/s$ and derivatives of the slowly varying
fields, this produces the constraints 
\begin{eqnarray}
\bphi(\tau,x) \cdot \bphi(\tau,x) = 1 \; \; \; \; \; \; \; 
\bphi(\tau,x) \cdot \bvec{l}(\tau,x) = 0
\end{eqnarray}
Setting $\Delta x=1$ (the fact that the fields vary slowly over 
this interval is justified a posteriori), 
we find that the leading contribution of the Hamiltonian to the action 
in the continuum limit is
\begin{eqnarray}
\frac{Js^2}{2} 
\int dx \; d\tau \; \left ( \left ( \partial_x \bphi \right )^2 + 
4\bvec{l}^2/s^2 \right )
\end{eqnarray}

We add up the phase terms by combining them in pairs: 
\begin{eqnarray}
-is{\cal A} = \frac{-is}{2} \int dx \; \left (
{\cal A}\left [ \bphi(\tau,x+1) + \bvec{l}(\tau,x+1)/s \right ]
+ {\cal A}\left [-\bphi(\tau,x) + \bvec{l}(\tau,x)/s \right ]
\right ) 
\label{area1}
\end{eqnarray}
Because ${\cal A}$ is an oriented area with respect to its 
argument, changing the sign of the argument also changes the
sign of the area. Eqn. (\ref{area1}) can be written,
\begin{eqnarray}
-is{\cal A} = \frac{-is}{2} \int dx \; \left (
{\cal A}\left [ \bphi(\tau,x) + \delta \bphi(\tau,x) \right ]
- {\cal A}\left [\bphi(\tau,x) \right ] \right )
\end{eqnarray}
where to leading order, $\delta \bphi(\tau,x) = \partial_x \bphi(\tau,x)
+ 2\bvec{l}/s$. This then gives,
\[-is{\cal A} = \frac{-is}{2} \int dx\; d\bphi \cdot \left (
\bphi(\tau,x) \times \delta \bphi(\tau,x) \right ) \]
\[= \frac{-is}{2} \int dx\; d\tau \; \bphi(\tau,x) \cdot \left (
\delta \bphi(\tau,x) \times \partial_{\tau} \bphi(\tau,x) \right ) \]
\begin{eqnarray}
= \frac{-is}{2} \int dx\; d\tau \; \left ( \bphi(\tau,x) \cdot 
( \partial_x \bphi(\tau,x) \times \partial_{\tau} \bphi(\tau,x) ) -
2\frac{\bvec{l}}{s}
\cdot ( \bphi(\tau,x) \times \partial_{\tau} \bphi(\tau,x) )
\right )
\end{eqnarray}
If we compactify $\bphi$ so that $\bphi \rightarrow$ constant
for $|x^2 + \tau^2| \rightarrow \infty$, and maintain the
constraint $\bphi^2 =1$ (valid to $1/s^2$), one can recognize the
integral
\begin{eqnarray}
Q = \frac{1}{4\pi} \int dt \; dx \; \bphi \cdot
\left ( \partial_{\tau} \bphi \times \partial_x \bphi \right )
\end{eqnarray}
as measuring the winding number of the sphere onto the sphere.
The integrand is
the Jacobian for the change of variables from compactified
coordinates on
the plane to those on $\bphi$-space (also a sphere).
Q is an integer corresponding to one of the countably many topologically
inequivalent ways there are to smoothly map the sphere onto the sphere;
thus the phase term can be written
as $-2\pi isQ$. For $s$ an integer, a sum over all possible
topological
configurations will not affect the path integral. For $s$
half-integer,
however, we can expect a drastic difference, since the path
integral will
be the difference between partition functions with even and odd
$Q$'s.
It is important to stress that this is a purely quantum
mechanical result
which has no analogue in the 2-D finite temperature classical
Heisenberg 
model (there is a well known equivalence between
$(d,1)$-dimensional
quantum field theory and $d+1$-dimensional finite temperature
classical
statistical mechanics \cite{polyakov}). A detailed discussion of
how
a half-integer $s$ will affect the physics will be omitted here;
the reader
is instead referred to \cite{leshouches} and references therein.

We can solve the equations of motion for $\bvec{l}$:
\begin{eqnarray}
\bvec{l} = -\frac{i}{gv} \left ( \bphi \times \partial_{\tau} \bphi \right )
\label{l_gen}
\end{eqnarray}
Not surprisingly, $\bvec{l}$ is the generator of rotations. After
integrating out the $\bvec{l}$ fields, the final action is,
\begin{eqnarray}
S =-2\pi isQ + \frac{Js^2}{2}
\int dx \; d\tau \; 
\left ( \partial_x \bphi \right )^2  +
\frac{Js^2}{2}
\int_0^{\beta}  d\tau \; dx \; \left ( \partial_{v\tau} \bphi
\right )^2
\label{nls_a}
\end{eqnarray}
Where we now define,
\begin{eqnarray}
v = 2Js  \; \; \; \; g = \frac{2}{s},
\end{eqnarray}
The action can be written 
\begin{eqnarray}
S  = 2\pi isQ + \frac{v}{2g} \int dx \; d\tau \;
\partial_{\mu} \bphi \partial^{\mu} \bphi
\label{action}
\end{eqnarray}
It is clear how $1/s$ corrections entered into the calculation
of the topological term. Moreover, we did not break parity to
derive (\ref{action}). 

We are interested in integer $s$ (in fact, $s=1$).
To this end we may ignore the topological
term in the action, as discussed, and consider the nonlinear 
${\bf \sigma}$-model:
\begin{eqnarray}
{\cal L} = \frac{Js^2}{2} \partial_{\mu} \bphi
\partial^{\mu} \bphi
\; \; \; \; \; \; \; \; \bphi^2 = 1
\label{lagr}
\end{eqnarray}
We now motivate the idea that,
contrary to spin-wave theory, this model features a gap
in its low energy spectrum. We first do this in the spirit of
reference \cite{polyakov}. We can deal with the constraint by
incorporating it
into the path integral as a Lagrange multiplier:
\begin{eqnarray}
{\cal Z} \propto 
\int {\cal D}\bphi {\cal D}\lambda e^{-\frac{Js^2}{2} \int d^2x
\;
\left ( \partial_{\mu} \bphi \partial^{\mu} \bphi 
+\lambda(\bphi^2 - 1)
\right ) }
\end{eqnarray}
The constraint is now enforced by the equation of motion for
$\lambda$. The
$\bphi$ fields can be integrated out in the usual way to give
\begin{eqnarray}
{\cal Z} \propto 
\int {\cal D}\lambda e^{\frac{Js^2}{2} \int d^2x \;
\lambda(\bold{x})
- \frac{N}{2} \log \mbox{det} || -\partial^2 + \lambda|| }
\end{eqnarray}
where $N$ is the number of components of $\bphi$.
As $N \rightarrow \infty$ the path integral is concentrated near
the smallest
value of the argument of the exponential. Minimizing this
argument with respect
to $\lambda$ we solve for the saddle point,
$\tilde{\lambda}$:
\begin{eqnarray}
\frac{Js^2}{2} = \frac{Nv}{2}  <x| \left \{ \frac{1}{-\partial^2 +
\tilde{\lambda}} \right \} |x> 
\label{saddle}
\end{eqnarray}
using the standard rules for functional differentiation. The RHS
of the above
is simply the Green's function for a boson field with mass
$v\sqrt{\tilde{\lambda}}$;
\begin{eqnarray}
<x| \left \{ \frac{1}{-\partial^2 + \tilde{\lambda}} \right \}
|x> = 
\int \frac{d^2k}{(2\pi)^2} \frac{1}{k^{\mu}k_{\mu} + \tilde{\lambda}}
\end{eqnarray}
\[=\frac{1}{4\pi} \log{\frac{\Lambda^2}{\tilde{\lambda}}} \]
where $\Lambda$ is an ultraviolet cutoff, $d^2k=dk\;d\omega/v$, and
$k^{\mu}k_{\mu} = k^2 + \omega^2/v^2$. Solving for the square of the mass,
$\tilde{\lambda}$:
\begin{eqnarray}
\tilde{\lambda} = \Lambda^2 e^{-4\pi Js^2/N}
\end{eqnarray}

Another way to see the presence of a mass gap is to integrate out
ultraviolet modes and apply the renormalization group.
We start with the Lagrangian Eqn. (\ref{lagr}) and parametrize
the fluctuations in terms of slow and fast modes. One then integrates
out the fast fields. This calculation is logarithmically
infrared divergent. One then renormalizes by subtracting out
the offending terms from the effective Lagrangian. 
Equivalently, one can achieve the same
effect to the same order in perturbation theory by redefining the
coupling constant in terms of its bare value. A calculation of this
sort (for the $O(N)$ model) is done in reference \cite{polyakov}.
The renormalized coupling constant becomes
\begin{eqnarray}
g(L) \approx \frac{g_0}{1- \frac{g_0}{2\pi}\ln L} 
\end{eqnarray}
With $g_0=2/s$, we now see that the coupling constant is of order
unity for length scales
\begin{eqnarray}
\xi \approx e^{\pi s}
\end{eqnarray}
Keeping in mind that this is a `Lorentz invariant' theory, there
must be a corresponding mass scale, $\Delta$: 
\begin{eqnarray}
\Delta \propto e^{-\pi s}
\end{eqnarray} 

There are other similar heuristic calculations that suggest a
mass gap; 
none are ironclad, but the sum of them together makes
for strong evidence that indeed the $s=1$ 1-D Heisenberg
antiferromagnet is disordered at all temperatures and is well
described by the NL$\sigma$ model. Better justification comes from
exact $S$-matrix results and numerical simulations.
The exact $S$-matrix results are due to work by Zamolodchikov and
Zamolodchikov \cite{zam}, and Karowski and Weisz \cite{karo}. The
$O(3)$ invariance of the NL$\sigma$ model allows for an infinite
number of conservation laws. These imply strong constraints on
$S$-matrix elements and, consequently, on on-shell Green's
functions. One characteristic of such an $S$-matrix is
factorizability. This means that $N$-particle scattering 
can be expressed as products of 2-particle scattering matrix elements.
The simplest such $S$-matrix consistent with the symmetries of
the NL$\sigma$ model has  a triplet of massive soliton states
with an effective repulsive local interaction. This conjecture
has been checked in perturbation theory in $1/N$ (for the $O(N)$
NL$\sigma$ model \cite{karo}) to
order $1/N^2$. 

Numerical results have been pursued since Haldane made his
conjecture in 1983 \cite{num,white,goll_n,sorensen}. They have
all essentially confirmed Haldane's picture and the validity of
the NL$\sigma$ model. To date, the best numerical work has been
due to White's method of the density matrix renormalization group
\cite{white} and recent exact diagonalization \cite{goll_n}. The
former predicts a gap $\Delta = .41050(2) J$, while the latter
has $\Delta = .41049(2) J$. Numerical investigations of the spin
operator structure factor \cite{sorensen}, $S(k)$, for the
isotropic chain show remarkable agreement with the `exact' 
$S$-matrix result for two magnon production over a region larger
than expected (two magnon production is
known to dominate at low momenta, $k\leq .3\pi$, from numerical 
studies \cite{sorensen}. This can be probed in neutron scattering 
experiments \cite{ma}.)
For higher momenta, one must include one magnon contributions 
which dominate as $k \rightarrow \pi$. The intermediate region in 
momentum space, $.3\pi \leq k \leq .8\pi$, is not expected to be
well represented by the NL$\sigma$ model; this is because the fields,
$\bphi$ and $\bvec{l}$ describe low energy (and therefore large
wavelength) excitations about $k=\pi$ and $k=0$, respectively.
The same study 
also determined the correlation length, $\xi = 6.03(1)$, the
velocity, $v/J \sim 2.5$ and the coupling constant, $g \sim 1.28$
in rough agreement with the $1/s$ expansion result \cite{khvesch}
$g \sim 1.44$ and the value derived above, $g = 2/s = 2$.

This ends the introductory discussion of the NL$\sigma$ model. A
more in-depth approach will be taken when we consider anisotropies
and develop the necessary tools to calculate the NMR relaxation
rate in Chapter 2.

\section{Boson Model}

Although the NL$\sigma$ model is convincingly accurate in
describing the low energy physics of the Heisenberg 1-D
antiferromagnet, it has several deficiencies. First, off-shell
Green's functions are not known; and second, anisotropies are not
easily tractable within the framework of the model (the $S$-
matrix is no longer factorizable, as earlier discussed,
since one loses the infinite
number of conservation laws). A happy compromise which contains
all of the qualitative aspects of the NL$\sigma$ model and yet
allows for more computability and generalization is the 
Landau-Ginsburg boson model \cite{ian_stamp}
\begin{eqnarray}
{\cal H}(x) = 
\frac{v}{2} \vec{\bold{\Pi}}^2 + \frac{v}{2} \left (
\frac{\partial \bphi}{\partial x} \right )^2 + \frac{1}{2v}
\Delta^2 \bphi^2 + \lambda |\bphi|^4
\label{H_Lan}
\end{eqnarray}
where the constraint $\bphi^2 = 1$ has been relaxed in the 
Lagrangian of the NL$\sigma$ model, and a $\phi^4$ interaction has
been added for stability. The Hamiltonian, (\ref{H_Lan}),
possesses the correct symmetries, three massive low energy
excitations, and a repulsive weak interaction. As with the
NL$\sigma$ model, the field $\bphi$ acts on the ground state to
produce the triplet of massive excitations or magnons. We note that
this model becomes exact in taking the $N \rightarrow \infty$ limit of
the $O(N)$ NL$\sigma$ model \cite{polyakov}
(recall that $N$ is the number of components
of the field $\bphi$). As in
Eqn. (\ref{l_gen}), the
generator of rotational  symmetry (the uniform part of the spin
operator) is 
\begin{eqnarray}
\vec{\bold{l}} = \bphi \times \vec{\bold{\Pi}}
\end{eqnarray}
where $\bPi = \frac{\partial \bphi}{\partial vt}$ (we absorbed the 
coupling constant,$g$, into the definition of $\bphi$ in (\ref{H_Lan}).)
Expanding $\bphi$ in terms of creation and annihilation operators, we see
that $\vec{\bold{l}}$ acts as a two magnon operator producing or
annihilating a pair, or else flipping the polarization of a
single magnon. This picture is obvious in this simpler model,
whereas the same analysis is only confirmed by the exact 
$S$-matrix results and the gratifying agreement with numerical work
in the case of the NL$\sigma$ model. 
The gap, $\Delta$ can be phenomenologically fitted to experiments
such as neutron scattering as can be the `velocity of light',
$v$. Including on-site anisotropy
\begin{eqnarray}
H_{\mbox{aniso}} = \sum_i \left ( D(S^z_i)^2 + E( (S_i^x)^2 -
(S_i^y)^2) \right )
\label{H_aniso}
\end{eqnarray}
simply amounts to introducing three phenomenological masses. This
will be discussed in more detail in Chapter 2.

On comparison of the predictions of both models one finds overall
qualitative agreement in studies of form factors
\cite{sorensen,wes}. As one moves away from zero wave vector
the agreement between the models weakens. This makes for one of
the disadvantages of the bosons. Also, in attempting to calculate
certain Green's functions, such as the staggered field
correlation function, one is forced to rely on perturbation
theory in $\lambda$. Although $\lambda$ can be phenomenologically
fitted, there is much ambiguity in choosing the interaction term.
One can equally put in by hand any positive polynomial term in
$\bphi^2$. This is because the fields carry no mass dimension
making all polynomial interaction terms relevant. It should be
understood that this model is phenomenological
and is introduced for its simplicity. In the final analysis, justification
for its use must come from numerical and real experiments.

\section{Fermion Model}

Before introducing the next model, 
we would like to begin by apologizing for the cryptic description of
the concepts
to be mentioned in this section. A deeper understanding would require a
diversion into conformal field theory
tangential to the main lines of the thesis. Instead, the
reader is invited to investigate the literature.

There is another model exhibiting some of the desirable
properties of the boson model. This is an analogue of the 
Landau-Ginsburg model but phrased in terms of a triplet
of relativistic fermions:
\[ {\cal H}(x) = \frac{1}{2}\bpsi_L iv\frac{d}{dx} \cdot \bpsi_L -
\frac{1}{2} \bpsi_R iv\frac{d}{dx} \cdot \bpsi_R + \]
\begin{eqnarray}
\frac{i}{2}\Delta(\bpsi_R \cdot \bpsi_L - \bpsi_L \cdot \bpsi_R)
+ \lambda (\bpsi_L \times \bpsi_L) \cdot (\bpsi_R \times \bpsi_R)
\label{ferms}
\end{eqnarray}
The fields $\bpsi$ are Majorana (Hermitean) fermions with equal
time anticommutation relations
\begin{eqnarray}
\{ \psi^i_S(x) , \psi^j_{S'}(y) \} =
\delta_{SS'}\delta^{ij} \; \delta(x-y) \; \; \; \; \; \; \; \;
S,S' = L,R
\label{fecomm}
\end{eqnarray}
The $L$ and $R$ label left and right moving fields, respectively.
This model is not trivially related to
either of the models described above; it was first introduced by Tsvelik
\cite{tsev} to achieve better agreement with experimental data on
the anisotropic Haldane Gap material NENP.
The motivation comes from 
a model sitting on the boundary between the Haldane
phase and a spontaneously dimerized phase \cite{dimer}, with the
Hamiltonian
\begin{eqnarray}
H = J\sum_i \left [ \bvec{S}_i \cdot \bvec{S}_{i+1} - (\bvec{S}_i
\cdot \bvec{S}_{i+1})^2 \right ]
\label{bethe}
\end{eqnarray}
This Bethe Ansatz integrable Hamiltonian features a gapless
spectrum and has a continuum limit equivalent 
to a $k=2$ Wess-Zumino-Witten (WZW)
NL$\sigma$ model. This, in
turn, is a conformal field theory \cite{ginsparg} equivalent to
three decoupled critical Ising models. The well known mapping of
the critical Ising model to a massless free Majorana fermion
\cite{itzy} brings us to write (\ref{bethe}) as
\begin{eqnarray}
{\cal H}(x) = \frac{iv}{2} \left ( \bvec{\psi}_L \cdot
\frac{\partial}{\partial x} \bvec{\psi}_L - \bvec{\psi}_R \cdot
\frac{\partial}{\partial x} \bvec{\psi}_R \right )
\end{eqnarray}
Reducing the biquadratic coupling in (\ref{bethe}) moves the
Ising models away from their critical point. Symmetry allows the
addition of interactions corresponding to
mass and four fermi terms, as in 
Eqn. (\ref{ferms}). The four fermi term proportional to $\lambda$
is the only marginal one allowed by $O(3)$ symmetry. It will generally
be ignored or treated perturbatively, in a similar phenomenological
spirit to that of the Landau-Ginsburg boson model (ultimate justification
for this, as for the boson model, comes from numerical and real
experiments). For weak
interactions (which is the case assumed) all Green's functions
will have simple poles at the phenomenological masses and will be
trivial on-shell. The off-shell behaviour depends on the
interaction terms chosen and is therefore very much model dependent.

It can easily be checked that (\ref{fecomm}) gives the right
commutation relations for the $SU(2)$ algebra, 
$[l^i(x),l^j(y)] = i\delta(x-y)
\epsilon^{ijk}l^k(x)$, with 
\begin{eqnarray}
\bvec{l} = \frac{-i}{2} \left ( \bpsi_L \times \bpsi_L + \bpsi_R
\times \bpsi_R \right ) 
\label{l_ferms}
\end{eqnarray}
This allows us to identify $\bvec{l}$ with the generator of global
rotations or the uniform part of the spin, $\bvec{S}$. Expanding
$\bpsi_R$ and $\bpsi_L$ in terms of creation and annihilation
operators, we see that, here too, $\bvec{l}$ is quadratic in such
operators.
Notice that this representation for $\bvec{l}$ {\em does not
couple} left and right movers. This is in sharp contrast to the
boson or NL$\sigma$ models (where one can write the boson
operator as a sum of left and right moving parts). We will later
see that this point can potentially give experimental predictions
which will contrast between the models.

The particles created by the fields, $\bpsi$, are identified with
massive magnons. The masses can be fixed by hand to agree with
the experimental dispersions so that on-site anisotropy terms 
coming from Eqn. (\ref{H_aniso}) can be easily parametrized,
as in the boson model. Other interaction terms which might arise
from breaking the symmetry are usualy ignored for ease of calculation.
As always, ultimate justification for this is found in numerical and real
experiment.

As mentioned above, $\bvec{l}$ is again a two
magnon operator. It is also possible to represent the staggered
magnetization (the analogue of $\bphi$) in this approach, but it
is considerably more complicated (one can use bosonization
techniques \cite{ginsparg}). Near the massless point, this
operator reduces to the fundamental field of the WZW model, or
equivalently to products of the order and disorder fields,
$\bvec{\sigma}$ and $\bvec{\mu}$, of the three Ising models
\cite{fateev,halaff}. These operators are highly non-local with
respect to the fermion fields. The corresponding correlation
functions can be expressed in terms of products of Painlev\'e
functions \cite{itzy,mccoy}. They exhibit poles at the fermion
masses together with additional structure at higher energy.
Unlike the free boson model, a simple interpretation of the
staggered magnetization density as a single magnon operator
doesn't hold. This complicates the use of this model.

One way to 
justify the use of the fermion model without resorting to complicated
explanations is to notice that in the long wave length limit of the
$O(3)$ symmetric case, all models are in agreement (see
Chapter 2). For smaller wave lengths, the different models correspond
to different continuum representations of the lattice model. 
$O(3)$ symmetry is broken differently in each model (for example, see
the different results for matrix elements of $\bvec{l}$ in Chapter 2).
The idea is that we have three (two, for lesser symmetry) workable
descriptions whose ultimate merits can only be decided phenomenologically.

\section{Nuclear Magnetic Relaxation Rate}

Experiments on condensed matter systems typically measure
observables which are directly related to Green's functions. This
is no surprise since most such experiments measure the response
of the system to an external probe. This is in contrast with
particle physics experiments which usualy examine the nature of
scattering into asymptotic states. Formally the difference is
that particle physicists measure time-ordered Green's functions
while their friends in condensed matter physics measure retarded
Green's functions. The nuclear magnetic relaxation rate, $1/T_1$,
measures the local correlations of the system at
low frequency. The probe is the nucleus of some atom in
the sample which has a non-zero nuclear magnetic moment weakly
coupled to the system of interest. In the case of the Heisenberg
1-D antiferromagnet, we assume that in addition to the spin
Hamiltonian, $H_S$, there is also Zeeman coupling, $H_Z$, to a
uniform magnetic field, $\bvec H$, by both the nuclear and
Heisenberg spins, and that there is
a hyperfine coupling between the two systems,
$H_{\mbox{Hyper}}$. We also assume for simplicity that the
nuclear spins do not directly couple to each other.
\begin{eqnarray}
H_{\mbox{Total}} = H_S + H_Z + H_{\mbox{Hyper}} \equiv H_0 +
H_{\mbox{Hyper}} \nonumber \\
=  H_S - \sum_i \mu_B \bvec{H} \cdot {\bf G}_e \cdot \bvec{S}_i -
\sum_j \mu_N \bvec{H} \cdot {\bf G}_N \cdot \bvec{I}_j + \sum_{i,j}
\bvec{S}_i \cdot {\bf A}_{ij} \cdot \bvec{I}_j
\end{eqnarray}
${\bf G}_e$ and ${\bf G}_N$ are the gyromagnetic tensors for the
electron and nuclear spins, respectively. ${\bf A}_{ij}$ is the
hyperfine tensor coupling the nuclear spin on site $j$ to the
electronic spin on site $i$. We now define the characteristic
frequency
\begin{eqnarray}
\omega_N \equiv \mu_N |\bvec{H}| 
\end{eqnarray}

In nuclear magnetic resonance (NMR) experiments
one strives to temporarily achieve a non-equilibrium
population difference
between nuclear spins with different spin eigenvalues along the
uniform field direction. This is normally achieved with pulses of
RF electromagnetic radiation possessing ac magnetic fields
perpendicular to the externally applied uniform magnetic field. As is well
known, a resonance phenomenon occurs at RF-frequencies near
$\omega_N$ (in reality it is easier to tune the uniform field to
resonate with a fixed RF field). 

In the presence of a non-equilibrium occupation of states,
the nuclear spins ``relax'' towards an equilibrium configuration
by making transitions between states of different spin eigenvalues.
This would not normally be possible if
the nuclear spins were completely free. Coupling to another
system is necessary in order to conserve energy during the
transitions. The energy given off or absorbed
must induce a corresponding transition into a
different energy state in the system which couples to the nuclear
spins. Let us illustrate the situation with an $s=1/2$ nuclear
spin. In the absence of hyperfine interactions, we assume that $I_z$ is
a good quantum number (where $z$ is the direction of the static magnetic
field), and that $\bold{G}_N$ is isotropic (these are
generally good assumptions).
The rate equation for the number of nucleii, $N_+$, with 
$I^z=+\frac{1}{2}$ is
\begin{eqnarray}
\frac{d \; N_+}{dt} = -N_+ \Omega_{+ \rightarrow -} +
N_-\Omega_{- \rightarrow +}
\end{eqnarray}
Where the transition probabilities per unit time are given by
$\Omega_{\pm \rightarrow \mp}$. We can rewrite this in terms of
the total number of spins, $N$ and the population difference,
$n$:
\begin{eqnarray}
\frac{d \; n}{dt} = N (\Omega_{- \rightarrow +} - 
\Omega_{+ \rightarrow -} ) - n (\Omega_{- \rightarrow +} + 
\Omega_{+ \rightarrow -} )
\label{rate}
\end{eqnarray}
Now, if we define 
\begin{eqnarray}
\frac{1}{T_1} \equiv (\Omega_{- \rightarrow +} + 
\Omega_{+ \rightarrow -} )
\end{eqnarray}
then we see that in the limit that the transition rates only
depend on time scales much shorter than those characteristic of the
experimental probe, and in the limit of linear response (ie.
$\Omega$ is independent of $n$) the solution to (\ref{rate}) is
\begin{eqnarray}
n(t) = n_0 + a\; e^{-t/T_1}
\end{eqnarray}
Where $n_0$ is the equilibrium population difference (at finite
temperature, states with an energy difference will necessarily
have a population difference). We see that the relaxation rate,
$1/T_1$, describes the evolution of the nuclear system towards
thermal equilibrium, or likewise, the decay of the population
inversion magnetization achieved by RF pulses in NMR.

We now derive an expression for the rate, $1/T_1$. For a system
with more general $I$, $1/T_1$ for a transition from an initial
state with $I^z=m$ to one with $I^z=m+1$ is normalized by the
factor $I(I+1)-m(m+1)$. To begin, we need an expression for the
transition rate $\Omega_{m \rightarrow m+1}$
describing a nuclear spin at site $j$ starting in the state with
$I^z=m$ and ending up with $I^z = m+1$. It does not matter which
nuclear spin we pick if we assume translational invariance; since
there is no nuclear spin-spin coupling\footnote{In NENP, the dipolar
nuclear spin-spin couplings are roughly 200 times smaller than the
hyperfine coupling.}, the relaxation rate for
one is the relaxation rate for the whole system.
Let us assume that the initial and final Heisenberg spin state 
are given by the labels $n$ and $n'$, respectively. Then 
Fermi's Golden Rule gives
\[ \Omega_{I^z_j=m,n \rightarrow I^z_j=(m+1),n'} = \]
\begin{eqnarray}
2\pi |<I^z_j=(m+1),n'|H_{\mbox{Hyper}}|I^z_j= m,n>|^2
\delta(E_{I^z_j=(m+1),n'} - E_{I^z_j=m,n})
\frac{e^{-E_{n}/T}}{{\cal Z}}
\end{eqnarray}
Notice that we multiplied the normal expression for the Golden
Rule by the Boltzmann probability that the Heisenberg spin system
is in the initial state, $|n>$. The only part of
$H_{\mbox{Hyper}}$ which will contribute is $S^{\nu}_i
{\bf A}^{\nu \; -}_{ij} I^+_j$:
\[ \Omega_{I^z_j=m,n \rightarrow I^z_j=(m+1),n'} =  \]
\begin{eqnarray}
2\pi (I(I+1) - m(m+1)) |\sum_{i,\nu} {\bf A}^{\nu \; -}_{ij}
<n'|S^{\nu}_i|n>|^2 \delta(E_{n'} - E_{n} - \omega_N) 
\frac{e^{-E_{n}/T}}{{\cal Z}}
\end{eqnarray}
The analogous expression for $\Omega_{(m+1)n' \rightarrow
I^z_j=m,n}$ is
\[ \Omega_{I^z_j=(m+1),n' \rightarrow I^z_j=m,n} = \]
\begin{eqnarray}
2\pi (I(I+1) - m(m+1)) |\sum_{i,\nu} {\bf A}^{\nu \; +}_{ij}
<n|S^{\nu}_i|n'>|^2 \delta(E_{n'} - E_{n} - \omega_N) 
\frac{e^{-E_{n'}/T}}{{\cal Z}}
\end{eqnarray}
Since ${\bf A}^{\nu \; +}_{ij}=({\bf A}^{\nu \; -}_{ij})^{\ast}$ we
get for the relaxation rate
\begin{eqnarray}
\left ( 1/T_1 \right )_{n \leftrightarrow n'} = 
2\pi |\sum_{i,\nu} {\bf A}^{\nu \; -}_{ij}
<n'|S^{\nu}_i|n>|^2 \delta(E_{n'} - E_{n} - \omega_N) 
\frac{(e^{-E_{n'}/T} + e^{-E_{n}/T})}{ {\cal Z}}
\end{eqnarray}
We now sum over all possible transitions to arrive at $1/T_1$:
\begin{eqnarray}
\frac{1}{T_1} = 2\pi\sum_{n,n'}|\sum_{i,\nu}
{\bf A}^{\nu \; -}_{ij}
<n'|S^{\nu}_i|n>|^2 \delta(E_{n'} - E_{n} - \omega_N) 
\frac{(e^{-E_{n'}/T} + e^{-E_{n}/T})}{ {\cal Z}}
\label{T_1_e}
\end{eqnarray}
Since the sum over the states $n$ and $n'$ is a trace over states
in the Heisenberg spin system, we can conveniently restrict 
ourselves to that system only and write
\begin{eqnarray}
\frac{1}{T_1} = \int_{-\infty}^{\infty} dt \; e^{-i\omega_N t} 
<\left \{ \sum_{i,\nu} {\bf A}^{\nu \; -}_{ij} S^{\nu}_i(t) ,
\sum_{m,\mu} {\bf A}^{\mu \; +}_{mj} S^{\mu}_m(0) \right \} >
\label{T_1}
\end{eqnarray}
where $< >$ denotes a thermal average. This is the famous
expression derived in \cite{moriya}. As promised,
when $\bold{A}_{ij}$ is well localized,
we see that $1/T_1$ is related to the low frequency local 
correlation function.


\chapter{Details of the Models}
\resetcounters

In this chapter we discuss in detail the three models introduced in the
last chapter. We will derive the necessary tools to
calculate the relaxation rate $1/T_1$ and mention some pertinent
issues which can be important in investigating 1-D Heisenberg 
antiferromagnets (1DHAF's) using other means.

\section{NL$\sigma$ Model: Temperature and Field Dependence of
the
Spectrum; Exact Results}

\label{sec_2.1}

\subsection{Temperature Dependence of the Gap}

\label{sec_2.1.1}

In this section and the next we will discuss how the excitation
energies of the lowest modes change with varying parameters. This
is especially important when one chooses to perform calculations
using the Landau-Ginzburg boson model. Since this model
implicitly adopts the gap parameters from the NL$\sigma$ model,
any dependence of the gaps on magnetic field or temperature must
first be calculated within the framework of the NL$\sigma$ model.
The results can become useful in interpreting experimental data
using the simple boson model.

We would like to begin by extending some recent work by Jolic{\oe}ur
and Golinelli \cite{joli} on the temperature dependence of the
low energy spectrum. It may seem strange or even contradictory
at first sight to speak of a spectrum as being temperature
dependent. What one must keep in mind is that the low energy
description of the NL$\sigma$ model as three massive bosons
with relativistic dispersion, is an effective one. The true
excitations of the model are collective, and if we insist on
maintaining a single particle description we should not be
surprised that the effective single particle interactions will be
temperature dependent (as, therefore, will be the effective 
one-particle spectrum). A similar approach is taken in BCS theory
where the BCS gap has a temperature dependence arising from a
consistency condition.

In Chapter One we introduced a consistency equation, 
Eqn. (\ref{saddle}), for the
classical or saddle point value of the Lagrange multiplier field,
$\lambda$, in the NL$\sigma$ model. This result always holds to
lowest order in the fluctuating field $\lambda(x)$, regardless of
the value of $N$ in the large $N$ expansion. Of course, it only
becomes exact for $N \rightarrow \infty$. We can also look at the
consistency equation as a constraint equation guaranteeing that
the $\bphi$ two point function is unity when evaluated at the
origin; when $N=3$,
\begin{eqnarray}
1=<\bphi(x) \cdot \bphi(x)> = G^2(0) = \frac{3v}{J}
\int \frac{d^2k}{(2\pi)^2}
\frac{1}{k^{\mu}k_{\mu}+ \frac{\Delta^2}{v^2}}
\label{constraint}
\end{eqnarray}
where we've assumed a renormalized mean value for $\lambda$
\cite{polyakov}. Notice, also, that we're choosing to work in
Euclidean space. One can likewise see that the constraint
equation is nothing more than a minimization of the zero point
energy of the system with respect to the fluctuating field
$\lambda$:
\begin{eqnarray}
E_0 = 3\int \frac{dk}{2\pi} \omega_k \; \;   -  \; \; \lambda J
\end{eqnarray}

where $\omega_k = v\sqrt{k^{\mu}k_{\mu} + \lambda}$.
If we choose to add on-site anisotropies to the model, as in
Eqn. (\ref{H_aniso}), then the
contribution to the Lagrangian (modulo irrelevant terms which
also break `Lorentz invariance') is

\begin{eqnarray}
-D(S_i^z)^2 - E( (S_i^x)^2 - (S_i^y)^2) \rightarrow
-D(\phi^z(x))^2 - E( (\phi^x(x))^2 - (\phi^y(x))^2) )
\end{eqnarray}
We can always find some axes, $xyz$, so that the addition to the
Lagrangian is in the above form. We can read off the new
constraint equation on the Green's function
\[ 1 = \frac{v^2}{J} \int \frac{dk \; d\omega}{(2\pi)^2} 
\left ( \frac{1}{\omega^2+ v^2k^2 + \Delta^2 + \Delta_D^2} \right. \]
\begin{eqnarray}
\left. + \frac{1}{\omega^2+ v^2k^2 + \Delta^2 + \Delta_E^2} +
\frac{1}{\omega^2+ v^2k^2 + \Delta^2 - \Delta_E^2}  
\right )
\label{eq2.3}
\end{eqnarray}
Where we've once more assumed renormalized masses with
the correspondence,
$v^2\lambda + 2vD \leftrightarrow \Delta^2 + \Delta_D^2$ and
$v^2\lambda \pm 2vE \leftrightarrow \Delta^2 \pm \Delta_E^2 $. As
usual, making this model temperature dependent consists of
replacing the integral over $\omega$ by a corresponding sum over
Matsubara frequencies.
Eqn. (\ref{eq2.3}) becomes,
\[ 1 = \frac{Tv^2}{J} \sum_n \int \frac{dk}{2\pi} \left (
\frac{1}{\omega_n^2+ v^2k^2 + \Delta^2(T) + \Delta_D^2} \right. \]
\begin{eqnarray}
\left. + \frac{1}{\omega_n^2+ v^2k^2 + \Delta^2(T) + \Delta_E^2} +
\frac{1}{\omega_n^2+ v^2k^2 + \Delta^2(T) - \Delta_E^2} \right )
\label{eq2.4}
\end{eqnarray}
In summing over the frequencies we use,
\begin{eqnarray}
\frac{1}{\beta} \sum_n \frac{1}{\omega_n^2 + v^2k^2 + m^2} =
-\frac{1}{2} \frac{\cot(\frac{i\beta \omega_k}{2})}{i\omega_k}
\nonumber \\
= \frac{1}{2\omega_k} \left ( 1 + \frac{2}{e^{\beta \omega_k} -1}
\right )
\end{eqnarray} 
where $\omega_k = \sqrt{v^2k^2 + m^2}$.
We also need the following two integrals
\begin{eqnarray}
\int dk \; \frac{1}{\omega_k}\left 
( \frac{2}{e^{\beta \omega_k} -1}
\right ) \approx 2\int dk \; \frac{ e^{-\beta m -\beta
k^2/2m}}{\sqrt{k^2 + m^2}} \nonumber \\
= \frac{2}{v} \sqrt{\frac{2\pi T}{m}} e^{-m/T}
\end{eqnarray}
\begin{eqnarray}
\int \frac{dk}{\omega_k} \approx \frac{2}{v}\log(2\Lambda v/m)
\end{eqnarray}
where $\Lambda$ is an ultraviolet cutoff and we made the
approximation $\beta m \gg 1$. Gathering all of the above, we can
set Eqn. (\ref{eq2.3}) equal to Eqn. (\ref{eq2.4}) to arrive at
\begin{eqnarray}
\log \left ( \frac{ \Delta^2_z(T) \Delta_+^2(T) \Delta^2_0(T) }
{\Delta^2_z \Delta^2_+ \Delta^2_- } \right ) \approx
2\sqrt{2\pi T} \left [ \frac{e^{-\Delta_z/T}}{\sqrt{\Delta_z}} +
\frac{e^{-\Delta_+/T}}{\sqrt{\Delta_+}} +
\frac{e^{-\Delta_-/T}}{\sqrt{\Delta_-}}  \right ]
\label{gaps_T}
\end{eqnarray} 
where $\Delta_z = \sqrt{\Delta^2 + \Delta_D^2}$ and
$\Delta_{\pm} = \sqrt{\Delta^2 \pm \Delta_E^2}$. Eqn.
(\ref{gaps_T}) implies a cubic equation for the square of the temperature
dependent gap, $\Delta^2(T)$. Once more making the approximation,
$\beta \Delta_- \gg 1$ we can linearize the equation and solve
for $\Delta(T)$:
\begin{eqnarray}
\Delta(T) - \Delta \approx
\frac{\sqrt{2\pi T}}{\Delta}
\left [ \frac{e^{-\Delta_z/T}}{\sqrt{\Delta_z}} +
\frac{e^{-\Delta_+/T}}{\sqrt{\Delta_+}} +
\frac{e^{-\Delta_-/T}}{\sqrt{\Delta_-}}  \right ] \left(
\frac{1}{\Delta_z^2} + \frac{1}{\Delta_-^2} +
\frac{1}{\Delta_+^2} \right )^{-1}
\label{Tgap}
\end{eqnarray}
In the $O(3)$ symmetric case this reduces to the formula derived in
\cite{joli}. We would like to say a few things about Eqn.
(\ref{Tgap}) before going on to the next section. First, notice
that we implicitly assumed that only the expectation value of the
fluctuating field $\lambda$ acquired a temperature
dependence. The renormalized values of the anisotropies,
$\Delta_D^2$ and $\Delta_E^2$ do not. This is because
renormalization occurs at $T=0$ first. At non-zero temperatures
the free energy may acquire a term linear in the fluctuations of
$\lambda$; the constraint equation, Eq. (\ref{constraint}),
amounts to cancelling that contribution in the Lagrangian. The
exponential terms logically appear as a result of calculating 
\begin{eqnarray}
<\lambda> = \mbox{Tr} \left [ e^{-\beta H} \lambda \right ] 
\end{eqnarray}
and then subtracting $<\lambda> \bphi^2$ from the Lagrangian.
We would also like to point out that the validity condition for
this analysis, $\beta \Delta_m \gg 1$, where $\Delta_m$ is the smallest
gap, is more robust than seems. It is
well known that for a value of the anisotropy, $D \sim  J$
\cite{schulz, khvesch}, the lower gap closes and the system goes
through a critical point, into a phase with a new singlet ground state
(the order parameter is a non-local operator in spins, and in
fact, this transition is not reproduced correctly via the
NL$\sigma$ model) and a gap. At large negative values, $D \sim -.4J$,
the system goes through an Ising transition into an antiferromagnetically
ordered phase.
The bottom line is that $|D|$ must be small
in comparison with $\Delta$. Similar, but more obvious,
cautionary remarks apply to $E$. Moreover, as discussed in
\cite{joli}, the NL$\sigma$ model is not expected to remain
valid at temperatures of order twice the gap. This is because the
model does not exhibit a maximum in the heat capacity and
in the magnetic susceptibility as shown in numerical studies
at these temperatures.

Finally, notice that the difference in gaps will close as $T$
increases. This is no surprise since at high temperatures the
mass scales are irrelevant and we expect a restoration of $O(3)$
symmetry. 

\subsection{Field Dependence of the Gap}

\label{sec_2.1.2}

      We will be interested in adding a magnetic field term to
the Hamiltonian. To do so consistently, we must couple the
magnetic field to the generator of rotations (the total spin
operator), $\sum_i \bvec{S}_i$. In terms of the continuum fields,
we couple the magnetic field, $\bvec{H}$, to $\bvec{l}$ via, 
$-g_e \mu_B \bvec{H} \cdot \int dx \; \bvec{l}(x)$,
and add this to the NL$\sigma$ Hamiltonian. 
The corresponding Euclidean space
Lagrangian is
\begin{eqnarray}
{\cal L} = -  \frac{1}{2g} \left ( 
| \partial \bphi /\partial t + \vec{h}
\times \bphi|^2 + v^2 (\partial \bphi /\partial x)^2 - 
2vD(\phi^z)^2 - 2vE( (\phi^x)^2 - (\phi^y)^2 ) \right )
\label{h_lag}
\end{eqnarray}
\[ \bphi^2 = 1 \]
where $\vec{h} = g_e \mu_B \bvec{H}$.
In the $O(3)$ and $U(1)$ symmetric cases, where the
field really couples to a conserved charge, no other terms are allowed
in (\ref{h_lag}). In the case of lower symmetry, we retain this as the
simplest form, realizing that other, more complicated terms may arise.
This time, when we integrate out the $\bphi$
fields, the eigenvalues of the $\bphi$ propagator are not as
trivial. However, if we assume that the field is placed along a
direction of symmetry, say the $z$-direction, then the $\mbox{Tr}
\log$ will be over eigenvalues of the matrix
\begin{eqnarray} \left (
\begin{array}{ccc}
\omega^2 + v^2k^2 + \Delta^2 - h^2 + \Delta^2_E & 2\omega h & 0 \\ 
-2\omega h & \omega^2 + v^2k^2 - h^2 + \Delta^2 - \Delta_E^2 & 0 \\
0 & 0 & \omega^2 + v^2k^2 + \Delta^2 + \Delta_D^2 
\end{array}
\right )
\end{eqnarray}
Where, again, we've assumed renormalized values for the masses.
The eigenvalues, $\eta_i$, are
\[ \eta_3 = \omega^2 + v^2k^2 + \Delta^2(h) + \Delta_D^2  \]
\begin{eqnarray}
\eta_{\pm} = \omega^2 + v^2k^2 - h^2 + \Delta^2(h) \pm
\sqrt{-4h^2 \omega^2 + \Delta_E^4}
\end{eqnarray}
We can write the constraint equation as
\[ \int \frac{d^2k}{(2\pi)^2} 
\left ( \frac{1}{\omega^2+ v^2k^2 + \Delta^2 + \Delta_D^2} +
\frac{1}{\omega^2+ v^2k^2 + \Delta^2 + \Delta_E^2} +
\frac{1}{\omega^2+ v^2k^2 + \Delta^2 - \Delta_E^2}  
\right ) \]
\begin{eqnarray}
= \int \frac{d^2k}{(2\pi)^2} 
\left ( \frac{1}{\eta_3} + \frac{1}{\eta_+} + \frac{1}{\eta_-}
\right )
\label{oops}
\end{eqnarray}
The integrand on the right hand side has poles at the negative 
solutions of the equations of motion
\[ \omega_3^2 = \left ( k^2 + \Delta^2(h) + \Delta_D^2 \right ) \]
\begin{eqnarray}
\omega_{\pm}^2 =  \left ( k^2 + h^2 + \Delta^2(h) \pm
\sqrt{ 4h^2(k^2 + \Delta^2(h)) + \Delta^4_E} \right )
\label{poles}
\end{eqnarray}
Integrating over these poles gives
\begin{eqnarray}
\int \frac{dk}{2\pi} \left ( \frac{1}{2\omega_3} +
\frac{\omega_+^2 - k^2  - \Delta^2(h) + h^2}{\omega_+
(\omega_+^2 - \omega_-^2)} -
\frac{\omega_-^2 - k^2  - \Delta^2(h) + h^2}{\omega_-
(\omega_+^2 - \omega_-^2)} \right )
\label{h_cons}
\end{eqnarray}
Before continuing, we mention that the $T$ dependence can easily
be worked in by multiplying the terms in the integrand above by 
$1 + \frac{2}{e^{\beta \omega_i} - 1}$, respectively.
It is possible to
simplify Eqn. (\ref{h_cons}) further to read
\begin{eqnarray}
\int \frac{dk}{4\pi} \left ( \frac{1}{\omega_3} +
\frac{1}{\omega_+} + \frac{1}{\omega_-} - \frac{4h^2}{\omega_+
\omega_- ( \omega_+ + \omega_-)} \right )
\label{h_integ}
\end{eqnarray}
where it is now clear that (\ref{oops}) is satisfied for
$h \rightarrow 0$. This can be shown to be the same result
obtained by minimizing the zero point energy.
To solve for $\Delta(h)$, one must decide on a sufficiently large
ultraviolet cutoff and resort to 
a numerical root finding routine. 
\begin{figure}
\centerline{
\epsfxsize=13.5 cm
\epsffile{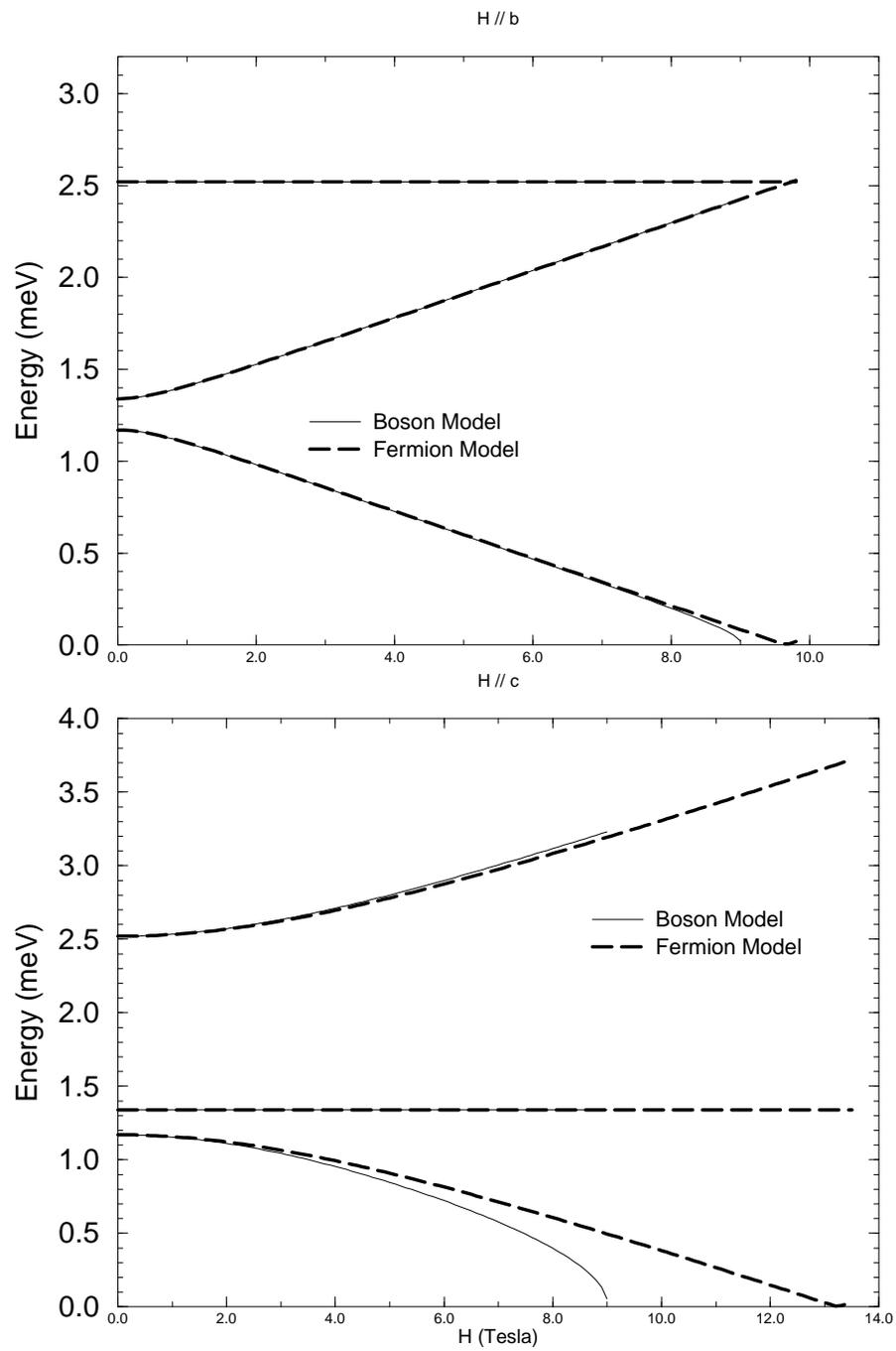}
}
\caption{Free boson and free fermion dispersions
with the gap parameters of NENP. Top graph: $\protect\bvec{H} \protect\| b$;
 bottom graph: $\protect\bvec{H} \protect\| c$.}
\label{fig:disp_1}
\end{figure}

\begin{figure}
\centerline{
\epsfxsize=13.5 cm
\epsffile{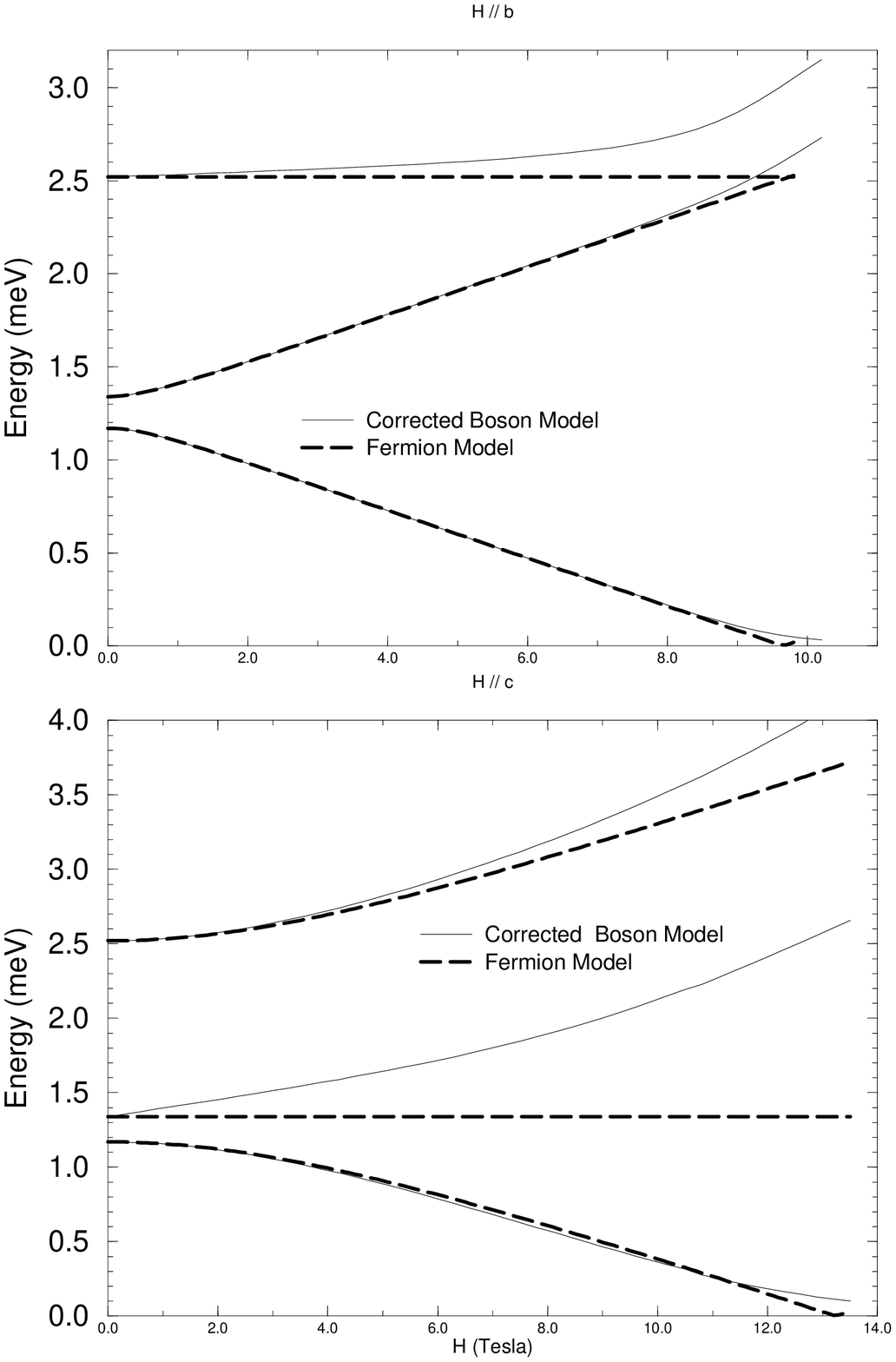}
}
\label{fig:disp_2}
\end{figure}

In Figure \ref{fig:disp_1} we plot the energy gaps of Eqn.
(\ref{poles}) unconstrained by Eqn. (\ref{oops}). To compare,
we also plot the field dependent gaps of the fermion model which
are generally considered in agreement with experiment \cite{aff}
(at least for the material NENP). The zero field gaps are fitted
to the gaps found in NENP: $\Delta_a = 1.17$meV, 
$\Delta_b = 2.52$meV and $\Delta_c = 1.34$meV. The subscripts, `$a,b$'
and `$c$' refer to appropriate crystal axes of NENP. The dispersions
differ most at higher fields, and for large $\Delta_E$.
In Figure \ref{fig:disp_2} we replot the gaps but this time correct 
for the constraint implied by Eqn. (\ref{h_cons}). The lower branches 
show good agreement right up to fields close to critical. There is,
however, a greater discrepancy in the gap corresponding to the 
field direction. 

We now turn our attention to a seeming infrared catastrophe which
occurs as $h$ approaches the critical value given by
\begin{eqnarray}
h_c = \sqrt{\Delta^2(h_c) - \Delta_E^2};
\label{h_crit}
\end{eqnarray}
this is where the lower gap closes and the integral
(\ref{h_integ}) diverges logarithmically in the thermodynamic
limit. At first sight one may hope that for $k=0$, the last two
terms in (\ref{h_integ}) conspire to eliminate the divergence for
some value of $h$ and $\Delta(h)$ satisfying (\ref{h_crit}).
This, however, requires that 
\begin{eqnarray}
h_c^2 = \Delta^2(h_c) \left ( \frac{5}{9} + \frac{4}{9} \sqrt{ 1
+ \frac{9 \Delta_E^4}{16 \Delta^4(h_c)} } \right )
\end{eqnarray}
be simultaneously satisfied; this is impossible unless $E=0$. In
fact, for $E=0$, $\Delta(h)$ is independent of $h$. This can be
seen directly from Eqn. (\ref{h_integ}) or by understanding that
the variation of the zero point energies 
$\omega_{\pm} = \sqrt{v^2k^2 + \Delta^2} \pm h$ with respect to 
$\Delta^2$ is independent of $h$. Let's try to get a deeper
feeling as to what's happening. Instead of starting with a finite
$E$ we can place the magnetic field near the $U(1)$ critical
value, $h_c = \Delta(h_c)$, and turn on the anisotropy. We write
\begin{eqnarray}
\omega_-^2(k=0) = |(\Delta(h) - h)^2 - \frac{\Delta_E^4}{4h\Delta} |
\end{eqnarray}
We now see that the limits $h \rightarrow h_c$ and 
$E \rightarrow 0$ don't commute after taking the derivative,
$\frac{\partial \omega_-}{\partial \Delta}$. Regardless of the
limit at which we start, we can expect trouble when
\begin{eqnarray}
E \sim \Delta_-(h)-h
\label{h_c_val}
\end{eqnarray}
This is true even for $h=0$, which is the
simplest case where this problem appears.

Essentially the trouble arises because, returning to
the NL$\sigma$ action, we integrated out low energy modes. In the
$U(1)$ case the day was saved by symmetry which prevented the
variation of $<\lambda>$ with magnetic field.
There is no such luck in the
$Z_2 \times Z_2 \times Z_2$ scenario and for a sufficiently large
value of $E$, it is no longer correct to integrate out all the
$\phi$ fields {\em even for large} $N$. In the case of large $N$,
one can integrate out all modes but the gapless one to arrive at
a critical theory. This should be done when $\log \left (
\frac{\Lambda}{\Delta-E} \right ) \sim N/g$. For $N=3$ this
criteria may be too restrictive, and instead one can adhere to
$E \ll \Delta(h) - h$ as the regime where the analysis of this
section is valid.

\subsection{Exact Results}

\label{sec_2.1.3}

In this section we note some important results 
which will be useful in the calculations of Chapter 3.
As was mentioned in the last chapter, the $O(3)$ model possesses
special properties that allow for some integrability. In
particular, at long wavelengths one can say much about the matrix
elements of the spin operator even after the $O(3)$ symmetry is
broken to $U(1)$ by a magnetic field. 

As will become clear in the next chapter, we are mostly concerned
with matrix elements $<k,a|S^i(0)|q,b>$, where $<k,a|$ denotes a
single magnon state created by the staggered magnetization
operator, $\phi^a_k$ (this acts as a free boson operator in the
large $N$ limit) with norm 
\begin{eqnarray}
<k,a|q,b> = 2\pi \delta_{ab} \delta(k-q)
\end{eqnarray}
$a$ and $b$ denote polarizations of the magnon states.
Clearly, $<k+q,a|\phi^i(0)|k,b> = 0$. This is obvious when the
bosons are completely free; interactions do not change this
picture since they must all be even in bosonic operators. The two
magnon operator, $l^i(0)$, is expected to contribute.\footnote{Strictly
speaking, since
$\bvec{l}$ is quadratic in $\bphi$, we expect it to be a two magnon
operator only in the large $N$ limit.}
The matrix
element is given by the Karowski and Weisz ansatz \cite{karo},
\begin{eqnarray}
<k,a|l^i(0)|q,b> = i\epsilon^{iab}
\frac{(\omega_k + \omega_q)}{2\sqrt{\omega_k \omega_q}}G(\theta)
\label{NLs_G}
\end{eqnarray}
where $\omega_k = \sqrt{k^2v^2 + \Delta^2}$ and the rapidity
variable, $\theta$, is defined via

\[ \sinh(\theta' - i\pi) = k \]
\[\sinh(\theta'' - i\pi) = q \]
\begin{eqnarray}
\theta = \theta' - \theta''
\end{eqnarray}

with
\begin{eqnarray}
G(\theta) = \exp \left( 2\int_0^{\infty} \frac{dx}{x}
\frac{ (e^{-2x}-1) \sin^2[x(i\pi-\theta)/2\pi]}{(e^x+1)\sinh(x)}
\right )
\end{eqnarray}
This ansatz is believed to be exact for the $O(3)$ NL$\sigma$ model, but
is only approximately true for the $s=1$ Heisenberg model;
however, numerical simulations \cite{sorensen} are in excellent
agreement with this form at least half way through the Brillouin
zone. Since we will largely be interested in $|k-q|v < \Delta \ll
\pi v$, this `exact' expression is more than sufficient. Some
particular limits of interest are $k \approx q$ and
$k \approx -q$, corresponding to forward
and back scattering, respectively; in the former case, 
$\theta \approx i\pi$ while in the latter, 
$\theta \approx i\pi +2kv/\Delta$. 
\begin{eqnarray}
G(i\pi + 2kv/\Delta) \approx  1 - \left ( \frac{4}{\pi^2} + \frac{1}{3}
\right ) \left ( \frac{kv}{\Delta} \right )^2
\end{eqnarray}
This expression is a different result than for free bosons
and reflects the effects of interactions. We will later
see how this might affect experimental predictions.

The above results change when a $D$ type anisotropy is added.
Essentially, the difference is that the function, $G(\theta)$
changes and the gap in the energy of states on the $S^z=0$ branch
will be shifted. There are no exact results for
this model which is why the phenomenological boson and fermion
models are important. We can, however, give some universal
(ie. model independent) results which only depend on the
conservation of total spin in the $z$-direction, when the
momentum exchange is small, 
$v|q-k| \ll \Delta$. The only part of $S^z(0)$ which will
contribute will be, essentially, 
$S^z_{q-k} \approx \int dx \; S^z(x)$. Since this is a conserved
operator, we can write down the one-particle matrix elements
immediately:
\begin{eqnarray}
<k,s^z \! = \! \pm 1| S^z(0)|q,s^z \! = \! \pm 1> \approx  \pm 1
\end{eqnarray}
Also true for all values of $v|q-k|$ are the following
\begin{eqnarray}
<k, s^z \! = \! 0| S^z(0)|q, s^z \! = \! a> = 0 \nonumber \\
<k, s^z \! = \! +1| S^z(0)|q, s^z \! = \! -1> = 0
\end{eqnarray}
In Chapter 3 we will show that it is largely this universal behaviour
which determines the relaxation rate, $1/T_1$ in anisotropic
media.

Note that adding a magnetic field to the $O(3)$ or $U(1)$ system
will break the symmetry in the $O(3)$ case (we naturally assume
that in the $U(1)$ case, the field is placed parallel to the $U(1)$
axis), but will hardly 
change any other results in either model.
This is because in adding a magnetic field all we have done is
add a term to the Hamiltonian proportional to the conserved
charge $l^z_0 = \int dx \; l^z(x)$.  Since $l^z_0$ commutes with
the Hamiltonian, they can be simultaneously diagonalized with
all low lying states labeled by their $l^z_0$ quantum number,
$+1,0$ or $-1$. Matrix elements of operators can only differ, at
most, by some cumulative phase which corresponds to turning on
the field sometime in the past.

\section{The Free Boson and Fermion Models}

\label{sec_2.2}

We now turn our attention to the phenomenological models
introduced in Chapter 1. After introducing the formalism
which we will require to perform calculations, we will compare
and contrast the energy spectra and fundamental matrix elements.

\subsection{Diagonalization}

\label{sec_2.2.1}

To do the necessary calculations we need to have a basis of
eigenstates for the non-interacting Hamiltonian and know the
expansion of the field operators in terms of
creation/annihilation operators for these states. 

\subsubsection{Bosons}

\label{sec_{2.2.1.1}}

Relaxing the constraint $\bphi^2 = 1$ in(\ref{h_lag}) we see that
we seek to diagonalize
\begin{eqnarray}
H = \int dx \left [ \bvec{\Pi}^2 + \frac{v^2}{2} (\frac{\partial
\bphi}{\partial x})^2  + \frac{1}{2} \bphi \cdot \bold{D} \bphi
- \vec{h} \cdot \bold{G} (\bphi \times \bvec{\Pi}) \right ]
\label{bos_H}
\end{eqnarray}

For now we assume that the mass and gyromagnetic tensors,
$\bold{D}$ and $\bold{G}$ respectively, are simultaneously
diagonalizable and work in this diagonal basis (this is rigorously
true when the crystal field symmetry is no lower than orthorhombic --
a sketch of a proof is found on p. 750 of \cite{abragam}).
\begin{eqnarray}
\bold{D} = \left ( \begin{array}{ccc}
\Delta_1^2 & 0 & 0 \\
0 & \Delta_2^2 & 0 \\
0 & 0 & \Delta_3^2 \end{array} \right ) \; \; \; \;
\bold{G} = \left ( \begin{array}{ccc}
g_1 & 0 & 0 \\
0 & g_2 & 0 \\
0 & 0 & g_3 \end{array} \right )
\end{eqnarray}
Also, we
mention that we've set $\hbar = 1 = a$, where $a$ is the lattice
spacing. This has the effect of measuring energy in units of
inverse seconds or {\em inverse} mass:
\begin{eqnarray}
[E] \sim [s]^{-1} \sim [M]^{-1} \sim [v] \sim [\phi]^{-2}
\end{eqnarray}
Diagonalizing (\ref{bos_H}) is tedious (especially when the field
does not lie in a direction of symmetry, for then all the
branches mix)
but the idea is to find the right Bogoliubov transformation. 
Working in momentum space, we define

\begin{eqnarray}  \bphi(k,t=0) = \frac{1}{\sqrt{2\omega_{0}}} 
[\bvec{a}^{\dagger}_{-k}
+ \bvec{a}_{k}] \end{eqnarray}
\begin{eqnarray}  \bvec{\Pi}(k,t=0) = 
i\sqrt{\frac{\omega_{0}}{2}} [\bvec{a}^{\dagger}_{-k}
- \bvec{a}_{k}] \end{eqnarray}
\begin{eqnarray}
[a_k^i, a_{k^\prime}^{j \dagger}] = 2\pi \delta_{ij}
\delta(k-k^\prime)
\end{eqnarray}
$\omega_0$ is an arbitrary quantity with the dimensions of energy. 
We need such a quantity to represent the fields $\bphi$ and $\bPi$
in terms of creation and annihilation operators. 
It turns out that when one writes 
$\bphi_k$ and $\bPi_k$ in terms of the creation/annihilation
operators which diagonalize the Hamiltonian,
the dependence on $\omega_0$ disappears. Furthermore,
the eigenvalues of the Hamiltonian are also independent of $\omega_0$,
as might be expected.
We will now restrict ourselves to the case where the field
lies in a direction of symmetry. This leaves $(\ref{bos_H})$ with
$Z_2 \times Z_2$ symmetry. Now only the excitations
transverse to the direction of the applied field mix and we need
only solve a $(4 \times 4)$ set of equations for the diagonalizing
creation/annihilation operators. Without loss of generality, we
take the field to lie in the $3$-direction, and set $g_3=1$.
All told, the Hamiltonian, Eqn. (\ref{bos_H}), for the mixed states is
\[ H = \int_{-\infty}^{\infty} H_k \; dk \]
\begin{eqnarray} 
H_{k} =  \vec{a}^{\dagger}_{k} {\bf A} \vec{a}_{k} +
\vec{a}^{\dagger}_{-k} {\bf A}^{\ast} \vec{a}_{-k} +
\vec{a}^{\dagger}_{k} {\bf B} \vec{a}^{\dagger}_{-k} +
\vec{a}_{k} {\bf B}^{\ast} \vec{a}_{-k} \end{eqnarray}
\begin{eqnarray}
{\bf A} = \frac{\omega_{0}}{4} {\bf I} + \frac{ {\bf K}
}{4\omega_{0}} -
\frac{1}{2} h \bold{\sigma}_{2}
\end{eqnarray}
\begin{eqnarray}
{\bf B} = -\frac{\omega_{0}}{4} {\bf I} + \frac{ {\bf K}
}{4\omega_{0}}
\end{eqnarray}
where $\bold{\sigma}_{2}$ is the usual Pauli matrix, and
\begin{eqnarray} 
{\bf K} = \left ( \begin{array}{cc}
\Delta_{1}^{2} + v^{2}k^{2} & 0 \\
0 & \Delta_{2}^2
+ v^{2}k^{2}) \end{array} \right )
\end{eqnarray}
The momentum space Hamiltonian can be written in terms of a
single matrix $\bold{M}$:
\begin{eqnarray}
H_k = \left ( \bvec{a}_k^{\dagger}, \bvec{a}_{-k} \right )
\bold{M} \left ( \begin{array}{c} 
\bvec{a}_k \\
\bvec{a}^{\dagger}_{-k} \end{array} \right )
\end{eqnarray}
As discussed in \cite{bla}, we seek the
eigenvectors of the (non-hermitian) matrix $\eta M$, where
\begin{eqnarray} \eta \bold{M} = \left ( \begin{array}{cc}
{\bf A} & {\bf B} \\
-{\bf B}^{\ast} & -{\bf A}^{\ast} \end{array}
\right ) \;\;\;\;\; \eta = \left ( \begin{array}{cc}
1 & 0 \\ 
0 & -1 \end{array} \right )  
\end{eqnarray}
This comes from requiring the new diagonal creation/annihilation
operators to have the standard commutation relations. This also
imposes the unusual normalization condition on the eigenvectors:
$\bvec{\zeta}^{\dagger} \eta \bvec{\zeta} = 1$.

Summarizing the above, we need to solve:
\begin{eqnarray}   0 = \left ( \begin{array}{cc}
\frac{\omega_{0}}{4} {\bf I} + \frac{ {\bf K} }{4\omega_{0}} -
\frac{1}{2} h \bold{\sigma}_{2} - \frac{\omega}{2} {\bf I} &
-\frac{\omega_{0}}{4} {\bf I} + \frac{ {\bf K} }{4\omega_{0}} \\
\frac{\omega_{0}}{4} {\bf I} - \frac{ {\bf K} }{4\omega_{0}} &
-\frac{\omega_{0}}{4} {\bf I} - \frac{ {\bf K} }{4\omega_{0}} -
\frac{1}{2} h \bold{\sigma}_{2} - \frac{\omega}{2} {\bf I}
\end{array} \right ) \end{eqnarray}

which can be manipulated to give

\begin{eqnarray}   0 = \left ( \begin{array}{cc}
(\omega_{0} - \omega) {\bf I} - h \bold{\sigma}_{2} &
-(\omega_{0} + \omega) {\bf I} - h \bold{\sigma}_{2} \\
\frac{ {\bf K} }{\omega_{0}} - \omega {\bf I} - h 
\bold{\sigma}_{2} &
\frac{ {\bf K} }{\omega_{0}} + \omega {\bf I} + h
\bold{\sigma}_{2}
\end{array} \right ) 
\label{matrix1}
\end{eqnarray}
The eigenvalues of $\eta \bold{M}$ are already known, as they are
the solutions to the classical equations of motion and come in
pairs $\pm \omega_{\pm}$. These are naturally the same
frequencies given in Eqn.(\ref{poles}) with the proper
substitutions made for the gaps
\[ \omega_3^2 =  k^2 +\Delta_3^2 \]
\begin{eqnarray}
\omega_{\pm}^2 =  k^2 + h^2 + \frac{\Delta_1^2
+ \Delta_2^2}{2} \pm
\sqrt{ 4h^2(k^2 + \frac{\Delta^2_1 + \Delta_2^2}{2}) + \left (
\frac{\Delta^2_1 - \Delta_2^2}{2} \right )^2 } 
\label{E_bos}
\end{eqnarray}

Furthermore, we need only work to find one eigenvector of each
pair because if $\left ( \begin{array}{c} X \\
Y \end{array} \right )$ is a right-eigenvector
of $\eta \bold{M}$ with eigenvalue $\omega$, then
$\left ( \begin{array}{c} Y^{\ast} \\
X^{\ast} \end{array} \right )$ is a right-eigenvector of 
$\eta \bold{M}$ with
eigenvalue $-\omega$ ($X$ and $Y$ are themselves two-component
vectors).

The bottom rows of (\ref{matrix1}) give the following set of
equations

\begin{eqnarray}  0 = (\Delta_{1}^{2} + v^{2}k^{2}) \chi_{\pm,1}
- \omega_{0} \omega_{\pm} \xi_{\pm,1} +ih\omega_{0} \xi_{\pm,2}
\end{eqnarray}
\begin{eqnarray}  0 = (\Delta_{2}^{2} + v^{2}k^{2}) \chi_{\pm,2}
- \omega_{0} \omega_{\pm} \xi_{\pm,2} -ih\omega_{0} \xi_{\pm,1}
\end{eqnarray}
where it turns out to be convenient to work with $\chi \equiv X +
Y$ and $\xi \equiv X - Y$. The top rows can be
manipulated to give
\begin{eqnarray} 0 = (h^{2}-\omega_{\pm}^{2}) \chi_{\pm,2} - ih
\omega_{0} \xi_{\pm,1} +
\omega_{0} \omega_{\pm} \xi_{\pm,2} \end{eqnarray}
\begin{eqnarray} 0 = -(h^{2}-\omega_{\pm}^{2}) \chi_{\pm,1} - ih
\omega_{0} \xi_{\pm,2} -
\omega_{0} \omega_{\pm} \xi_{\pm,1} \end{eqnarray}

These can then be worked to give
\begin{eqnarray} (\Delta_{1}^{2} + v^{2}k^{2} + \omega_{\pm}^{2}
- h^{2}) \chi_{\pm,1} =
2\omega_{0} \omega_{\pm} \xi_{\pm,1} \end{eqnarray}
\begin{eqnarray} (\Delta{_2}^{2} + v^{2}k^{2} + \omega_{\pm}^{2}
- h^{2}) \chi_{\pm,2} =
2\omega_{0} \omega_{\pm} \xi_{\pm,2} \end{eqnarray}
\begin{eqnarray} (h^{2} + \Delta_{1}^{2} + v^{2}k^{2} -
\omega_{\pm}^{2}) \chi_{\pm,1} =
-2ih\omega_{0} \xi_{\pm,2} \end{eqnarray}
\begin{eqnarray} (h^{2} + \Delta_{2}^{2} + v^{2}k^{2} -
\omega_{\pm}^{2}) \chi_{\pm,2} =
2ih\omega_{0} \xi_{\pm,1} \end{eqnarray}
Note that if we fix the phase of $\chi_{1}$ to be real then $\xi_{1}$
must also be real as $\chi_{2}$ and $\xi_{2}$ must be pure imaginary.
The normalization condition,  $X^{\dagger}X - Y^{\dagger}Y = 1$, 
now allows us to solve for the eigenvectors which form the
columns of the transformation matrix between the old and diagonal
bases of creation/annihilation operators. In
terms of $\chi$ and $\xi$ this is
\begin{eqnarray} \chi_{\pm,1} \xi_{\pm,1} - \chi_{\pm,2}
\xi_{\pm,2} = 1 \end{eqnarray}
The solution is

\begin{eqnarray} \chi_{\pm,1} = \left (  \frac{ \omega_{0}
\omega_{\pm}
(h^{2} + \Delta_{2}^{2} + v^{2}k^{2} - \omega_{\pm}^{2}) }
{(\Delta_{1}^{2} + v^{2}k^{2} + \omega_{\pm}^{2} - h^{2})
(h^{2} + \frac{\Delta_{1}^{2} + \Delta_{2}^{2}}{2} + v^{2}k^{2}
- \omega_{\pm}^{2}) } \right )^{1/2} 
\label{eq2.2.1}
\end{eqnarray}
\begin{eqnarray} \xi_{\pm,1} = \frac{\Delta_{1}^{2} + v^{2}k^{2}
+ \omega_{\pm}^{2} - h^{2}}
{2\omega_{0} \omega_{\pm}} \chi_{\pm,1} 
\label{eq2.2.2}
\end{eqnarray}
\begin{eqnarray} \chi_{\pm,2} = \frac{2ih \omega_{0}}
{h^{2} + \Delta_{2}^{2} + v^{2}k^{2} - \omega_{\pm}^{2}}
\xi_{\pm,1} 
\label{eq2.2.3}
\end{eqnarray}
\begin{eqnarray} \xi_{\pm,2} = \frac{\Delta_{2}^{2} + v^{2}k^{2}
+ \omega_{\pm}^{2} - h^{2}}
{2\omega_{0} \omega_{\pm}} \chi_{\pm,2} 
\label{eq2.2.4}
\end{eqnarray}

With the inclusion of the trivially diagonal unmixed component
(ie. the three component), we can define the three by three
matrices $\chi$ and $\xi$ with the columns labeled by the
eigenvalues $(+,-,3)$ and the rows labeled by the original masses
$(1,2,3)$: $\chi_{33} =
\sqrt{\frac{\omega_{0}}{\sqrt{\Delta_{3}^2 + v^{2}k^{2}} } }$ and
$\xi_{33} = \sqrt{ \frac{\sqrt{\Delta_{3}^2 + v^{2}k^{2}}
}{\omega_{0}}} $.
One easily verifies that 
\begin{eqnarray}  
\bvec{a}_k = \frac{1}{2} \left [ 
\left ( \chi_k + \xi_k \right ) \bvec{b}_k +
\left ( \chi_k^{\ast} - \xi_k^{\ast} \right )
\bvec{b}_{-k}^{\dagger} \right ]
\label{eq2.2.5}
\end{eqnarray}

Where the $b$'s are the operators which diagonalize $H$.
Equations (\ref{eq2.2.1})--(\ref{eq2.2.5}) are the main results of
this section. Before ending, we give some limiting forms for
$\chi$ and $\xi$. In the limit $h \rightarrow 0$,
\begin{eqnarray} \chi = \xi^{-1  \dagger} = \left (
\begin{array}{ccc}
\sqrt{ \frac{\omega_{0}}{\sqrt{\Delta_{1}^2 + v^{2}k^{2}} } } & 0
& 0 \\
0 & i\sqrt{ \frac{\omega_{0}}{\sqrt{\Delta_{2}^2 + v^{2}k^{2}} }}
& 0 \\
0 & 0 &
\sqrt{ \frac{\omega_{0}}{\sqrt{\Delta_{3}^2 + v^{2}k^{2}} } }
\end{array} \right ) \end{eqnarray}
While in the limit $\Delta_{2} \rightarrow \Delta_{1}$,
\begin{eqnarray} \chi =\xi^{-1 \dagger}
= \left ( \begin{array}{ccc}
\sqrt{ \frac{\omega_{0}}{2\sqrt{\Delta^2 + v^{2}k^{2}} } } &
\sqrt{ \frac{\omega_{0}}{2\sqrt{\Delta^2 + v^{2}k^{2}} } } & 0 \\
-i\sqrt{ \frac{\omega_{0}}{2\sqrt{\Delta^2 + v^{2}k^{2}} } } &
i\sqrt{ \frac{\omega_{0}}{2\sqrt{\Delta^2 + v^{2}k^{2}} } } & 0
\\
0 & 0 &
\sqrt{ \frac{\omega_{0}}{\sqrt{\Delta_{3}^2 + v^{2}k^{2}} } }
\end{array}  \right ) \end{eqnarray}

\subsubsection{Fermions}

\label{sec_2.2.1.2}

We would now like to repeat the diagonalization procedure for the
fermion model. The free Hamiltonian with minimal coupling to
the magnetic field (ie. coupling only to $\bvec{l}$) and simplest
parametrization of the mass terms (corresponding to anisotropies
and giving zero field dispersion branches $\omega_k^a = \sqrt{\Delta_a^2
+ k^2v^2}$) is

\[{\cal H}(x)  =
\frac{1}{2} \left [  \vec{\psi}_L \cdot iv\partial_x \vec{\psi}_L
- \vec{\psi}_R \cdot iv\partial_x \vec{\psi}_R +  \right. \]
\begin{eqnarray}
\left. i\sum_{i=1}^{3}
\Delta_i (\psi_{R,i} \psi_{L,i} - \psi_{L,i}\psi_{R,i}) +
i\vec{h} \cdot
(\vec{\psi}_L \times \vec{\psi}_L+\vec{\psi}_R \times
\vec{\psi}_R) \right ]
\label{H_1_ferms}
\end{eqnarray}
with, setting $v=1$,
\begin{eqnarray} \vec{\psi}_R = \int_{0}^{\infty} \frac{dk}{2\pi}
\left (
e^{-ik(t-x)} \vec{a}_{R,k} + e^{ik(t-x)} \vec{a}_{R,k}^{\dagger}
\right ) \end{eqnarray}
\begin{eqnarray} \vec{\psi}_L = \int_{0}^{\infty} \frac{dk}{2\pi}
\left (
e^{-ik(t+x)} \vec{a}_{L,k} + e^{ik(t+x)} \vec{a}_{L,k}^{\dagger}
\right ) 
\end{eqnarray}
\begin{eqnarray}
\{a_k^i, a_{k^\prime}^{j \dagger} \} = 2\pi \delta_{ij}
\delta(k-k^\prime)
\end{eqnarray}

Notice that we coupled the magnetic field to the generator of
global rotations, \\  $\int dx \; \bvec{l}(x)$, given by
(\ref{l_ferms}). 
the Hamiltonian density in $k$-space becomes
\begin{eqnarray} H_k = \bvec{\alpha}_k^{\dagger} \bold{M}_k
\bvec{\alpha}_k
\end{eqnarray}
where
\begin{eqnarray} \bold{M} = \left ( \begin{array}{cc}
\bf{I} k - i\vec{h} \times & i\bf{\Delta} \\
-i\bf{\Delta} & -\bf{I} k - i\vec{h} \times
\end{array} \right ) \end{eqnarray}
\begin{eqnarray} \bvec{\alpha}_k = \left ( \begin{array}{c}
\bvec{a}_{R,k} \\
\bvec{a}_{L,k}^{\dagger} \end{array} \right ) \end{eqnarray}
The idea now is to diagonalize this matrix and find the
eigenvalues and
eigenvectors. In other words, find the unitary transformation
which diagonalizes $H$.
Once more, we assume the field is in a direction of
symmetry so that we need only diagonalize a $4 \times 4$ matrix.
Given that the field is in the $3$ direction, the eigenvalues 
are:
\[\omega_3^2 =  k^2 +\Delta_3^2 \]
\begin{eqnarray}
\omega_{\pm}^2 =  k^2 + h^2 + \frac{\Delta_1^2
+ \Delta_2^2}{2} \pm
\sqrt{ 4h^2(k^2 + \frac{(\Delta_1 + \Delta_2)^2}{4}) + \left (
\frac{\Delta^2_1 - \Delta_2^2}{2} \right )^2 } 
\label{E_ferms}
\end{eqnarray}

It may be more illuminating to write out $\bold{M}$ in a basis that is
more natural
to the $U(1)$ problem. Using

\begin{eqnarray} \left( \begin{array}{c} a^1_{R,k} \\ a^2_{R,k}
\end{array}
\right ) = \frac{1}{\sqrt{2}} \left ( \begin{array}{cc}
1 & 1 \\
i & -i \end{array} \right ) \left ( \begin{array}{c}
a^+_{R,k} \\
a^-_{R,k} \end{array} \right ) \end{eqnarray}

\begin{eqnarray} \left ( \begin{array}{c} a^{1 \; \dagger}_{L,k}
\\
a^{2 \; \dagger}_{L,k} \end{array} \right )
= \frac{1}{\sqrt{2}} \left ( \begin{array}{cc}
1 & 1 \\
-i & i \end{array} \right ) \left ( \begin{array}{c}
a^{+ \; \dagger}_{L,k} \\
a^{- \; \dagger}_{L,k}  \end{array} \right ) \end{eqnarray}

In this basis, $\bold{M}$ becomes
\begin{eqnarray} \bold{M} = \left ( \begin{array}{cc}
k \bf{I}  - h\bold{\sigma}_3 & i\Delta \bold{\sigma}_1 + i\delta
\bf{I} \\
-i\Delta \bold{\sigma}_1-i\delta \bf{I} & -k \bf{I} +
h\bold{\sigma}_3
\end{array} \right ) 
\label{M_ferms}
\end{eqnarray}
Where $\Delta = \frac{\Delta_1 + \Delta_2}{2}$ and $\delta =
\frac{\Delta_1 -
\Delta_2}{2}$.
The equations for the components of the eigenvectors possess the
symmetries
\begin{eqnarray} u_1 \leftrightarrow u_2,  u_3 \leftrightarrow
u_4, h \leftrightarrow -h \end{eqnarray}

\begin{eqnarray} u_1 \leftrightarrow u_3,  u_2 \leftrightarrow
u_4, \omega \leftrightarrow
-\omega \end{eqnarray}
where $\omega$ is the eigenvalue.
After some algebra,
\begin{eqnarray} u_4 = \frac{2i\Delta (k-\omega)}{\omega^2 +
\Delta^2 -(k+h)^2 -
\delta^2}u_1
\end{eqnarray}
\begin{eqnarray} u_3 = \frac{2i\delta (k-\omega)}{(\omega-h)^2 -
k^2 + \delta^2
-\Delta^2}u_1
\end{eqnarray}
\begin{eqnarray} u_2 = \frac{2i\Delta (k+\omega)}{\omega^2 +
\Delta^2 -(k+h)^2 -
\delta^2}u_3
\end{eqnarray}
Using the normalization condition,
\begin{eqnarray} \sum_{i=1}^{4} |u_i|^2 = 1 \end{eqnarray}
we set the phase of $u_1$ to be real for positive eigenvalues;
the above symmetries allow us
the freedom to choose a convenient phase for the $u_1$'s corresponding
to negative eigenvalues.
\begin{eqnarray} u_1 & = &  2\delta (k+\omega)
(\omega^2 + \Delta^2 -(k+h)^2 - \delta^2) \div \\ &  &
[  4\delta^2 (k+\omega)^2 (  (\omega^2 + \Delta^2 -(k+h)^2 -
\delta^2)^2 +
4\Delta^2 (k-\omega)^2 ) + \nonumber \\ & &
((\omega+h)^2 - k^2 + \delta^2 -\Delta^2)^2 (
(\omega^2 + \Delta^2 -(k+h)^2 - \delta^2)^2 + 4\Delta^2
(k+\omega)^2 ) ]^\frac{1
}{2}
\nonumber
\end{eqnarray}

We define the $6\times 6$ diagonalizing matrix with columns given
by the eigenvectors
$\vec{u}_{\omega}$ as
\begin{eqnarray} X_{i,\omega} = ( u_{\omega_+}^i, u_{\omega_-}^i,
u_{\omega_3}^i
, u_{-\omega_+
}^i,
u_{-\omega_-}^i, u_{-\omega_3}^i ) \end{eqnarray}

\begin{eqnarray} \bvec{\alpha}^i_k = 
U X^i_{\omega} \bvec{\beta}^{\omega}
\end{eqnarray}

\begin{eqnarray}
U = \frac{1}{\sqrt{2}} \left ( \begin{array}{cccccc}
1 & 1 & 0 & 0 & 0 & 0 \\
i & -i & 0 & 0 & 0 & 0 \\
0 & 0 & \sqrt{2} & 0 & 0 & 0 \\
0 & 0 & 0 & 1 & 1 & 0 \\
0 & 0 & 0 & -i & i & 0 \\
0 & 0 & 0 & 0 & 0 & \sqrt{2}
\end{array} \right ) \equiv \left ( \begin{array}{cc}
V & 0 \\
0 & V \sigma_1  \end{array} \right )
\end{eqnarray}

The diagonal operators, $\bvec{\beta}_{k}$ are defined as:
\begin{eqnarray} \bvec{\beta}_k = \left ( \begin{array}{c}
\bvec{c}_k \\
\bvec{d}_{k}^{\dagger} \end{array} \right ) \end{eqnarray}

Our freedom in
choosing the phases for the eigenvectors corresponding to
negative
eigenvalues allows us to write

\begin{eqnarray}
X = \left ( \begin{array}{cc}
R & T \\
T & R \end{array} \right )
\end{eqnarray}

Each index of this matrix runs over six states; the first and
last three
correspond to right and left movers respectively. In the case of
$U(1)$ symmetry or higher, each set would correspond to states of
definite spin.

The $d$'s and $c$'s correspond to left and right moving fermions,
respectively.
This becomes clear in the limit $\Delta_1 \rightarrow \Delta_2
\rightarrow 0$.
Some limiting forms of $R$ and $T$ are:

\begin{eqnarray}
R(h \rightarrow 0 ) = \frac{1}{2} \left ( \begin{array}{ccc}
\sqrt{\frac{\omega_1+k}{\omega_1}} &
-\sqrt{\frac{\omega_2+k}{\omega_2}} & 0 \\
\sqrt{\frac{\omega_1+k}{\omega_1}} &
\sqrt{\frac{\omega_2+k}{\omega_2}} & 0 \\
0 & 0 & \sqrt{2\frac{\omega_3+k}{\omega_3}}
\end{array} \right )
\end{eqnarray}

\begin{eqnarray}
T(h \rightarrow 0 ) = \frac{1}{2} \left ( \begin{array}{ccc}
-i\sqrt{\frac{\omega_1-k}{\omega_1}} 
& -i\sqrt{\frac{\omega_2-k}{\omega_2}} & 0 \\
-i\sqrt{\frac{\omega_1-k}{\omega_1}} 
& i\sqrt{\frac{\omega_2-k}{\omega_2}} & 0 \\
0 & 0 & -i\sqrt{2\frac{\omega_3-k}{\omega_3}}
\end{array} \right )
\end{eqnarray}

\begin{eqnarray}
R(\delta \rightarrow 0 ) = \frac{1}{\sqrt{2}} \left (
\begin{array}{ccc}
0 & -\sqrt{\frac{\omega_{\Delta}+k}{\omega_{\Delta}}} & 0 \\
\sqrt{\frac{\omega_{\Delta}+k}{\omega_{\Delta}}} & 0 & 0 \\
0 & 0 & \sqrt{\frac{\omega_3+k}{\omega_3}}
\end{array} \right )
\label{Rlim}
\end{eqnarray}

\begin{eqnarray}
T(\delta \rightarrow 0 ) = \frac{1}{\sqrt{2}} \left (
\begin{array}{ccc}
-i\sqrt{\frac{\omega_{\Delta}-k}{\omega_{\Delta}}} & 0 & 0 \\
0 & i\sqrt{\frac{\omega_{\Delta}-k}{\omega_{\Delta}}} & 0 \\
0 & 0 & -i\sqrt{\frac{\omega_3-k}{\omega_3}}
\end{array} \right )
\label{Tlim}
\end{eqnarray}

\subsection{Discussion: Comparison of Spectra and Spin Operator
Matrix Elements}

\label{sec_2.2.2}

The spectra for the boson and fermion models are given by
Eqns. (\ref{E_bos}) and (\ref{E_ferms}), respectively. In the
case of $U(1)$ symmetry, $\Delta_1 = \Delta_2$, the two sets of
formulae agree. However, with the lower orthorhombic symmetry,
the two models are in agreement only for low magnetic field, 
$h \ll \mbox{Min}(\Delta_1,\Delta_2)$.
The difference is most significant at the critical field where
the lower gap vanishes. The boson model predicts $h_c =
\mbox{Min}(\Delta_1, \Delta_2)$, while the fermion model gives
$h_c = \sqrt{\Delta_1 \Delta_2}$ (see Fig. \ref{fig:disp_1} and
\ref{fig:disp_2}). Experimental evidence seems to
favour the fermion model, but there are some subtleties which
have previously been ignored. The data supporting the fermion
dispersion comes from neutron scattering and NMR relaxation rate
experiments performed on the anisotropic 1DHAF material, NENP. In
analyzing the data, however, crucial structural properties were
neglected in the interpretation (namely, the fact that the local
chain axes did not coincide with the crystallographic axes).
This, we believe, also led to
a seeming contradiction with ESR data on the same substance which
seemed to side with the boson dispersion \cite{aff}.\footnote{For
more details on these matters, please see sections 5.1 and 6.3.}
Aside from material
properties, the possible temperature and field dependence of the mass
parameters, $\Delta_i$, has also been ignored so far. Since the
boson model derives from the NL$\sigma$ model, one should
incorporate such field and temperature dependence into these
basic parameters. We showed that considering field dependent masses
brought closer agreement on the lower branch dispersion
between the models up to fields given by Eqn. (\ref{h_c_val}).

We mention in passing that the Hamiltonian, Eqn. (\ref{bos_H}),
is not the only quadratic one possible when the magnetization
density is no longer conserved. It is possible to construct a
modified boson Hamiltonian including extra terms designed to
reproduce gaps identical with the fermion model \cite{halp}. The
only constraint on such terms is that they do not mix the $s^z=0$
modes corresponding to the degree of freedom parallel to the
field. It is not obvious, however, what justifies such a
modification other than a more convenient spectrum which replicates
the fermion model at low energies.

The fermion model is expected to become more accurate
close to the critical field. The nature of the critical point was
established in Ref. \cite{Aff}. With $U(1)$ symmetry, the phase
transition is in the two dimensional $xy$ universality class. The
lowest lying mode of the Landau-Ginsburg boson model can be
reduced to a single free boson (a phase field corresponding to the
Goldstone mode), but the parameters of the resulting low energy
Lagrangian must be renormalized to give the correct critical
exponents of the $xy$-model.  One does not have to resort to such
lengths with the fermion model which correctly describes the
transition without interactions. This is expected on several
grounds. First, the many body ground state wave-function for a
dilute gas of repulsive bosons is simply that of {\em free}
fermions multiplied by a sign function to correct for the
statistics. Second, the $U(1)$ fermionic modes
can be represented as particles and holes using a single Dirac 
fermion with chemical potential $h$ (this can be seen in
the matrix equation (\ref{M_ferms})); this means that at $h >
\Delta$ the ground state will be occupied by fermion states, each
with $s^z=1$, and hence non-zero magnetization. The simplicity
of the coupling to $h$ guarantees that interactions
will be as important near $h_c$ as they are near $h=0$.
In particular, they will be negligible in the dilute gas limit.
We thus see that interactions become progressively more important close to
criticality in a boson theory, while the opposite takes place in
an equivalent fermion theory.

In the $Z_2 \times Z_2$ case we expect an Ising-like transition
corresponding to the breaking of one of the $Z_2$ symmetries
remaining. Here things are even clearer. Mean field theory for
the boson model is completely hopeless as is evidenced by the
unphysical behaviour of the lowest lying gap at $h >
\mbox{Min}(\Delta_1, \Delta_2)$. This function always possesses a
zero even at non-vanishing $k$-vectors. Moreover, it is imaginary
for fields $\sqrt{k^2v^2 + \mbox{Min}(\Delta_1^2,\Delta_2^2)}
< h_c < \sqrt{k^2v^2 + \mbox{Max}(\Delta_1^2, \Delta_2^2)}$.
The spectrum for
the low lying fermion, in contrast, shows all the desirable
properties, vanishing at $h_c = \sqrt{\Delta_1 \Delta_2}$ only
for $k = 0$; in addition, the effective gap, $\Delta_e \sim |h-
h_c|$, is as expected in the Majorana fermion representation of
the critical Ising model and so is the relativistic dispersion 
for long wavelengths. Finally, when we integrate out the more
massive fermionic modes we are left with a {\em strictly} non-interacting
free Majorana fermion theory regardless of any zero-field interactions
in (\ref{H_1_ferms}); this is because all interactions will be polynomial
in the one Majorana field left, and will vanish by fermi statistics.
Thus we see that in contrast with the boson description, the free fermion
theory is actually {\em best} near $h=h_c$.

To summarize, on general grounds, one
can expect qualitative agreement between both models up to
magnetic fields close to $h_c$ where the fermion model is
expected to be a better description of the system.

We now wish to look at some important matrix elements as phrased
in the two models. We start by defining,
\begin{eqnarray}
\bvec{l}_{a,b}(k,q) \equiv <a,k| \bvec{l}(0) |q,b>
\label{l_def}
\end{eqnarray}
we use $\bvec{l} = \bphi \times \bvec{\Pi}$ for bosons to write
\begin{eqnarray}
\bvec{l}_{a,b}(k,q) = -\frac{i}{2} \left ( \xi^{\dagger}(k) \bvec{\Sigma}
\chi(q) + \chi^{\dagger}(k) \bvec{\Sigma} \xi(q) \right )_{a,b}
\end{eqnarray}
where we define the cross product matrix with the Levi-Civita
symbol by $\Sigma^i = \epsilon^{ijk}$.
Using, $\bvec{l} = \frac{-i}{2} ( \bpsi_L \times \bpsi_L +
\bpsi_R \times \bpsi_R )$ for fermions, we write the analogous
expression
\[ \bvec{l}_{a,b}(k,q) = \] 
\begin{eqnarray}
-i \left ( \begin{array}{cc}
R^{\dagger} V^{\dagger} \bvec{\Sigma} V R + T^{\dagger} \sigma_1
V^{\dagger} \bvec{\Sigma} V \sigma_1 T &
R^{\dagger} V^{\dagger} \bvec{\Sigma} V^{\ast} T^{\ast} +
T^{\dagger} \sigma_1 V^{\dagger} \bvec{\Sigma} V^{\ast} \sigma_1
R^{\ast}  \\
T^{T} V^{T} \bvec{\Sigma} V R + R^{T} \sigma_1
V^{T} \bvec{\Sigma} V \sigma_1 T &
T^{T} V^{T} \bvec{\Sigma} V^{\ast} T^{\ast} + R^{T} \sigma_1
V^{T} \bvec{\Sigma} V^{\ast} \sigma_1 R^{\ast} \end{array}
\right )
\label{hugeguy}
\end{eqnarray}
$\chi, \xi, V, \sigma_1, R$ and $T$ were all defined in the
sections on diagonalizing the models.
As can be explicitly checked, the $O(3)$ free bosons are analogous to the
NL$\sigma$ model with the function, $G(\theta)$, defined in Eqn.
(\ref{NLs_G}), set to one. This is the general result for the
$O(N)$ model for large $N$, and makes sense, since the Landau-
Ginsburg model is a large $N$ approximation to the NL$\sigma$
model. In case of axial symmetry, one need only substitute the
correct gaps into the energy factors:
\begin{eqnarray}
<k,a|l^i(0)|q,b> = i\epsilon^{iab} \frac{\omega^a_k +
\omega^b_q}{2\sqrt{\omega^a_k \omega^b_q}}
\label{shit}
\end{eqnarray}
with $\omega_k^a = \sqrt{ k^2v^2 + \Delta_a^2}$. 

The $O(3)$ fermion model exhibits a non-trivial $G(\theta)$-function.
We can use the results from Eqns. (\ref{Rlim}) and (\ref{Tlim})
in Eqn. (\ref{hugeguy}) to calculate that
\begin{eqnarray}
G(\theta) = - \mbox{sech}(\theta/2) = \left [ 
\sqrt{(\omega_k - k)(\omega_q - q)} + \sqrt{(\omega_k +
k)(\omega_q + q)} \right ] \frac{1}{\omega_k + \omega_q}
\end{eqnarray}
To obtain the $U(1)$ results we, again, make the gap
substitutions as done in Eqn. (\ref{shit}). This result is quite
different than the boson prediction. It, in fact, vanishes with
the gap for backscattering, $k \sim -q$. This is because the
$\bvec{l}$ does not couple left and right moving fermions
while the opposite holds true with the bosons (and NL$\sigma$
model). For small momentum exchange, all the models give
universal predictions for matrix elements of $S^z(0)$.
However, matrix elements of
$S^{\pm}(0)$ at small momentum exchange are somewhat sensitive to
the ratio of the gaps, in the boson model, while not at all so in
the fermion model. In Chapter 6 we discuss experiments which
might investigate this behaviour further.
 
When the symmetry is orthorhombic there are few conservation laws to
restrict the form of matrix elements of spin operators.
Furthermore, when a magnetic field is added, Lorentz invariance
is explicitly broken. We can, however, say that correlations
among spin operators are still diagonal:
This is true by virtue of the $Z_2 \times
Z_2$ symmetry. We also know that the new energy eigenstates,
labeled by $+$ and $-$, are mixtures of eigenstates of $S^3$ with
eigenvalues $s^z = \pm1$. This guarantees that 
\begin{eqnarray}
<k,+|S^i|q,-> \propto \delta_{i3}
\end{eqnarray}
It isn't terribly illuminating to write down the actual matrix
elements. We can say, however, that in the boson model,
for $h \rightarrow h_c$, all
matrix elements of form, $<-,k|l^i(0)|b,q>$, which are not zero
by arguments given above, diverge at $k=0$ as fractional powers
of $(h_c-h)$. Everything is nice and
finite with the fermions. This is another symptom of the sickness
of the {\em free} boson model near criticality. Again we see that
interactions are expected to play a crucial role in the boson
description.

We finish by describing some matrix elements near zero magnetic
field.  We expect that the intrabranch matrix elements,
$<\pm,k|S^3(0)|\pm,q>$, vanish at $k=q$ with the field. For
$k=q=0$ and $h \rightarrow 0$,  \\
\begin{eqnarray}
<-,0|l^3(0)|-,0> = \frac{h}{\Delta_2} \left ( \frac{ \Delta_1^2 +
3\Delta_2^2}{\Delta_1^2 - \Delta_2^2} \right ) \; \; \; \;
\mbox{bosons} \nonumber \\
= \frac{2h}{\Delta_1 - \Delta_2} \; \; \; \; \mbox{fermions}
\label{lowh_l}
\end{eqnarray}
where we've assumed, $\Delta_1 > \Delta_2$.
The result for $<+,0|l^3(0)|+,0>$ is obtained by exchanging $2$
and $1$. Notice that this limit does not commute with the $U(1)$
limit, $\Delta_2 \rightarrow \Delta_1$. This is to be expected
since these matrix elements are constant in the axially symmetric
case.


\chapter{Model Predictions for $T_1^{-1}$}
\resetcounters

Let's recall the expression for $1/T_1$, Eqn. (\ref{T_1}),
derived in Chapter 1:
\begin{eqnarray}
\frac{1}{T_1} = \int dt \; e^{-i\omega_Nt} 
\sum_{j,k,\mu,\nu} \bold{A}^{\nu -}_{ij} \bold{A}^{\mu +}_{ik}
< \left \{ S_j^{\nu}(t), S_k^{\mu}(0) \right \} >
\end{eqnarray}
where $h$ is taken to be in the $\hat{z}$ direction.
As discussed in (\ref{sec_2.2.2}), only diagonal components of
the spin correlation function will contribute. We also assume
that the hyperfine tensor, $\bold{A}_{ij}$, is local
\begin{eqnarray}
\bold{A}_{ij} = \bold{A} \delta_{ij}
\end{eqnarray}
Thus we can write
\begin{eqnarray}
\frac{1}{T_1} = \sum_{\nu} |\bold{A}^{+ \nu}|^2
\int dt \; e^{-i\omega_Nt} 
< \left \{ S^{\nu}(x=0,t), S^{\nu}(0) \right \} >
\end{eqnarray}
where we've used translational invariance to evaluate the
correlation function at the origin.
We can now take a step back to Eqn. (\ref{T_1_e}) and write the
above as
\begin{eqnarray}
\frac{1}{T_1} = 2\pi\sum_{n,n',\nu}|\bold{A}^{+ \nu}|^2 
|<n'|S^{\nu}(0)|n>|^2 \delta(E_{n'} - E_n - \omega_N)\frac{(e^{-
E_{n'}/T} + e^{-E_n/T})}{{\cal Z}}
\label{T_2_e}
\end{eqnarray}
We will concern ourselves largely with the limit, $\omega_N \ll T
\ll \Delta_{\mbox{min}}$ (note that $\omega_N \sim 1mK$ for
$H \sim 1.5$ T), so
that the last factor in (\ref{T_2_e}) can be set to $2e^{-E_n/T}$.

\begin{figure}
\centerline{
\epsfxsize=5 cm
\epsffile{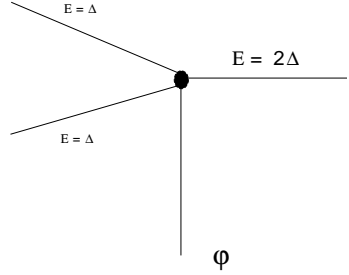}
}
\caption{First non-vanishing contribution to relaxation due to
the staggered part of the spin}
\label{fig:fig1}
\end{figure}

Consider now the operator in question, $\bvec{S}(0) = \bphi(0) +
\bphi(0) \times \bPi(0)$. We wish to investigate whether dominant
contributions to (\ref{T_2_e}) come from the staggered field,
$\bphi(0)$, or the uniform part of the spin, $\bphi(0) \times
\bPi(0)$. Let us first use the boson model to analyze the
staggered contributions. This will be\footnote{
There will not be contributions from cross terms between
the staggered and uniform fields. These vanish because for 
$<n|\bphi|m> \neq 0$, one needs the number of magnons, $n+m$,
to be odd, while this in turn implies $<n|\bvec{l}|m> = 0$. 
Certain types of structural perturbations, such as discussed
in Chapter 4, may change this analysis at high temperature and/or
high fields.}
\begin{eqnarray}
\frac{1}{T_1}_{\mbox{Stagg}} \propto \sum_{n,n'}
|<n'|\phi^{i}(0)|n>|^2 \delta(E_{n'} - E_n - \omega_N) e^{-E_n/T}
\end{eqnarray}
We assume, that we are in the regime $T \ll \Delta_{\mbox{min}}(h)$,
where $\Delta_{\mbox{min}}(h)$
is the lowest (possibly field dependent) gap, or in
other words, that the magnetic field is well below $h_c$, and we
are therefore well justified in using the boson model (or
NL$\sigma$ model) to describe the situation. Since $\bphi(0)$ is
a single magnon operator in the noninteracting theory, it only
has matrix elements between states whose energies differ by a
single magnon energy, $\omega(k,h)$; since $\phi$ is evaluated at
the origin, $k$ can be arbitrary. In particular, {\em there are
no matrix elements with energy difference}, $\omega_N$ (which is
essentially zero, compared to the other energy scales around).
Including interactions, there will be contributions at finite
$T$. The simplest process is shown in Fig. \ref{fig:fig1}.
It involves a $\phi^4$-type interaction, as might occur in the
Landau-Ginsburg or NL$\sigma$ model. The vertical line represents
the field $\phi$. The incoming line from the right represents a
thermally excited magnon of non-zero momentum, $k$, and energy
$2\Delta$ (for simplicity, we use the isotropic model at zero
field to make this argument; extending this to the anisotropic models
is straight forward). The two outgoing lines to the left
represent magnons at rest 
(recall that
the wave vector, $k$, is actually shifted by $\pi$, and so the 
lowest energy antiferromagnetic spin
excitations vary spatially as $e^{i\pi x}$).
This diagram gives a non-zero matrix
element proportional to $\lambda/\Delta^2$ (where $\lambda$
parametrizes the $\phi^4$ interaction). Note, however, that since
the initial and final state energies must be at least $2\Delta$,
there will be a Boltzmann suppression factor of $e^{-2\Delta/T}$
to this contribution. Thus 
\begin{eqnarray}
\frac{1}{T_1}_{\mbox{Stagg}} \propto \lambda^2 e^{-2\Delta/T}
\end{eqnarray}
Including anisotropy and a finite field will give various
contributions of this type. The greatest will be suppressed by
$\exp(-2\Delta_{\mbox{min}}(h)/T)$. It is also
consistent to interpret this result as giving the single magnon a
finite width at $T\neq 0$. This, however, cannot change the
conclusion that there is a double exponential suppression factor
contrary to the model proposed by Fujiwara et. al.  \cite{fuji}.

\begin{figure}
\centerline{
\epsfxsize=13.5 cm
\epsffile{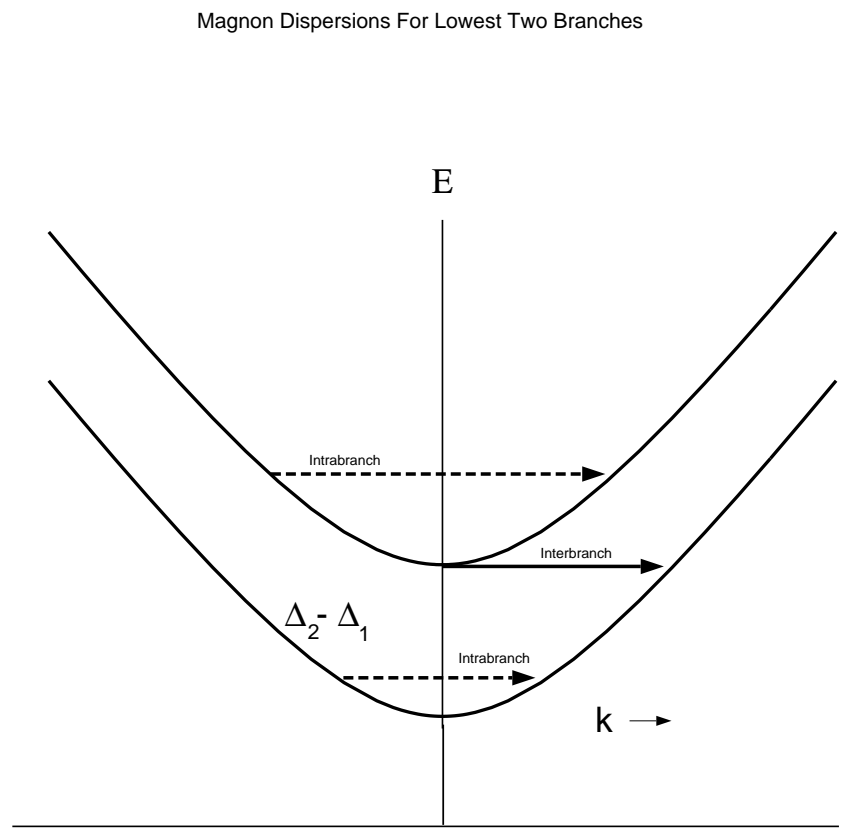}
}
\caption{Inter- vs Intrabranch transitions}
\label{fig:fig2}
\end{figure}

Let us now consider the contributions to $1/T_1$ from the uniform
part of the spin:
\begin{eqnarray}
\frac{1}{T_1}_{\mbox{Unif}} \propto \sum_{n,n'}
|<n'|\l^{i}(0)|n>|^2 \delta(E_{n'} - E_n - \omega_N) e^{-E_n/T}
\label{T_unif}
\end{eqnarray}
As discussed in Chapter 2, the 1-particle
matrix elements selected above are
non-zero in general, even in the non-interacting boson or fermion
model. This is because $l^b(0)$ is a two magnon operator, able to
create one magnon and annihilate another. In the presence of
anisotropy and magnetic field, the three magnon branches are
split, so we must distinguish between interbranch and 
intrabranch  transitions (see Fig. \ref{fig:fig2}). 
This is possible
since contributions may come from all wave vectors. One set of important
processes (ie. the ones corresponding to transitions between the
lowest energy magnon states) will come from single particle
intrabranch transitions along the lowest mixed branch
(intrabranch transitions are not allowed along the branch
corresponding to $s^z=0$ since $h$ is parallel to $z$). 
This will be one of the leading effects with
Boltzmann suppression of $e^{-\Delta_-(h)/T}$, where $\Delta_-(h)$ is
the lowest field dependent gap. It is important to
realize that these will only be present if the hyperfine
coupling, $\bold{A}^{+ z}$, is non-zero. In fact the zero
field gap structure and the choice of direction for placing the
magnetic field will affect whether there are competing
transitions.
The next contribution, possibly as significant as the one just
described, can come from intrabranch transitions along the second
lowest branch and/or interbranch transitions between the lowest
mixed branch and one of the other two branches. The Boltzmann
suppression factor will, again, favour the lowest energy processes
which occur at the gap to the highest branch involved in the
transition. The important point is that the Boltzmann suppression
factor for any of these processes is larger than that associated with
contributions from the staggered field. To summarize, the dominant
contribution to $1/T_1$ at $T \ll \Delta_{\mbox{min}}$, will come
from Eqn. (\ref{T_unif}).

As we approach the critical field, the above analysis breaks
down. As discussed previously, interactions are expected to
become large in the boson model. Moreover, as the gap closes, the
Boltzmann factors will fail to discriminate between the
contributions of the uniform and staggered fields to $1/T_1$.
Arguments involving the fermion model are tricky because the
staggered field has no simple representation in terms of
fermionic operators. However, we
explicitly show later in this chapter that the
staggered component will dominate sufficiently close to $h_c$. 

Above the critical field, the analysis depends on the symmetry.
For $U(1)$ or higher symmetry, the system remains critical and
the staggered correlator remains dominant. For lower
symmetry, the gap opens up once more, and sufficiently far above
the critical field, we expect the uniform correlator to dominate
once more.

\section{$T_1^{-1}$ for $h \ll h_c$}

In this section we concern ourselves with the regime discussed
above, $\omega_N \ll T \ll \Delta_{\mbox{min}}(h)$. We calculate
the relaxation rate in the isotropic, $U(1)$ and $Z_2$ scenarios.
We begin by deriving a general result valid in this regime, and
proceed to discuss its application in the different cases of
symmetry.

Consider a contribution to $1/T_1$ coming from transitions
between branches $r$ and $s$. Without loss of generality, we
assume that $r$ has a higher or equal gap to $s$ ($r$, in fact,
could be the same branch as $s$). We call the corresponding
contribution to Eqn. (\ref{T_2_e}), $\frac{1}{T_1}_{rs}$. This
will be a sum over single particle states on $r$ and $s$:
\begin{eqnarray}
\frac{1}{T_1}_{rs} = 4\pi\sum_{i} |\bold{A}^{+ i}|^2 \int
\frac{dk \; dq}{(2\pi)^2} \delta(\omega_s(q) - \omega_r(k) -
\omega_N) e^{-\omega_r(k)/T} |<k,r|l^i(0)|q,s>|^2 
\label{T_rs}
\end{eqnarray}
Note that there will be a similar contribution with the labels
$s$ and $r$ exchanged, if $s$ and $r$ are different branches.
This corresponds to scattering an initial particle on the $s$
branch through the hyperfine interaction to a final magnon on the
$r$ branch, or the reverse process. We take account of both of
these possibilities later. Also, we are keeping $\omega_N$ finite
to cut off infrared divergences which crop up in the intrabranch
processes. We now do the integral over $q$ to get
\[ \frac{1}{T_1}_{rs} =4\sum_{i} |\bold{A}^{+ i}|^2 \int_0^{\infty}
\frac{dk}{2\pi} \frac{(\omega_r(k) + \omega_N)}{Q(k)} \times \]
\begin{eqnarray}
e^{-\omega_r(k)/T}  \left ( |l^i_{r,s}(k,Q(k))|^2 +
|l^i_{rs}(k,-Q(k))|^2 \right ) \left ( \frac{\partial
\omega_s^2}{\partial q^2} \right )^{-1}_{q=Q(k)}
\label{T_rs2}
\end{eqnarray}
where $Q(k)$ is defined by $\omega_s(Q(k)) = \omega_r(k) +
\omega_N$, and $l^i_{r,s}$ is as defined in (\ref{l_def}).
When $\Delta_r(0) \gg T$, the above integral will be strongly
peaked at $k=0$. Moreover, we can neglect $\omega_N$ in
$\omega_r(k) + \omega_N$. The only factors in the integrand for
which we should retain a $k$-dependence are the exponential and
the possibly infrared divergent denominator, $Q(k)$. We therefore
write
\[ \frac{1}{T_1}_{rs} =\sum_{i} \frac{2|\bold{A}^{+ i}|^2
\omega_r(0)}{\pi} 
\left ( |l^i_{r,s}(0,Q(0))|^2 + |l^i_{rs}(0,-Q(0))|^2 \right )
\left ( \frac{\partial \omega_s^2}
{\partial q^2}\right )^{-1}_{q=Q(0)} \times \]
\begin{eqnarray}
\int_0^{\infty} dk \; \frac{e^{-\omega_r(k)/T}}{Q(k)}
\label{better}
\end{eqnarray}
It is not too difficult to show that to first order in small
quantities, $k^2$ and $\omega_N$, one has
\begin{eqnarray}
Q^2(k^2) \approx Q^2(0) + 
\left ( \frac{\partial \omega_s^2}
{\partial q^2}\right )^{-1}_{q=Q(0)}  
\left ( \frac{\partial \omega_r^2}
{\partial q^2}\right )_{q=0} k^2 
\end{eqnarray}
We can also expand the exponent:
\begin{eqnarray}
\omega_r(k)/T \approx \left ( \omega_r(0)/T + 
\frac{ \left ( \frac{\partial \omega_r^2}
{\partial q^2}\right )_{q=0} k^2}{2\omega_r(0) T} \right )
\label{exponent}
\end{eqnarray}
The relevant integral over $k$ then becomes
\begin{eqnarray}
\int_0^{\infty} dk \; \frac{e^{-\frac{ \left ( \frac{\partial
\omega_r^2}{\partial q^2}\right )_{q=0} k^2}{2\omega_r(0) T}} }{
\sqrt{Q^2(0) + \left ( \frac{\partial \omega_s^2}
{\partial q^2}\right )^{-1}_{q=Q(0)}  
\left ( \frac{\partial \omega_r^2}
{\partial q^2}\right )_{q=0} k^2 } } \label{big_int}
\end{eqnarray}
By changing variables, $\frac{ \left ( \frac{\partial
\omega_r^2}{\partial q^2}\right )_{q=0} k^2}{2\omega_r(0) T}
\rightarrow k^2$, we can write (\ref{big_int}) as
\begin{eqnarray}
\left ( \frac{\partial \omega_r^2}{\partial q^2}
\right )^{-\frac{1}{2}}_{q=0} \left ( \frac{\partial
\omega_s^2}{\partial q^2} \right )^{\frac{1}{2}}_{q=Q(0)}
\int_0^{\infty} dk \; \frac{e^{-k^2}}{\sqrt{k^2 +
\alpha_{rs}(T,h) } }
\end{eqnarray}
where
\begin{eqnarray}
\alpha_{rs}(T,h) = \frac{ Q^2(0) \left ( \frac{\partial
\omega_s^2}{\partial q^2} \right )_{q=Q(0)} }{2\omega_r(0)T}
\end{eqnarray}
The $h$-dependence of $\alpha_{rs}$ will largely come from its
dependence on $\omega_r(0)$. The integral can be expressed in
terms of special functions:
\begin{eqnarray}
\left ( \frac{\partial \omega_r^2}{\partial q^2}
\right )^{-\frac{1}{2}}_{q=0} \left ( \frac{\partial
\omega_s^2}{\partial q^2} \right )^{\frac{1}{2}}_{q=Q(0)}
e^{\alpha_{rs}(T,h)/2} K_0 \left ( \alpha_{rs}(T,h)/2 \right )
\end{eqnarray}
where $K_0$ is the zero order modified Bessel Function. When the
gap, $\omega_r(0)$, is very large compared to the typical
momentum, $Q(0)$, exchanged in the transitions, (this is the case
for intrabranch transitions), $\alpha_{rs} \rightarrow 0$. In
this limit
\begin{eqnarray}
e^{\alpha_{rs}(T,h)/2} K_0 \left (\alpha_{rs}(T,h)/2 \right )
\rightarrow
-\log \left ( \alpha_{rs}(T,h)/4 \right ) - \gamma
\end{eqnarray}
where $\gamma = 0.577216...$ is Euler's constant. 
We can now summarize
\[ \frac{1}{T_1}_{rs} =\sum_{i} \frac{2|\bold{A}^{+ i}|^2
\omega_r(0)}{\pi}
\left ( |l^i_{r,s}(0,Q(0))|^2 + |l^i_{rs}(0,-Q(0))|^2 \right )
e^{-\omega_r(0)/T} \]
\begin{eqnarray}
\left ( \frac{\partial \omega_s^2}
{\partial q^2}\right )^{-\frac{1}{2}}_{q=Q(0)}
\left ( \frac{\partial \omega_r^2}{\partial q^2}
\right )^{-\frac{1}{2}}_{q=0} 
e^{\alpha_{rs}(T,h)/2} K_0 \left ( \alpha_{rs}(T,h)/2 \right )
\label{t_rs_final}
\end{eqnarray}
The full expression for the relaxation rate is 
\begin{eqnarray}
\frac{1}{T_1} = \sum_{r,s} \frac{1}{T_1}_{rs}
\label{t_rs_final2}
\end{eqnarray}
The effect of interchanging $s$ and $r$ in (\ref{T_rs}) 
is therefore included in the above. 

An important thing to learn from the above calculation is that
contributions from transitions between states involving small
momentum exchange ($Q \rightarrow 0$) will dominate due to the
logarithmic divergence in (\ref{t_rs_final}). This is
particularly the case with intrabranch versus interbranch
transitions. In intrabranch transitions one is allowed momentum
exchanges as small as $Q \sim \sqrt{2\Delta_r(0) \omega_N}/v$.
This will typically be much smaller than the smallest allowed
interbranch momentum exchange, $Q \sim (\Delta_r(0) -
\Delta_s(0))/v$. The conclusion is that, unless the branches in
question are extremely close to each other, interbranch
transitions will play a secondary role to intrabranch processes,
even ignoring the more obvious suppression due to different
Boltzmann factors. Of course, if the hyperfine interaction has
high symmetry, one will not see intrabranch transitions at all.
This suggests that an NMR relaxation study could provide
information as to the nature of the hyperfine tensor.

Before expounding on this result in the individual cases of
different symmetry, we would like to mention the effects of higher
temperature, or correspondingly, including the $k$-dependence of
the various terms approximated at $k=0$. In the more general $Z_2
\times Z_2$ situation\footnote{One is more fortunate in the
$U(1)$ case; because of the simplicity of the gaps, the expansion
is good for $k \leq \Delta_{\perp}$, regardless of the value of
$h$.}, we are strictly justified in expanding the exponent in
Eqn. (\ref{exponent}) only for $T \ll \Delta_r(h)$. For higher
temperatures, one expects contributions from $k > \Delta_r(h)/v$,
where the expansion is not convergent; in this case one is better
off numerically integrating (\ref{better}) (making the $k=0$
approximation for the other terms is still valid, as we will
see). In either case, we can estimate the error in neglecting
terms of order $k^{2n}$. First, notice that all gaps are always
greater than $\sqrt{v^2k^2 + \Delta_m^2}-h$, where $\Delta_m$ is
the smallest zero field gap. We therefore write
\begin{eqnarray}
\int dk k^{2n} \frac{e^{-\omega_r(h,k)/T}}{\sqrt{k^2 + Q^2(k)}} <
e^{h/T} \int dk k^{2n-1} e^{-\sqrt{v^2k^2 - \Delta_m}/T} \nonumber \\
= \frac{e^{h/T}}{2 v^{2n}} \int_{\Delta_m}^{\infty} d\omega
e^{-\omega/T}
\omega (\omega^2 - \Delta_m^2)^{n-1} < \frac{e^{-(\Delta_m-h)}}{4}
\left ( \frac{2 \mbox{Max}(T,\Delta_m)}{v} \right )^{2n}
\end{eqnarray}
The last estimate is actually quite generous, especially for
large $n$. In the worse case scenario of the Haldane phase,
$\Delta/v \sim 1/4$. This allows us to expect an error of at most
$10 \%$ in neglecting the $k$-dependence of the terms in
(\ref{better}). 

We still have to estimate the error incurred in
making the expansion in the exponential at $T \ll \Delta_m$. The
next term in the expansion is $-\frac{k^4v^4_r}{8\Delta_r^3 T}$,
with $v^2_r=\left ( \frac{\partial \omega_r^2}{\partial q^2} \right )_{q=0}$.
This will give a
contribution,
\begin{eqnarray}
e^{-\omega_r(0)/T}
\int dk \; \frac{k^4v^4_r}{8\Delta_c^3 T} \frac{e^{-k^2 v^2_r/2 \omega_r(0)
T}}{\sqrt{k^2 + Q^2}} < \frac{Te^{-\omega_r(0)/T}}{4\Delta_r}
\end{eqnarray}
This is potentially more serious as $T \rightarrow \Delta_r$ or 
$\Delta_r \rightarrow 0$. To
summarize, Eqn. (\ref{better}) is generally a very good
approximation; when $T \ll \Delta_r$, one can safely expand the
exponential, while for $T \geq \Delta_r$, one is better off
numerically integrating (\ref{better}).

\subsection{Isotropic Symmetry}

\begin{figure}
\centerline{
\epsfxsize=8 cm
\epsffile{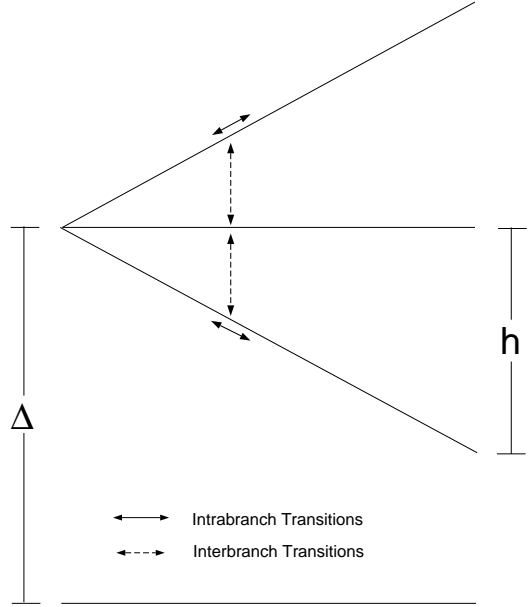}
}
\caption{The gap structure for $O(3)$ symmetry.}
\label{fig:o3gaps}
\end{figure}

In the case of $O(3)$ symmetry, the field dependent gap structure is
as in Fig. \ref{fig:o3gaps}. The lowest and highest branches
correspond to magnons with $s^z = \pm 1$ respectively. The middle
branch corresponds to $s^z = 0$. The interbranch gap is $h$.
Using the result of the previous section and that of
\ref{sec_2.1.3} we can immediately write down the intrabranch
contributions to $1/T_1$:
\begin{eqnarray}
\frac{1}{T_1}_{\mbox{Intra}} = |\bold{A}^{+ z}|^2
\frac{4\Delta}{\pi v^2} [\log(4T/\omega_N) - \gamma] (e^{-
(\Delta-h)/T} + e^{-(\Delta+h)/T} )
\end{eqnarray}
where in this case, $Q_0^2 = 2\omega_N \Delta /v^2$. It is quite
likely that higher dimensional effects may cut off this
contribution at energy scales larger than $\omega_N$. For
example, weak interchain couplings, $J_I$, would replace
$\omega_N$ in the above expression by a quantity of order
$J_I$.

The interbranch contributions between the lower two and higher
two branches can be likewise calculated to give
\begin{eqnarray}
\frac{1}{T_1}_{\mbox{Inter}} = (|\bold{A}^{+x}|^2 + |\bold{A}^{+
y}|^2) \frac{4\Delta}{\pi v^2} e^{(h + \frac{h}{2\Delta})/2T}
K_0((h + \frac{h}{2\Delta})/2T)
(e^{-\Delta/T} + e^{-(\Delta+h)/T})
\end{eqnarray}

Following Fujiwara et. al. \cite{fuji}, we write
\begin{eqnarray}
\frac{1}{T_1} = \frac{1}{T_1}_{\mbox{Intra}} +
\frac{1}{T_1}_{\mbox{Inter}} \equiv F(h,T) e^{-(\Delta - h)/T}
\label{F_fuji}
\end{eqnarray}
We see that the nature of $F(h,T)$ depends largely on the form of
the hyperfine coupling.
A less general and somewhat more qualitative version of this formula
was given by Jolic{\oe}ur and Golinelli \cite{joli}, and by Troyer
et. al. \cite{troy}, independently of our work. Jolic{\oe}ur and
Golinelli discussed the isotropic NL$\sigma$ model and derived
only the leading exponential dependence on temperature; Troyer
et. al. considered the Heisenberg ladder problem, which has a
low energy one-magnon
excitation spectrum identical to that in the
isotropic NL$\sigma$ model, and only included the leading interbranch
transition in their expression.

\subsection{Axial Symmetry}

Here we are faced with two possible situations: the $s^z = 0$
branch can lie above or below the doublet. In the former case,
the larger the interbranch gap between the doublet and the
singlet branches, the more suppressed will be the interbranch
contributions to $1/T_1$. On the other hand, in the latter
scenario, inter- and intrabranch contributions will always be on
the same footing (see Fig \ref{fig:u1gaps}). 
\begin{figure}
\centerline{
\epsfxsize=13.5 cm
\epsffile{fig_u1gaps}
}
\caption{The gap structure for $U(1)$ symmetry. 
(a) $\Delta_3 > \Delta_{\perp}$; (b) $\Delta_3 < \Delta_{\perp}$}
\label{fig:u1gaps}
\end{figure}
The expression for
the intrabranch transitions will be essentially identical to the
one in the $O(3)$ case:
\begin{eqnarray}
\frac{1}{T_1}_{\mbox{Intra}} = |\bold{A}^{+ z}|^2
\frac{4\Delta_{\perp}}{\pi v^2} [\log(4T/\omega_N) - \gamma]
(e^{-(\Delta_{\perp}-h)/T} + e^{-(\Delta_{\perp}+h)/T} )
\end{eqnarray}
where $\Delta_{\perp}$ is the gap to the $s^z=\pm 1$ branches 
and $\Delta_3$ is the gap to the $s^z=0$ branch.
The corresponding formula for the interbranch transitions is
somewhat more subtle and depends on the position of the singlet
with respect to the doublet as well as on the size of the
magnetic field. This is because $\frac{1}{T_1}_{rs}$ in
(\ref{t_rs_final}) depends on which branch lies higher and the
size of the gap between them. When the singlet sits higher,
\[ \frac{1}{T_1}_{\mbox{Inter}} = (|\bold{A}^{+x}|^2 + |\bold{A}^{+
y}|^2) \frac{2}{\pi v^2} \]
\[ \left \{ e^{\alpha_{1\; rs}/2}
K_0(\alpha_{1\; rs}/2) e^{-\Delta_3/T} \Delta_3 \left (
|l^x_{0 -}(0,Q_1(0))|^2 + |l^x_{0 -}(0,-Q_1(0))|^2 \right ) \right . \]
\begin{eqnarray}
\left. + e^{\alpha_{2\; rs}/2}
K_0(\alpha_{2\; rs}/2) e^{-M(h)/T} \tilde{M}(h)  \left (
|l^x_{0 -}(0,Q_2(0))|^2 + |l^x_{0 -}(0,-Q_2(0))|^2 \right ) \right \}
\end{eqnarray}
where
\[ \alpha_{1\; rs} = \frac{(\Delta_3 + h)^2 -
\Delta_{\perp}^2}{2\Delta_3 T} = \frac{Q_1^2(0) v^2}{2\Delta_3 T} \]
\begin{eqnarray}
\alpha_{2\; rs} = \left \{ \begin{array}{lr}
\frac{(\Delta_3 - h)^2 - \Delta_{\perp}^2}{2\Delta_3 T} = \frac{Q_2^2(0) v^2}{
2\Delta_3 T} &
h<\Delta_3 - \Delta_{\perp} \\
\frac{(\Delta_{\perp} + h)^2 - \Delta_3^2}{2\Delta_{\perp} T}=
\frac{Q_2^2(0) v^2}{2\Delta_{\perp} T} &
h>\Delta_3 - \Delta_{\perp} \end{array} \right. \nonumber \\
M(h) = \left \{ \begin{array}{lr}
\Delta_3 = \tilde{M}(h) & h<\Delta_3 - \Delta_{\perp} \\
\Delta_{\perp}+h = \tilde{M}(h) + h 
& h>\Delta_3 - \Delta_{\perp} \end{array} \right.
\end{eqnarray}
In the other case, when the singlet sits lower than the doublet,
one has
\[ \frac{1}{T_1}_{\mbox{Inter}} = (|\bold{A}^{+x}|^2 + |\bold{A}^{+
y}|^2) \frac{2}{\pi v^2} \]
\[ \left \{ e^{\alpha_{1\; rs}/2}
K_0(\alpha_{1\; rs}/2) e^{-(\Delta_{\perp}+h)/T} \Delta_{\perp} \left (
|l^x_{0 -}(0,Q_1(0))|^2 + |l^x_{0 -}(0,-Q_1(0))|^2 \right ) \right. \]
\begin{eqnarray}
\left. + e^{\alpha_{2\; rs}/2}
K_0(\alpha_{2\; rs}/2) e^{-M(h)/T} \tilde{M}(h)  \left (
|l^x_{0 -}(0,Q_2(0))|^2 + |l^x_{0 -}(0,-Q_2(0))|^2 \right ) \right \}
\end{eqnarray}
where this time,
\[ \alpha_{1\; rs} = \frac{(\Delta_{\perp} + h)^2 -
\Delta_3^2}{2\Delta_{\perp} T} = \frac{Q_1^2(0) v^2}{2\Delta_{\perp} T} \]
\begin{eqnarray}
\alpha_{2\; rs} = \left \{ \begin{array}{lr}
\frac{(\Delta_{\perp} - h)^2 - \Delta_3^2}{2\Delta_{\perp} T} 
= \frac{Q_2^2(0) v^2}{
2\Delta_{\perp} T} &
h<\Delta_{\perp} - \Delta_3 \\
\frac{(\Delta_3 + h)^2 - \Delta_{\perp}^2}{2\Delta_3 T}
=\frac{Q_2^2(0) v^2}{2\Delta_3 T} &
h>\Delta_{\perp} - \Delta_3 \end{array} \right. \nonumber \\
M(h) = \left \{ \begin{array}{lr}
\Delta_{\perp}-h = \tilde{M}(h)-h & h<\Delta_{\perp} - \Delta_3 \\
\Delta_3 = \tilde{M}(h) 
& h>\Delta_{\perp} - \Delta_3 \end{array} \right.
\end{eqnarray}
A seeming catastrophe occurs when two of the branches cross at
$h=|\Delta_3 - \Delta_{\perp}|$. The interbranch contribution to
$1/T_1$ diverges logarithmically. There are essentially two
effects that would cut off this divergence. Higher dimensional
couplings can be counted on once more to replace $Q_2^2(0)$
as it approaches zero, with a quantity of order
$10J_I/J$ as derived in Eqn. (\ref{cutoff_c}) in Chapter Four. 
Also, the divergence in the integrand
leading to this problem is $\sim \frac{1}{\sqrt{h - |\Delta_3 -
\Delta_{\perp}|} }$. This will be cured by a field with finite
width. One still expects a peak in the relaxation rate, but this
will be smoothed by the mentioned effects. 

\subsection{$Z_2 \times Z_2 \times Z_2$ Symmetry}

There isn't much more to say which would be illuminating in this
case. We can, however, easily give the results for intrabranch
contributions. These behave as the analogous expressions from the
more symmetric situations.
\[ \frac{1}{T_1}_{\mbox{Intra}} = \frac{4|\bold{A}^{+z}|^2}{\pi}
(\log(4T/\omega_N) - \gamma) \times \] 
\begin{eqnarray}
\left ( |l^z_{- -}(0,0)|^2 \omega_-(0)
\left (\frac{\partial \omega_-^2}{\partial k^2} \right )^{-1}_{k=0}
e^{-\omega_-(0)/T} + ( - \leftrightarrow +)
\right )
\label{z2_intra}
\end{eqnarray}
$|l^z_{- -}(0,0)|^2$ depends on $h$ as per Eqn. (\ref{lowh_l}).
The formulae for interbranch transitions will, again, depend on
the positions of the branches and the relative gaps between
branches. Note that if there are intrabranch transitions allowed
by the hyperfine coupling, then there will also be transitions
between the $+$ and $-$ branches.  Finally, from Eqn. (\ref{lowh_l}),
we see that (\ref{z2_intra}) vanishes quadratically with the field.

\section{Close to the Critical Field}

In this section we give qualitative results on the behaviour of
the relaxation rate and in the process prove that the
contribution from the staggered correlator becomes crucial as $h
\rightarrow h_c$. We assume that we are now in the regime $|h_c -
h| \ll T$. In this limit, intrabranch processes along the lowest
branch will dominate. Even if the hyperfine tensor possesses high
symmetry (thereby ruling out intrabranch contributions from the
uniform spin operator), we expect intrabranch contributions from
the {\em staggered} part of the spin. Since the fermion model
becomes exact in this limit, we will rely on its predictions.
Long wavelength modes are now expected to play the most
important role; we therefore write the dispersion relation of the
lowest branch as
\begin{eqnarray}
\omega(k,h) = \left \{ \begin{array}{lr}
(h_c - h) + \frac{v^2k^2}{2h_c} & O(3) \; \; \; \mbox{and} \; \;
\; U(1) \; \; \;  \mbox{cases} \\
\sqrt{v_e^2k^2 + \Delta_e^2} & (Z_2)^3 \; \; \; \mbox{case}
\end{array} \right.
\end{eqnarray}
where $v_e^2 = v^2 \frac{(\Delta_1-\Delta_2)^2}{(\Delta_1 +
\Delta_2)^2}$, and the effective gap is $\Delta_e =
2(h-h_c)h_c/(\Delta_1 + \Delta_2)$. Now that the gap is actually
smaller than the temperature, we must include multiparticle
processes. This is simply done by replacing the Boltzmann weight
by the appropriate occupation factors, $f_f(\omega) (1 -
f_f(\omega))$. The derivation is straight forward and can be
found in standard texts on many body physics (for example, see
\cite{mah}). The uniform contribution to the relaxation rate is
given by
\begin{eqnarray}
\left ( T_1^{-1} \right ) = \frac{4|\bold{A}^{+ z}|^2}{\pi}
|l^z_{- -}(0,0)|^2 \int_0^{\infty} dk \frac{(\omega+ \omega_N)
f_f(\omega) (1 - f_f(\omega))}{Q(k)}
\left ( \frac{\partial \omega^2}{\partial k^2} \right )^{-1}
\end{eqnarray}
Due to the simple form of the density of states in the isotropic
or axially symmetric case, one still obtains logarithmic
behaviour for the above formula. In the anisotropic case, things
are a bit different. We can combine the last expression with the
results from \ref{sec_2.2.2} to get
\begin{eqnarray}
\left ( T_1^{-1} \right )_{\mbox{Unif}} = 
\frac{4|\bold{A}^{+ z}|^2}{\pi v^2}
\frac{\Delta_1 \Delta_2}{(\Delta_1 - \Delta_2)^2}
\int_0^{\infty} v_e \; dk \frac{(\omega+ \omega_N)
\mbox{sech}^2(\frac{\omega}{2T}) }{\sqrt{(\omega+\omega_N)^2 -
\Delta^2_e} }
\end{eqnarray}
At criticality we set $\Delta_e = 0$. We may simply rescale the
integration variable to obtain
\begin{eqnarray}
\left ( T_1^{-1} \right )_{\mbox{Unif}} \propto T
\label{unif_cor}
\end{eqnarray}
This is expected from the Ising model where the uniform part of
the spin corresponds to the Ising energy density
operator\footnote{The rest of the analogy goes as follows: the
magnetic field plays the role of temperature as is obvious from
the form of the spectrum; the inverse temperature is analogous to
the size of the system in the Euclidean time direction; finally,
the staggered magnetization corresponds to the disorder field,
$\sigma$.}, $\epsilon$, of scaling dimension 1. In terms of
Majorana fermions this operator is $\psi_L \psi_R = \epsilon$.
The correlator of the energy density
operator on the infinite Euclidean plane is known from its
scaling dimension and the restrictions of conformal field theory
to be \cite{itzy}
\begin{eqnarray}
<\epsilon(z) \epsilon(0)> = 1/|z|^2
\end{eqnarray}
If periodic boundary conditions
are placed in the time direction (corresponding to
finite temperature), the correlator can be obtained by making a
conformal transformation from the Euclidean plane into the
cylinder (see \cite{itzy,car}), $z_p = e^{2\pi iz_sT}$:
\begin{eqnarray}
<\epsilon(z) \epsilon(0)> = \frac{(\pi T/v_e)^2}{|\sin(T\pi z)|^2}
\end{eqnarray}
Setting $z = it + \delta$, we can get the contribution to 
$T_1^{-1}$ by integrating over $\int dt e^{-i\omega_N t}$:
\begin{eqnarray}
\left ( T_1^{-1} \right )_{\mbox{Unif}} \propto
\int dt \; e^{-i\omega_N t} \frac{(\pi T/v_e)^2}{|\sinh(T\pi[t -
i\delta] )|^2}
\end{eqnarray}
Changing variables, and assuming the integral is analytic as
$\omega_N \rightarrow 0$ (this can actually be proven by contour
techniques), we see that by rescaling the time variable we
reproduce Eqn. (\ref{unif_cor}).

We now turn our attention to the behaviour of the staggered
correlator at criticality. We know the form of this function in both
$U(1)$ (and therefore $O(3)$) and Ising cases from Ref.
\cite{Aff}. On the infinite Euclidean plane we have
\begin{eqnarray} \begin{array}{lr}
<\phi^{\dagger}(z) \phi(0)> \sim |z|^{-\frac{1}{2}} & U(1) \\
<\sigma(z) \sigma(0)> \sim  |z|^{-\frac{1}{4}} &  \mbox{Ising}
\end{array}
\end{eqnarray}
The field $\phi = \phi^x + i \phi^y$ is the charged $U(1)$ field
of the boson model; there are no problems in using this (in the
$U(1)$ case) as long as we account for the interactions. $\sigma$
is the disorder operator of the Ising model. It is highly non-local
in fermionic language, and aside from its dual, the order
operator, and the energy density operator, is the only primary
operator in the model. Once more, making a conformal
transformation into the cylinder of circumference $1/T$,
\begin{eqnarray}  \begin{array}{lr}
<\phi^{\dagger}(z) \phi(0)> \sim \frac{(2\pi T \Delta/v^2)^{\frac{1}{2}}}
{|\sin(\pi z T)|^{\frac{1}{2}}} & U(1) \\
<\sigma(z) \sigma(0)> \sim \frac{(\pi T/v_e)^{\frac{1}{4}}}
{|\sin(\pi z T)|^{\frac{1}{4}}} &  \mbox{Ising} \end{array}
\end{eqnarray}
Once more, setting $z = it + \delta$ and integrating over time,
we get that in the experimentally important limit, $T \gg
\omega_N$, 
\begin{eqnarray} 
\left ( T_1^{-1} \right )_{\mbox{Stag}} \propto 
(2\pi T\Delta/v^2)^{-\frac{1}{2}}
+ O(\omega_N) \; \; \; \; \; U(1) \nonumber \\
\left ( T_1^{-1} \right )_{\mbox{Stag}} \propto (\pi T/v_e)^{-\frac{3}{4}}
+ O(\omega_N) \; \; \; \; \; \mbox{Ising} \nonumber \\
\label{T_1_crit}
\end{eqnarray}
For both symmetries, this implies a
significantly stronger contribution from the staggered part than
the uniform part. In fact, as long as we are sufficiently close
to the critical regime, perturbation theory tells us that the
above result will only be suppressed by factors of order 
$O(|h-h_c|/T)$. In order to observe this behaviour experimentally,
one must have $T$ sufficiently large (having a large anisotropy,
$\Delta_1 - \Delta_2$, would also help), so that the decrease in
relaxation with temperature is obvious. This would require that the
experiment be done over a broad range of temperatures so that any
constant contributions to $1/T_1$ could be subtracted. In any case,
the above should at least serve to clarify that the staggered contribution
becomes important in this regime.

Farther still from criticality, the analysis breaks down. We do,
however, expect the staggered spin contribution to influence
$T_1^{-1}$ through to the region $T \sim |h-h_c|$.

\section{Above the Critical Field}

Far above the critical field, $h - h_c \gg T$, the situation
becomes even simpler. In the $O(3)$ and $U(1)$ case the system
remains critical. The relaxation rate will be dominated by the
staggered part of the spin. The fact that $\bphi$ develops a
vacuum expectation value (or likewise, the non zero magnetization
of the ground state) will have no effect on the relaxation rate
since $\omega_N$ isn't strictly zero:
\begin{eqnarray}
\int dt <0|S^i(t,0)|0><0|S^i(0,0)|0> e^{-i\omega_Nt} =
2\pi (M^i)^2\delta(\omega_N)/L^2
\end{eqnarray}
where $\bvec{M}$ is the magnetization. We can therefore say that
for the $O(3)$ and $U(1)$ models, the relaxation rate 
(assuming $\omega_N \ll T$) has the simple
temperature dependence
\begin{eqnarray}
\frac{1}{T_1} \propto (2\pi \Delta_{\perp} T/v^2)^{\eta-1}
\label{T_u1_crit}
\end{eqnarray}
where $\eta$ is the critical exponent of the staggered  spin
correlator. Haldane argued $\eta = \frac{1}{2} + O(\rho)$, where
$\rho = |\bvec{M}|/L$ \cite{hal_crit}. Thus the field dependence 
of $1/T_1$ is only through $\eta$.

When axial symmetry is broken, the gap reappears for $h>h_c$. In
this case, one can use the fermion model to calculate the
relaxation rate. This is made much simpler since the gaps to the
two upper branches are presumably much higher than the lower gap
(by at least $2\Delta_-(h)$). Therefore only intrabranch processes
along the lower branch need be considered. The result is 
\begin{eqnarray}
\frac{1}{T_1}=\frac{4 |\bold{A}^{+3}|^2}{\pi} e^{-\Delta_-/T} \Delta_-
\left ( \frac{\partial \omega_-^2}{\partial k^2} \right
)_{k=0}^{-1} |l^3_{-,-}(0,0)|^2 (\log(4T/\omega_N) - \gamma)
\end{eqnarray}
At sufficiently large magnetic field, $h \gg \Delta_1-\Delta_2$,
some of the expressions simplify:
\begin{eqnarray}
|l^3_{-,-}(0,0)|^2 \rightarrow 1 \; \; \; \; \; \; \; \Delta_-
\rightarrow h - (\Delta_2 + \Delta_1)/2 \nonumber \\
\left ( \frac{\partial \omega_-^2}{\partial k^2} \right )_{k=0}
\rightarrow \frac{2v^2}{\Delta_2 + \Delta_1} \left ( h -
\frac{\Delta_2 + \Delta_1}{2}  \right )
\end{eqnarray}
The rate will drop exponentially with increasing magnetic field.

\section{Summary}

We would like to summarize the main results of this section. At 
temperatures much lower than the lowest 
gap\footnote{We refer to the lowest gap corresponding to a polarization
direction perpendicular to the field.}, $\Delta_m$
(this can be below the critical field
or far above it in the case of $Z_2$ symmetry), $T^{-1}_2 \sim 
e^{-\Delta_m/T}$. This is due to the dominance of two magnon 
intrabranch relaxation processes. In the case of axial or higher symmetry,
the only temperature and field dependence comes from 
\begin{eqnarray}
\frac{1}{T_1}_{\mbox{Intra}} \propto [\log(4T/\omega_N) - \gamma]
e^{-(\Delta-h)/T}
\end{eqnarray}
Given such symmetry, this is a model independent result. 
Including anisotropies and interbranch
processes, the relaxation rate is given by Eqns. (\ref{t_rs_final})
and (\ref{t_rs_final2}). 

When the lowest gap is much smaller than the temperature, the dominant
contributions come from one magnon processes (due to the staggered part
of the spin). When the field reaches its critical value, whereupon the 
gap vanishes, the relaxation rate is given by Eqn. (\ref{T_1_crit}).
Above the critical field,
the system remains critical with axial symmetry, and the rate is then
given by Eqn. (\ref{T_u1_crit}). With Ising symmetry, the gap reopens and
eventually becomes large once more.


\chapter{Material Properties and Possible Effects on Experiment}
\resetcounters

\section{Hyperfine Tensor}
Here we briefly discuss the effect of the nature of the Hyperfine
tensor, $\bold{A}^{\mu \nu}_{ij}$, on the NMR relaxation rate. As
a reminder, $\mu$ and $\nu$ are spin indices while $i$ and $j$
are spatial indices coupling the nuclear spin at site $j$ to the
atomic spin on site $i$. 

Assuming the magnetic field lies in the $z$ direction,
if $\bold{A}^{\mu \nu}_{ij}$ is isotropic in its spin indices,
only $|\bold{A}^{+ -}|^2$ will contribute in Eqn. (\ref{T_1}). In
particular, there will be no intrabranch contributions as these
require a coupling to $S^z_i$. Interbranch transitions will not
be limited, but no contribution to them will come
from the term proportional to $\bold{A}^{+ +} \bold{A}^{- -}$.
These statements still
hold true for a hyperfine tensor diagonal in the Heisenberg spin
basis. Note that this implies that for the $O(3)$ model,
intrabranch transitions are prohibited so long as one assumes
that the nuclear gyromagnetic tensor is simultaneously
diagonalizable with the hyperfine tensor.

In general, especially if the NMR nucleus does not coincide with
the magnetic ion giving rise to the effective spin in the 1DHAF,
the anisotropies on the spin chain will not be simultaneously
diagonalizable in the hyperfine tensor basis. Moreover, if there
is more than one nuclear moment per spin contributing to the
signal, it is unlikely that the effective 
$\bold{A}^{\mu \nu}_{ij}$ will have the same symmetry as the
nuclear Zeeman interaction. Thus conditions have to be quite
convenient for intrabranch transitions to be missing from the
rate. This can be important at very low temperatures where we can
experimentally distinguish between the processes. First of all,
intrabranch transitions along the `$-$' branch will increase
exponentially with field (for example, in the $O(3)$ symmetric case,
the behaviour is $e^{-(\Delta-h)/T}$). This can be a most obvious difference
at low temperatures.  However, as is clear from the discussion in
the last chapter, the `$-$' gap can lie above the $s^z=0$ gap
(for example, if one places the field along the $a$-chain
direction in NENP); thus both inter- and intrabranch processes
will feature the same exponential rise with field. Also, it is
possible that the lower gap depends very weakly on the field for
certain magnetic field directions. This is true for anisotropic
materials where it is difficult to place the field along a
direction of symmetry local to the chain. There are two ways to
distinguish the transitions in these cases. The simplest solution
is to repeat the experiment changing the magnetic field direction
until one clearly sees the $e^{-\Delta_-(h)/T}$ behaviour.
Alternatively, one can try to extract information out of the low
temperature behaviour, rather than field dependence. For $T \sim
h$, F(h,T) in Eqn. (\ref{F_fuji}) will behave as $\log(T/\omega_N) -
\gamma$ if intrabranch transitions are allowed. If they are
prohibited, $F(h,T)$ will more likely behave as $\sim
\sqrt{T/(\delta + h)}$ where $\delta$ is roughly the smallest
interbranch gap at zero field. 

In principle, if one has enough information about the gap
structure of the chain, it is possible to deduce the relative
values of the hyperfine matrix elements from the relaxation rate.
This may be done by comparing the ratios of the rates measured
with magnetic field along each of the effective gap directions
(or in case of high symmetry, any three perpendicular
directions); assuming one has extracted the intrabranch
contributions from the measurements of $1/T_1$, one can then 
work backwards using
\begin{eqnarray}
\frac{1}{T_1}_{\mbox{Intra}} \propto |\bold{A}^{+ 3}|^2
\end{eqnarray}
(3 corresponds to the field direction) to arrive at ratios of the
hyperfine matrix elements. This will be, presumably, model
dependent even in the low field limit. In Chapter 6, we suggest
experiments which would distinguish between the models.

Finally, we discuss nearest neighbour effects of 
$\bold{A}^{\mu \nu}_{ij}$. When these exist, contributions to
$1/T_1$ will come from the correlation function,
\begin{eqnarray}
<S^{\mu}_i(t) S^{\nu}_{i+1}(0)> = \frac{1}{N} \sum_n
e^{-2\pi in/N} <S_n^{\mu}(t) S_{-n}^{\nu}(0)>
\end{eqnarray}
where $S_n$ is the $n$th fourier mode. Taking $N \rightarrow \infty$,
we write
\begin{eqnarray}
<S^{\mu}_i(t) S^{\nu}_{i+1}(0)> = <S^{\mu}_i(t) S^{\nu}_i(0)>
 + \int \frac{dk}{2\pi} \left ( e^{-ik} - 1 \right ) 
<S_k^{\mu}(t) S_{-k}^{\nu}(0)>
\end{eqnarray}
Expanding in $k$, the odd contributions in $k$ from the second term
will vanish when integrated over all states.
From the work done in the last chapter, we 
know that powers of $k^2$ will give small perturbations of order
$2\Delta^2/v^2$. We see that including nearest neighbour
contributions, the relaxation will be essentially given by making the
substitution,
\begin{eqnarray}
\bold{A}^{+ \mu} \rightarrow \bold{A}^{+ \mu} + \bold{A}^{+ \mu}_{nn}
\end{eqnarray}
in all previous expressions where, $\bold{A}^{+ \mu}_{nn}$,
is the nearest neighbour hyperfine coupling. 

\section{Impurities}

So far, we have dealt with a single, infinitely long spin chain.
In real experiments, however, chains are always finite, and they
come in three dimensional crystals, and so there are many chains
of varying lengths in each sample. In this section we deal with
the fact that these chains often end or have defects. This is
what we mean by impurities. Of course one can introduce doping
(for example, replace some $Ni$ sites in NENP with $Cu$) to
explore the issue further; we will restrict ourselves to `pure'
samples, although our treatment can be extended to doped samples.

We start by describing a `finite' chain. Exact work on a related
$S=1$ Hamiltonian -- the `valence bond solid' \cite{valence},
which also features a unique ground state (in the thermodynamic
limit) and a gap -- has indicated that at the ends of a finite
chain there are free spin-$\frac{1}{2}$ degrees of freedom. It is
expected that the same holds true for finite chains of the
Heisenberg antiferromagnet. Indeed, there has been convincing
numerical and experimental evidence for this conjecture 
\cite{spin_end, white_2}. The same evidence also supports the
notion that these nearly free end spins can interact with each
other, essentially exchanging virtual NL$\sigma$ bosons. The
interaction is exponentially decaying with the size of the chain: 
\begin{eqnarray}
H_I \propto (-1)^L e^{-L/\xi} \bvec{S}_0 \cdot \bvec{S}_L
\end{eqnarray}
Here, the correlation length, $\xi$, is roughly six lattice
spaces. For experimentally realizable lengths, this interaction
is negligible and the finite chain has a fourfold degenerate
ground state. 

In a `pure' sample, these finite chains will lie end to end, or
be separated by some non-magnetic defect. It is therefore
reasonable to presume that two adjacent end spins will interact
with an effective exchange coupling, $J'$, that will vary in
strength from zero to something of order $J$. Recent work
\cite{imps} has explored this situation in depth. For weak
coupling between adjacent end spins the effective Hamiltonian can
be written,
\begin{eqnarray}
H_E = J' \bvec{S}_1 \cdot \bvec{S}_2 = \alpha^2 J' \bvec{S}_1'
\cdot \bvec{S}_2'
\label{c_ham}
\end{eqnarray}
where $\bvec{S}'$ is a spin-$\frac{1}{2}$ operator and $\alpha$
is
the projection of a spin-1 end spin into a spin-$\frac{1}{2}$
subspace. From numerical work \cite{white_2, sorensen}, we know
that $\alpha \sim 1$. This immediately gives a low lying triplet
above a singlet with a gap $\Delta E = \alpha^2 J'$ (we assume
that $J'$ is antiferromagnetic. The ferromagnetic case is 
expected to give similar results, reversing the order of the singlet
and triplet, but numerical work has not yet been done to support
the analysis in this limit). This
triplet will sit  inside the Haldane gap. The triplet corresponds
to bound states at the chain ends; this has been seen numerically
in \cite{imps}, which also demonstrated that the above first
order perturbation theory result for $\Delta E$ is accurate up to
$J_E \sim .3J$. The embedded states become delocalized and join
the continuum at about $J' \sim .7J$. This picture is
unchanged up to about $J' \sim 2J$, after which the triplet
returns as a bound state in the Haldane gap to merge  with the
singlet state as $J' \rightarrow \infty$. In the type of chain
we are considering, $J'$ is unlikely to become much greater than
$J$; it is much more likely that defects in a pure sample will
serve to reduce the effective coupling between sites rather than
enhance it.

\begin{figure}
\centerline{
\epsfxsize=7 cm
\epsffile{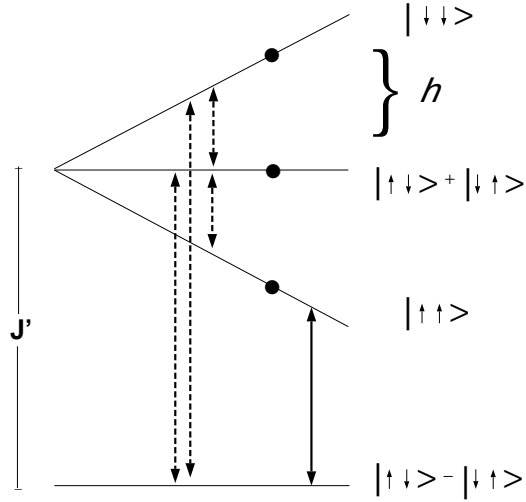}
}
\caption{Impurity level diagram when $D'=0$.}
\label{fig:nuc}
\end{figure}

\begin{figure}
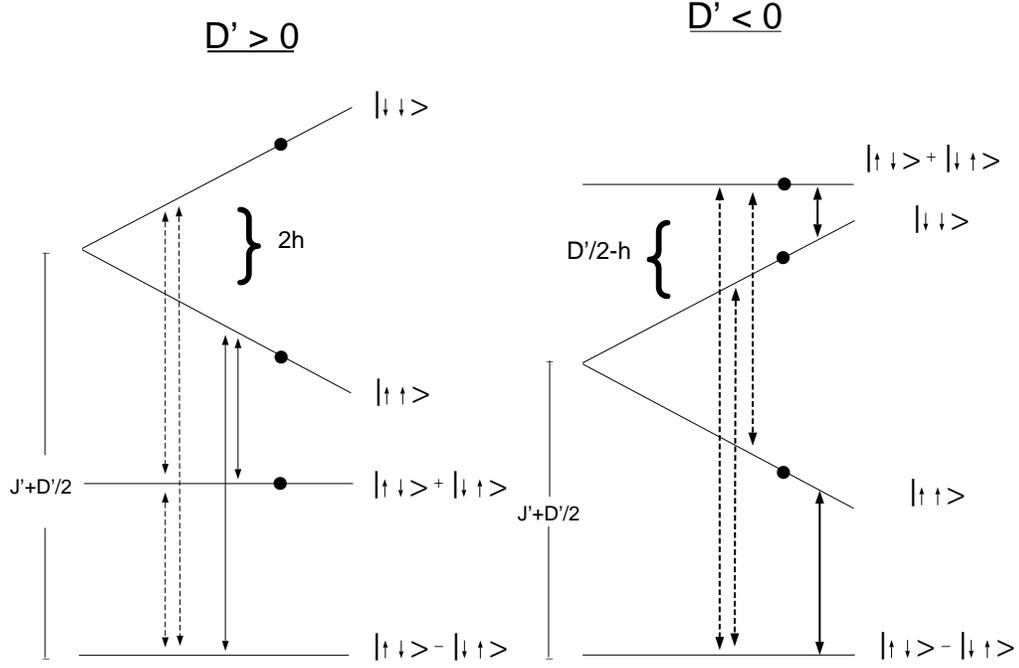

\centerline{
\epsfxsize=6.75 cm
\epsffile{fig_nuc_2g}
\epsfxsize=6.75 cm
\epsffile{fig_nuc_2s}
}
\caption{Impurity level diagrams for $D' \neq 0$.}
\label{fig:nuc_2}
\end{figure}

To understand the effect on NMR relaxation we explore the
environment of NMR nucleii near the end spins\footnote{We would like to
thank D. MacLaughlin for privately communicating his suggestions on
the effects of end spin excitations in NMR.}. Take, for
example, the nuclear spin coupled to $\bvec{S}_1'$. It sees the
Zeeman split level diagram shown in Fig. \ref{fig:nuc}. 
If we assume that the `free end spins' are not completely free,
but are weakly coupled to the magnons on the chain, then
relaxation can occur in two ways: when two levels are $\omega_N$
apart the chain end spin could make a transition -- this will
happen for a fields $h \sim J'$. Alternatively, a thermal magnon
coupled to the chain end could decay into another magnon with or
without the accompaniment of an end spin transition -- again, the
energy difference between initial and final states must be
$\omega_N$. Marked on the diagram are the transitions that
can be induced by $\bvec{S}'$: solid arrows represent possible
transitions that potentially do not require coupling to the rest
of the chain; dashed arrows represent transitions that could
occur only if the impurity is coupled to the rest of the chain;
solid circles represent spectator transitions which, again,
require magnon assistance.  It is easy to see that in all the
above scenarios, the transition will be broadened by thermal magnons.
This implies that the characteristic width will have a 
temperature dependence exponential in the Haldane gap.
There is an additional mechanism, which we now discuss,
that can, in general, affect this picture. 
If the material is anisotropic, with for example, a $D$-type
anisotropy, we must add the following term to the chain-end
effective Hamiltonian, Eqn. (\ref{c_ham}),
\begin{eqnarray}
H_E \rightarrow H_E + \alpha^2 D \left ( (S^{z\prime}_1)^2 +
(S^{z\prime}_2)^2
\right ) + D' S^{z\prime}_1 S^{z\prime}_2
\end{eqnarray}
The first term is the familiar on-site anisotropy; it will only
contribute a c-number to the effective Hamiltonian. The second
term is allowed by symmetry, and we presume that it is a
consequence of the defect (which arguably, would manifest itself
in accordance with the available symmetry). We assume $D' > -J'$,
so that the exchange interaction is still antiferromagnetic in the
$z$-direction. As a result of this, the two $s^z = 0$ levels will
shift by $-\frac{D'}{4}$, while the $s^z = \pm 1$ levels will be
shifted by $\frac{D'}{4}$. The transitions induced by 
the hyperfine interaction are shown in the new level diagrams
in Fig. \ref{fig:nuc_2}. 
It is this more general case for
possible transitions which we now carefully analyze
(further anisotropic purturbations will not qualitatively change the
picture).

We begin by characterizing the interaction between the bulk
magnons and the end spin. Since we have no information as to the
nature of this coupling we will parametrize it in the spirit of
Mitra et. al. \cite{mit} using free bosons which after scattering
with the impurity obtain a phase shift. We make two assumptions:
first, that leakage
across the impurity site is negligible, and second, that the impurity 
spin coupling to the boson does not allow for exchange of spin (this is 
consistent so long as
$J'$ is sufficiently weak);
bosons on each side of the impurity will have wave-functions of
form:
\begin{eqnarray}
\phi_{k,i}^{\pm}(x) \sim C^i_{\pm}(k) \left ( e^{ikx} - e^{-
2i\delta_{\pm}^i(k)} e^{-ikx} \right )
\label{b_wfn}
\end{eqnarray}
where $i$ refers to the boson branch, $k$ is assumed positive,
$|C_{\pm}(k)|^2 = 1/2L$ and $\pm$ refers to the sector of the
impurity spin with $s^{z \prime} = \pm  \frac{1}{2}$.
The boundary condition, $\phi_{k,i}^{\pm}(L) = 0$, gives
\begin{eqnarray}
k_n^{i,\pm} = (n\pi - \delta_{\pm}^i(k))/L
\end{eqnarray}
We assume that the phase shifts are small and grow with
increasing energy. This is tantamount to assuming a large step potential
barrier of infinite extent (thus allowing no leakage for states below
the barrier). This is certainly true in the limit $J' \rightarrow 0$.
A heuristic ansatz which has this behaviour is
\begin{eqnarray}
\delta_{\pm}^i(k) \sim \lambda_{\pm}^ik 
\label{phases}
\end{eqnarray}
Notice that the $\phi^+$'s need not be orthogonal to the $\phi^-$'s, since
these states are in separate Hilbert Spaces. 
The energy of one of these bosons at low temperatures
is given by the free form
\begin{eqnarray}
E_n^{i,\pm} =  \Delta_i + \frac{v^2(k_n^{i,\pm})^2}{2\Delta_i}
\end{eqnarray}
Before going on we note that $\omega_N$ is typically much smaller
than the energy level spacings due to finite size, for typical
chain lengths. For example, in NENP, $\delta E \sim .04$ Kelvin
for chains $L \sim 1000a$. One may therefore question the
validity of boson assisted transitions when the bosons lack the
ability to `fine tune' a transition so that the difference
between initial and final states is $\omega_N$. What saves the
day, in this case, are higher dimensional effects. For sufficiently
long chains or sufficiently strong interchain couplings, these will
densely fill the spacings in energy levels along the chain
direction. Assuming this is the case, (we show the conditions
for this explicitly 
in the next section) we will not worry about
this point further.

Consider the coupling of a nuclear spin to one of the impurity spins,
say $S_1'$ (from here on we will implicitly write  $S_1'=S'$, for ease of
notation).
The familiar formula for the transition rate can be cast as
\[ \frac{1}{T_1}_I = \frac{\alpha^2}{{\cal Z}}
\int_{-\infty}^{\infty} dt \; e^{-i\omega_Nt}
\sum_{\mu} |\bold{A}^{- \mu}|^2  \sum_{n,l,n',l'} \]
\begin{eqnarray}
\left \{ <n_l;l|e^{(-\beta + it)(H_b + H_E)} 
S^{\mu \prime} e^{-it(H_b+H_E)} |n^{\prime}_{l'};l'>
<n^{\prime}_{l'};l'|S^{\bar{\mu} \prime}|n_l;l>
\right. \nonumber \\ \left. +
<n_l;l|e^{-\beta (H_b+H_E)}
S^{\bar{\mu} \prime}|n^{\prime}_{l'};l'><n^{\prime}_{l'};l'|
e^{it(H_b+H_E)} S^{\mu \prime} e^{-it(H_b+H_E)}|n_l;l> \right \}
\end{eqnarray}
where $\mu = -\bar{\mu} = \pm,0$. $H_b$ is the free boson
Hamiltonian; $|l;m_l>$ denotes a state of the impurity spin (ie. an
eigenstate of $H_E$), which for brevity, we denote as $|l>$;
$n_l$ denotes the boson content of the state; in general, $|n_l;l>$ will
contain two different
multiparticle free
boson states that have projections onto the
$S_1^{z \prime}= \pm\frac{1}{2}$ subspaces
of $|l;m_l>$ (with appropriate phase shifts). 
To elaborate and make
things a bit clearer, take the $s^z = 0$ state from the triplet.
\begin{eqnarray}
|n_{1,0};1,0> \equiv \frac{1}{\sqrt{2}} \left (
|\uparrow,\downarrow>\otimes |n^+> + |\downarrow, \uparrow>
\otimes |n^-> \right )
\end{eqnarray}
where the states $|n^{\pm}>$ correspond to the phase shifted bosons with
$S_1^{z \prime}= \pm\frac{1}{2}$, respectively.
Making the approximation, $\alpha =1$,
we return to the relaxation rate:
\[ \frac{1}{T_1}_1 =\frac{2}{{\cal Z}}
\int_{-\infty}^{\infty} dt \; e^{-i\omega_Nt}
\sum_{\mu} |\bold{A}^{- \mu}|^2  \]
\begin{eqnarray}
\sum_{n,l,n',l'} e^{-
\beta(E_{n_l} + \varepsilon_l)} \cos(t[E_{n_l} -
E_{n^{\prime}_{l'}} + \varepsilon_l - \varepsilon_{l'}])
|<l|S^{\mu \prime}|l'>|^2 |<n_l|n^{\prime}_{l'}>|^2
\end{eqnarray}
Let's pick a particular transition and work it through. Consider
$\mu=+$, $|l> = |1,1>$ and $|l'> = |1,0>$. This transition could
be of the type we've been discussing where for $h \sim D'/2$, we
expect strong resonance if $D'>0$. The expression for the rate
becomes
\begin{eqnarray}
\frac{1}{T_1}_I(l,l',\mu) =
\frac{1}{{\cal Z}} e^{-\beta(J' + D'/2 -h)}
\int_{-\infty}^{\infty} dt \; e^{-i\omega_Nt}
|\bold{A}^{- +}|^2  
\nonumber \\ \mbox{Re} \left \{ e^{it(D'/2-h)}
\sum_{n,n'} e^{-\beta E_{n^+}} e^{it(E_{n^+} - E_{n^{- \prime})}} 
|<n^+|n^{-\prime}>|^2 \right \}
\end{eqnarray}
Since the boson multiparticle states, $<n|$, are direct products
of symmetrized free $N$-particle states, and since the energy of
such a state is the sum of single particle energies, we can write
the last equation in terms of single particle states:
\begin{eqnarray}
\frac{1}{T_1}_I(l,l',\mu) = 
\frac{1}{{\cal Z}_E} e^{-\beta(J' + D'/2 -h)}
\int_{-\infty}^{\infty} dt \; e^{-i\omega_Nt}
|\bold{A}^{- +}|^2  \nonumber \\
\mbox{Re} \left \{ e^{it(D'/2-h)}
\mbox{Exp}\left (-{\cal Z}_1 + \sum_{n,n',i,j} 
e^{-\beta E^i_{n^+}} e^{it(E^i_{n^+} - E^j_{n^{- \prime}})} 
|<n^+;i|n^{- \prime};j>|^2 \right ) \right \}
\label{stuff}
\end{eqnarray}
where we've used the fact that the grand partition function for
noninteracting bosons is the exponentiated partition function for
a single boson. $n$ and $n'$ now index single boson states, and
$i$ and $j$ denote boson branches.
The overlap of the boson states can be calculated from the form
given by Eqn. (\ref{b_wfn}),
\begin{eqnarray}
|<n^+;i|n^{- \prime};j>|^2 = \frac{1}{L^2}\delta_{ij} \left (
\frac{\sin(\delta_+^j(k) + \delta_-^j(k'))}{k+k'} -
\frac{\sin(\delta_+^j(k) - \delta_-^j(k'))}{k-k'} \right )^2
\end{eqnarray}
where we have parametrized the momenta of $n$ and $n'$ with $k$
and $k'$, respectively. 
Finally, we write the one particle
partition function (we can ignore the phase shifts for this purpose)
as
\begin{eqnarray}
{\cal Z}_1 = \sum_{n,i} e^{-\beta E^i_{n^+}} <n^+;i|n^+;i>
\nonumber \\
= \sum_{n,i} e^{-\beta E^i_{n^+}} |<n^+;i|n^-;i>|^2
\end{eqnarray}
Combining all of this allows us to write the exponential in
Eqn. (\ref{stuff}) as 
\[ \sum_i e^{-\Delta_i(h)/T} \int \frac{dk \; dk'}{\pi^2} 
4\left ( \frac{k'k}{k^2 - k^{\prime 2}} \right )^2 (\lambda_+^i
- \lambda_-^i)^2 \]
\begin{eqnarray}
\times e^{-\beta v^2k^2/2\Delta_i} 
(e^{it\frac{(k^2-k^{\prime 2})v^2}{2\Delta_i}}-1)
\end{eqnarray}
\begin{eqnarray}
=  -\sum_i e^{-\Delta_i(h)/T} \frac{8\Delta_iT}{v^2 \pi^2}
(\lambda_+^i - \lambda_-^i)^2 \int_0^{\infty} dx \; dx'
e^{-x} \left [ \frac{1-e^{i(x-x')Tt}}{(x-x')^2} \right ] 
\sqrt{x'x} \equiv \Theta(t)
\label{stuff_2}
\end{eqnarray}
The effect of the phase shifts is contained in the factor,
$ (\lambda_+^i - \lambda_-^i)^2$, as seen from Eqn. (\ref{phases}).
Corrections to this due to $O(k^3)$ contributions to 
$\delta_{\pm}^i(k)$ will be suppressed by factors of $2\Delta_iT/v^2$.
The imaginary part of $\Theta(t)$ will shift the resonance 
from $h=D'/2$. This shift, at low temperatures, will be negligible.
We are interested in the long time behaviour of $\Theta(t)$.
In this limit, the real part of $\Theta(t)$ becomes:
\[ 
\mbox{Re} \Theta(t)= -\sum_i e^{-\Delta_i(h)/T}\frac{16\Delta_iT}{v^2 \pi^2}
(\lambda_+^i - \lambda_-^i)^2 \int_0^{\infty} dx \; dx'
e^{-x} \frac{\sin^2((x-x')Tt/2)}{(x-x')^2} \]
\begin{eqnarray}
\rightarrow -\sum_i e^{-\Delta_i(h)/T} \frac{8\Delta_iT^2}{v^2 \pi}
(\lambda_+^i - \lambda_-^i)^2 |t| \equiv -\Gamma(T) |t|
\end{eqnarray}

The expression for the relaxation rate becomes 
\[ \frac{1}{T_1}_I(l,l',\mu) \approx
\frac{ |\bold{A}^{- +}|^2}{{\cal Z}_E}
e^{-\beta(J' + D'/2 -h)} \]
\begin{eqnarray}
\times \int_{-\infty}^{\infty} dt \; 
e^{-i\omega_Nt} 
\cos\left( t(D'/2-h) \right ) e^{-|t| \Gamma(T)}
\end{eqnarray}
\begin{eqnarray}
= \frac{2|\bold{A}^{- +}|^2}{{\cal Z}_E}
e^{-\beta(J' + D'/2 -h)}  \frac{\Gamma(T)}{ \Gamma^2(T) + (h-D'/2)^2}
\label{t_imp_sf}
\end{eqnarray}
This is the most important equation of this section. The other
transitions can be treated the same way to arrive at analogous
results. The key issue to note is that the impurity relaxation rate
is an extremely sensitive function of the temperature and field.
At temperatures
well below the gap it is essentially a delta-function of $h$.
As the temperature
increases and becomes comparable to the gap, the rate broadens rapidly.

Before summing up,
we discuss the other possible transitions. First, notice that 
changing $\mu$ has the same effect as reversing the
sign of $h$  and exchanging $l$ and $l'$:
\begin{eqnarray}
\mu \rightarrow \bar{\mu} \; \; \; \; \longleftrightarrow
\; \; \; \; h\leftrightarrow -h
 \; \; \; \;
\longleftrightarrow \; \; \; \; l \leftrightarrow l'
\end{eqnarray}
Note that in this simple model of boson-impurity
coupling {\em there are no transitions via} $S^{z \prime}$, and therefore
no transitions between the singlet and the $s^z=0$ state of the
triplet. In other words, the solid circles in Figs. \ref{fig:nuc}
and \ref{fig:nuc_2} are ignorable as are the dashed lines from
$|1,0>$ to $|0>$. This is expected in all but the most extreme of
anisotropic exchange impurity models. Furthermore, the effect of
reversing the spin states on the triplet is the same as reversing
the sign of magnetic field:
\begin{eqnarray}
m_l,m_{l'} \rightarrow -m_{l},-m_{l'}
\; \; \; \; \longleftrightarrow \; \; \; \; 
h \rightarrow -h 
\end{eqnarray}
Finally, the result of exchanging the $|0,1>$ state with the
singlet amounts to adding $J'$ to the associated energy factor in the
Lorentzian. A final expression for the impurity relaxation rate involving
all eight possible impurity level transitions is 
\begin{eqnarray}
 \frac{1}{T_1}_I(J',D') = 
4\sum_{i=1}^4\sum_{f=1}^2
\frac{|\bold{A}^{- \sigma_1}|^2}{{\cal Z}_E}
e^{-\beta E_i} 
\frac{\Gamma(T)}{ \Gamma^2(T) + 
E_{fi}^2}
\label{imp1}
\end{eqnarray}
where $E_i$ denotes one of the four possible initial impurity states,
and $E_{fi}$ is the difference in energy between the initial state
and one of the two possible consequent final states.
The factor of two 
represents the contribution of both end spins on each chain (we
neglect surface effects).
Now it can be seen more clearly that all but two of the elements
in the sum above will contribute little due to the narrow
gaussian form. The important terms are those where the energy in
the gaussian is small; this can happen for certain magnetic
fields: $h \sim |D'/2|$ and $h \sim J' + D'/2$. In Heisenberg
chains we might expect $J' \gg |D'|$ for most defects. Furthermore,
the impurity contribution should be most evident at lower fields
where the gap still lies high. Consequently, in experiment, one
expects the $h \sim |D'/2|$ transition to dominate the picture of
impurity contributions to $1/T_1$.

In a real sample, the NMR signal from the impurity will be
proportional to the density of the impurities. Moreover, since
defects will vary from chain to chain, one would be wise to
average over a random distribution of couplings, $J'$ and $D'$.
In practice, experimental data could be analyzed for the `peak'
values of $J'$ and $D'$. One could then model the distribution of
couplings with the appropriate peak values. This
could, in principle be checked against low temperature ESR
measurements which ought to concur with the impurity model.

A final expression for the relaxation rate due to impurities is
\begin{eqnarray}
\left ( \frac{1}{T_1} \right )_{\mbox{Imp}} = 
\bar{n} \int dJ' \; dD'\; \rho(\bar{J}' - J') 
\rho(\bar{D}' - D')  \frac{1}{T_1}_I(J',D')
\label{T_imp_final}
\end{eqnarray}
where $\bar{n}$ is the density of impurities (or inverse length
of the average chain);
$\rho$ is some distribution function.

\section{Interchain Couplings}

In previous sections we mentioned the effects of interchain
couplings on various aspects of the physics. We now examine these
in more detail. 

Nearest neighbour interchain couplings will enter the Hamiltonian
as
\begin{eqnarray}
H \rightarrow H + J_I\sum_{<i,j>} \bvec{S}_i \cdot \bvec{S}_j
\end{eqnarray}
where $<i,j>$ index nearest neighbour spins {\em not} on the same
chain. We can return to the derivation of the NL$\sigma$ model to
see the effect of this additional term. Taylor expanding the
continuum representation for $\bvec{S}_j$ and assuming reflection
symmetry about a site, Eqn. (\ref{nls_a}) will
change to
\[ S = 2\pi isQ + \frac{Js^2}{2\Delta x \Delta y} \int d^4x
(\partial_z \bphi)^2 + \frac{J_Is^2}{2 \Delta x \Delta y} 
\int d^4x \sum_i (\bvec{\zeta}_i\cdot \bvec{\nabla} \bphi)^2 \]
\begin{eqnarray}
+ \frac{Js^2}{2 \Delta x \Delta y} \int d^4x
(\partial_{v\tau} \bphi)^2 
\end{eqnarray}
where we chose the $z$-direction to be along the chain, and the
vector, $\bvec{\zeta}_i$, is the displacement vector
to the $i$th nearest
neighbour of a spin not on the same chain (again, we assume that
$|\bvec{\zeta}|$ is smaller than the correlation length. 
Note that the correction
to dynamical part of the Lagrangian will correspond to $J \rightarrow
J + J_I$.
Presumably, $J_I \ll J$, meaning that
we were justified in ignoring this term.
Setting the lattice
spaces, $\Delta x, \Delta y,$ to 1,
we can now write an effective Landau-Ginsburg
Hamiltonian to describe the physics; Eqn. (\ref{H_Lan}) will read
\begin{eqnarray}
{\cal H}(\bvec{x}) = \frac{v}{2} \bPi^2 + \frac{v}{2}\left (
\frac{\partial \bphi}{\partial x} \right)^2 + \frac{2J_Is}{2}
\sum_i (\bvec{\zeta}_i \cdot \bvec{\nabla} \bphi)^2 + 
\frac{1}{2v}\Delta^2 \bphi^2 + O(\phi^4)
\end{eqnarray}
The leading relevant interaction terms will always be local.
Ignoring these, the resulting equations of motion are
\begin{eqnarray}
\frac{1}{v} \ddot{\bphi} = \left ( v\partial^2_x  -
\frac{\Delta^2}{v} + 2J_Is\sum_i (\bvec{\zeta}_i \cdot
\bvec{\nabla})^2 \right ) \bphi
\end{eqnarray}
The dispersion relation becomes,
\begin{eqnarray}
\omega^2 = v^2k^2 + \Delta^2 + v_{\perp}^2 \sum_i (\bvec{\zeta}_i
\cdot \bvec{k})^2
\end{eqnarray}
where $v_{\perp} \propto \sqrt{J_IJ}$. It is from this last formula
which we now extract qualitative information about interchain
coupling effects.

First, recall that we claimed that for finite chains of certain
lengths in real experimental situations, we no longer need to
concern ourselves with 1-D finite size effects. In other words,
we said that energy levels arising from interchain couplings will
densely fill the small gaps between magnon energy levels, $\Delta
E \sim \frac{v^2 \pi^2}{2\Delta L^2}$. Let's calculate this
length in terms of $J$ and $J_I$. We start by assuming a simple
form for the interchain contribution to the dispersion:
\begin{eqnarray}
v_{\perp}^2 \sum_i (\bvec{\zeta}_i \cdot \bvec{k})^2
= v_{\perp}^2 a^2_{\perp} (k_x^2 + k_y^2)
\end{eqnarray}
where $a_{\perp}$ is some typical interchain distance, and
expected to be $O(1)$. The size of the interchain band will be 
$\frac{v_{\perp}^2 a^2_{\perp} \pi^2}{\Delta}$. Setting this
equal to the gap in the magnon levels we get
\begin{eqnarray}
L^2 \sim \frac{v^2}{v^2_{\perp}} = \frac{J}{J_I}
\end{eqnarray}
In NENP, for example, this corresponds to lengths of
approximately 100 lattice units.

There is also the issue of cutting off divergent integrals which
we discussed in Chapter 3. In calculating transition rates, one
often encounters integrals such as
\begin{eqnarray}
\int dk \; dq \; \delta(\omega^i_k - \omega^j_q - E) f(k)
\label{cutoff}
\end{eqnarray}
When $E$ is close to the gap between the branches, $\omega^i_k$
and $\omega^j_q$, this integral can diverge logarithmically in
the infrared. If one introduces interchain couplings, the
integral over the delta function becomes
\[ \frac{1}{(2\pi)^4} \int dq \; d^2k_{\perp} \; d^2q_{\perp}
\delta(\omega^i_k - \omega_q^j -E) \approx \]
\begin{eqnarray}
\int dq \; \frac{dk_{\perp}^2 \; d^2q_{\perp}}{(4\pi)^2 a^4_{\perp}}
\delta(\frac{v^2k^2}{2\Delta_i} +
\frac{v_{\perp}^2k_{\perp}^2}{2\Delta_i} -
\frac{v^2q^2}{2\Delta_j} -
\frac{v_{\perp}^2q_{\perp}^2}{2\Delta_j} )
\end{eqnarray}
where we have assumed a simple form for the interchain
dispersion. For ease of calculation, we now assume that $\Delta_i
= \Delta_j \equiv \Delta$. The integral becomes
\\
\[ \frac{\Delta}{8\pi^2 a^4_{\perp}v} 
\int \frac{dk_{\perp}^2 \; dq_{\perp}^2 }
{ \sqrt{v^2 k^2 +  v_{\perp}^2k_{\perp}^2 - v_{\perp}^2
q_{\perp}^2 } } \theta(k^2 +  v_{\perp}^2k_{\perp}^2 -
v_{\perp}^2 q_{\perp}^2) \]
\begin{eqnarray}
= \frac{\Delta}{4 \pi^2 v_{\perp}^2 a^4_{\perp}v} \int dk_{\perp}^2
\sqrt{v^2 k^2 +  v_{\perp}^2k_{\perp}^2 }
\end{eqnarray}
At low momentum, $k$, where we need a cutoff, this integral is
approximately
\begin{eqnarray}
\sim \frac{\Delta \pi}{6 v_{\perp} va_{\perp}}
\end{eqnarray}
If we write the integral in Eqn. (\ref{cutoff}) as
\begin{eqnarray}
\frac{2\Delta}{v^2} \int dk \; \frac{f(k)}{\sqrt{k^2 + C}}
\end{eqnarray}
then the cutoff, $C$, is seen to be
\begin{eqnarray}
C \sim \frac{ 144 v_{\perp}^2a^2_{\perp}}{\pi^2 v^2} 
\sim 10 \frac{J_I}{J}
\label{cutoff_c}
\end{eqnarray}
We recall that
the for intrabranch transitions, $Q_0^2 \sim 2 \Delta \omega_N
/v^2$. Comparing this to $C$, we find that the cutoff becomes
important for  $\omega_N < 70 J_Ia^2_{\perp}$. 
For example, in NENP, where
$J_I \sim 25 \; \; m$K, we expect the cutoff to
significantly dominate over the Larmour frequency.

\section{Crystal Structure}

When analyzing experimental data in terms of the idealized
Heisenberg model with on-site anisotropies, one must keep in mind
that the symmetry of the proposed spin-chain Hamiltonian may be
constrained by the symmetry of the crystal and the local symmetry
about the magnetic ion. Additional terms may be added or
subtracted to accommodate the structure of the substance, and
these can have a great effect on the interpretation of data. Some
important questions which must be asked before deciding on a
model Hamiltonian for the material are: is the local crystal
field symmetry about the magnetic ion commensurate with the
symmetry of the unit cell? Is there more than one chain per unit
cell? If so, are all chains identical? Is there more than one
magnetic ion of a single chain per unit cell? If so, is there
true translational symmetry from one spin site to another?

In the next chapter, we analyze experiments performed 
on NENP. In so doing, we will address such considerations.


\chapter{NENP: Direct Comparison with Experiment}
\resetcounters

\section{The Structure of NENP and Experimental Ramifications}

A schematic diagram of $Ni(C_2H_8N_2)_2NO_2(ClO_4)$ (NENP) is
shown in Fig \ref{fig:nenp}.
\begin{figure}
\vspace{5.0 in}
\caption{NENP}
\label{fig:nenp}
\end{figure}
Each chain is comprised of Ethylenediamine-Nickel chelates
separated by 
nitrite groups. The magnetic ion is $Ni^{2+}$; experiments
indicate that these ions interact antiferromagnetically along the
chain with coupling $J \sim 55$K. There is a large single ion
anisotropy, $D \sim 12$K, as well as a small axial symmetry
breaking anisotropy $E \sim 2$K. Interchain couplings are
estimated at $J_I/J \approx 10^{-4}$ \cite{renard}. 

It is important to realize
that two neighbouring $Ni^{2+}$ along the $b$-direction 
are not equivalent; rather, one
is related to
the other by a $\pi$ rotation about the $b$ axis. Also, the angle
along the $N-Ni-O$ bond is not exactly $\pi$, meaning that the
$Ni$ site is not truly
centrosymmetric. Most importantly, the local symmetry axes of
each $Ni$ ion are
rotated with respect to the $abc$ (crystallographic) axes. To
demonstrate this
we now note the
coordinates of the Nitrogen atoms in the Ethylenediamine chelate
surrounding
the Nickel (placing the Nickel at
the origin): \cite{mey}

\vspace*{.2in}

\begin{center}
\begin{tabular}{|c|c|c|c|}
 Atom  &  $a$ $(\AA)$ & $b$ $(\AA)$ & $c$ $(\AA)$  \\ \hline
 $N(1)$ &  $2.053$ $(3)$  &  $.162$ $(3)$  &  $.338$ $(3)$ \\
     &   &   &   \\
     $N(2)$ &  $.619$ $(3)$  &  $-.184$ $(3)$  &  $-1.971$ $(3)$
\\
       &   &   &   \\
       \end{tabular}
\end{center}

\vspace*{.2in}

\begin{figure}
\centerline{
\epsfxsize=10 cm
\epsffile{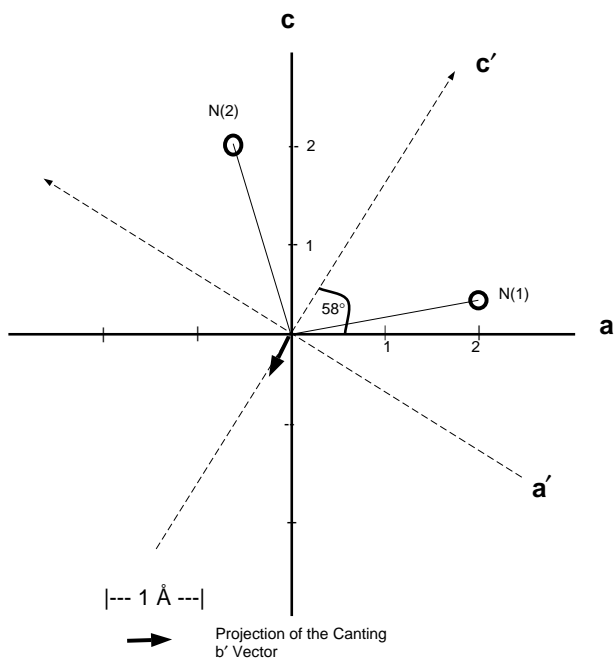}
}
\caption{Local and crystallographic axes projected onto the $ac$-plane in NENP}
\label{fig:proj1}
\end{figure}

The other Nitrogen atoms in the chelate can be obtained by
reflection
through the Nickel. One easily sees that projecting this
structure onto
the $b$ plane yields symmetry axes (in the $b$ plane) rotated
$\sim 60^{\circ}$
from the $ac$-axes. This is shown in Fig. \ref{fig:proj1}. 
The inclination
of the local Nickel axes from the $abc$ system can be obtained by
taking
the cross product of the two Nitrogen vectors (ie. the normal to
the plane
described by the four Nitrogen atoms in the chelate:
\begin{eqnarray}
\hat{n} = \left ( -.06 \mbox{ } (1) ,.98 \mbox{ } (1) ,-.11
\mbox{ } (1) 
\right ) 
\end{eqnarray}

The local $Ni$ $b^{\prime}$-axis
makes a $\sim 10^{\circ}$
angle with the $b$-axis, while the azimuthal angle in the $ac$
plane is
$\sim -28^{\circ}$ from $c$. 
The $10^{\circ}$ tilt is roughly about the $a^{\prime}$ direction
of the local
symmetry axes.

One may worry that the $NO_2^-$ group may distort the local
symmetry axes, but remarkably enough,
when projected onto the $ac$ plane, the three atoms in the
molecular ion sit on the $c^\prime$ axis. This
reinforces our suspicion that the local symmetry axes are indeed
the above.

\begin{figure}
\centerline{
\epsfxsize=13.5 cm
\epsffile{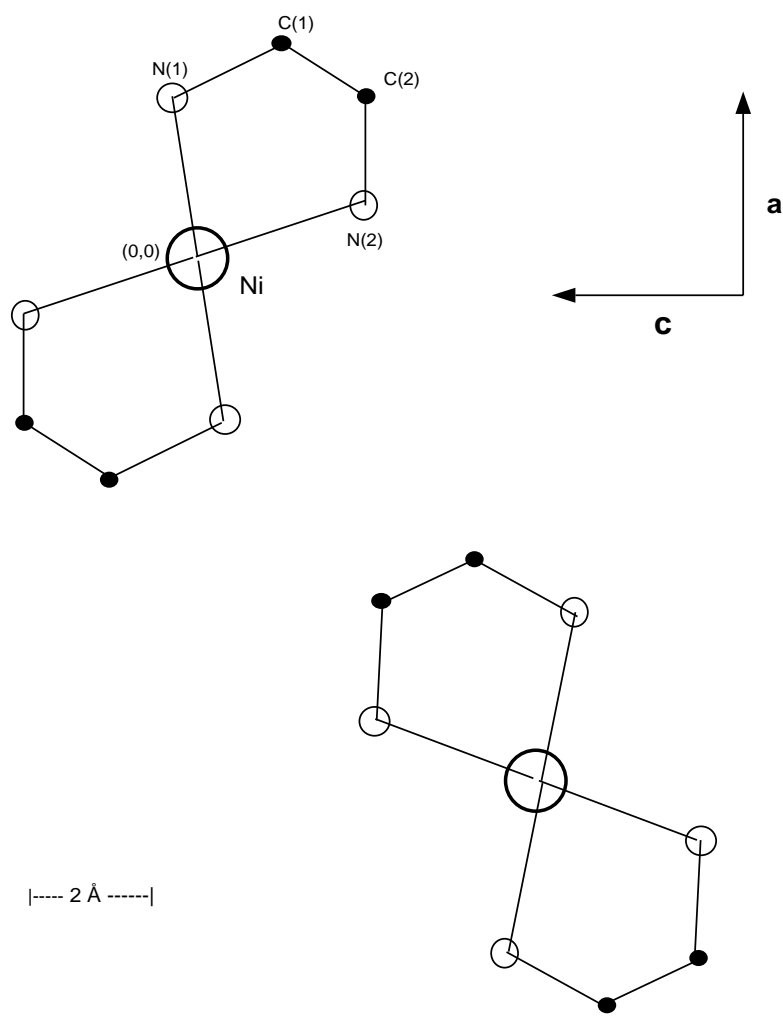}
}
\caption{A projection of the NENP unit cell onto the $ac$-plane showing two
chains per unit cell}
\label{fig:proj2}
\end{figure}

Next, we consider the whole space group of NENP. The most recent
attempt to
solve for the crystal symmetries has concluded that the true
space group of
the material is $Pn2_{1}a$ \cite{mey}; this is a
non-centrosymmetric space group with a
screw $2_{1}$ symmetry about the $b$ axis, diagonal glide plane
reflection symmetry
along the $a$ axis, and an axial glide plane reflection symmetry
along $c$.
Experimentally, attempts to
solve the structure in $Pn2_{1}a$ have not been successful;
rather, it seems that
$Pnma$ gives a better fit. The main difference between the two is
the presence in $Pnma$
of a mirror plane parallel to $b$ at $\frac{1}{4}b$, centers of
symmetry at various locations
in the unit cell, and two-fold screw axes separating these
centers of symmetry.
The reason for the experimental discrepancy is attributed to
disorder
in the orientation of the nitrite group, the perchlorate anions,
and the existence
of a local or pseudo center of symmetry lying very close to the
$Ni$ (thousandths
of an Angstrom) \cite{mey}. A crucial point is that both space
groups 
share the axial glide planes along $a$, the diagonal glide planes
along 
$c$ and the $2_{1}$ screw symmetry about $b$. These 
generate a total of 4 $Ni$ sites per primitive cell and two
chains through
each cell. The two chains are such that the $Ni$ chelates on one
are the mirror image
of the other. Figure \ref{fig:proj2} shows a projection of this
picture onto
the $ac$ plane.
The presence of the $2_{1}$ screw symmetry about each chain axis
introduces staggered
contributions to the local anisotropy and gyromagnetic tensors.
This is because, as motivated above,
these are not diagonal in the crystallographic coordinate system.
The resulting spin
Hamiltonian is
  \begin{eqnarray}
  H = J\sum_{i}  \left [
\vec{S}_{i} \cdot \vec{S}_{i+1}  +
\vec{S}_{i} \cdot {\bf D} \vec{S}_{i} - \mu_{B} \vec{S}_{i} \cdot
{\bf G} \vec{H
} +
(-1)^{i} (\vec{S}_{i} \cdot {\bf d} \vec{S}_{i} - \mu_{B}
\vec{S}_{i} \cdot {\bf
 g} \vec{H})
 \right ]  
 \end{eqnarray}

We make the assumption that the symmetry of the anisotropy and
$g$-tensors
is the same (ie. that at each site they can be simultaneously
diagonalized). This is rigorously true when the crystal field
symmetry about the magnetic ion is no lower than orthorhombic (a
sketch of a proof is found on p. 750 of \cite{abragam} ).
We can get the required parametrization for the $g$-tensors from
high
temperature uniform susceptibility data \cite{mey}. This is based
on the idea that at
high temperatures the $Ni$ atoms will behave as an ensemble of
uncoupled spins ($s=1$)
with the same gyromagnetic tensor as in the antiferromagnetic
case.
With this in mind we get 

\begin{eqnarray}
\bf{G} = \left ( \begin{array}{ccc}
G_{c^{\prime}} \cos^{2} (\theta) + G_{b^{\prime}} \sin^{2}
(\theta) & 0 & 0 \\
            0 & G_{a^{\prime}} & 0   \\
              0 & 0 & G_{b^{\prime}} \cos^{2} (\theta) +
G_{c^{\prime}} \sin^{2} (\theta)
                \end{array} \right ) 
\end{eqnarray}

  \begin{eqnarray}
  \bf{g} = \left ( \begin{array}{ccc}
               0 & 0 & \sin(\theta) \cos (\theta) (G_{b^{\prime}}
- G_{c^{\prime}}) \\
               0 & 0 & 0 \\
               \sin(\theta) \cos (\theta) (G_{b^{\prime}} -
G_{c^{\prime}}) & 0 & 0
                         \end{array} \right ) 
\end{eqnarray}

Here $\theta \sim 10^\circ $, and $G_{a^{\prime}} = 2.24$,
$G_{b^{\prime}} = 2.15$, 
$G_{c^{\prime}} = 2.20$
are the values for the local $G$-tensor that give the observed
high temperature
$g$-tensor when averaged over the unit cell.
Correspondingly, the anisotropy tensors must have the following form:
 \begin{eqnarray}
\bf{D} = \left ( \begin{array}{cccc}
D_{c^{\prime}} & 0 & 0 \\
0 & D_{a^{\prime}} & 0 \\
0 & 0 & D_{b^{\prime}}
\end{array} \right ) 
\end{eqnarray}

  \begin{eqnarray}
  \bf{d} = \left ( \begin{array}{cccc}
         0 & 0 & \frac{\tan(2\theta)}{2} (D_{b^{\prime}} -
D_{c^{\prime}}) \\
                  0 & 0 & 0 \\
                    \frac{\tan(2\theta)}{2} (D_{b^{\prime}} -
D_{c^{\prime}}) & 0 & 0
                 \end{array} \right ) 
\end{eqnarray}

The parameters $D_{a^{\prime}},D_{c^{\prime}},D_{b^{\prime}}$ 
are to be fitted by experiment to the model used to describe
the system.
The boson Hamiltonian can now be written
\begin{eqnarray}
H = \int dx  [ \frac{v}{2} \bvec{\Pi}^2 + \frac{v}{2}
(\frac{\partial
\bvec{\phi} }{\partial x})^2 + \frac{1}{2v} \bvec{\phi} \cdot {\bf
D}
\bvec{\phi} - \mu_{B} \bvec{H} \cdot {\bf G} (\bvec{\phi} \times
\bvec{\Pi}) + \\
\frac{1}{v} \bvec{\phi} \cdot {\bf d} (\bvec{\phi} \times
\bvec{\Pi})
  - \mu_{B} \bvec{\phi} \cdot {\bf g} \bvec{H} + \lambda
(\bvec{\phi}^2)^2  ] \nonumber
   \end{eqnarray}

The term containing {\bf d} breaks the $Z_2$ symmetry along the
$a^{\prime}$ (lowest mass)
direction. It will also renormalize the masses. The second effect
can be ignored 
in the approximation that the $\phi^4$ term is ignored if we
assume the masses are
physical. Symmetry breaking, however, leads to the presence of a
static staggered field even
below a critical magnetic field. A gap will always persist. The
staggered field term
will break the $Z_2$ symmetry along the $c^{\prime}$ or $b$ axis,
depending on whether the field is
applied in the $b$ or $c^{\prime}$ direction, respectively. A
static staggered moment will
likewise appear due to this term. The effect on the relaxation
rate will be small, although there may be consequences in other
experiments \cite{halp, sakai}.

We would now like to discuss the effect of having two
inequivalent chains per unit cell,
with local axes different from the crystallographic axes. 
We label the two chains found in a unit cell of NENP
`chain 1' and `chain 2' corresponding to the chains in the upper
left and
lower right corners of Figure \ref{fig:proj2} respectively. The
dispersion
branches of chain 1 are given by Eqn. (20) of \cite{aff} (the 
expressions are roots of a complicated cubic equation and we
feel that citing them will not prove illuminating) only the
field is
$\alpha - 30^\circ$ from the $c^{\prime}$ axis where $\alpha$ is
the angle
of the field from the crystallographic $c$-axis. Similarly, the
dispersion
branches of chain 2 are calculated with the field $\alpha
-150^\circ$
from the $c^{\prime}$ axis.

Experiments which average over signals, like susceptibility or
NMR $T_1^{-1}$
measurements, must consider their results an average of two
different measurements
(corresponding to the two different chains with their relatively
different applied field
configurations). On the other hand, experiments such as ESR,
should show a separate
signal for each chain. The NMR relaxation calculations performed
in Chapter 3 assume the field is placed along one of the crystal
axes. In this special case, the dispersions for the two different
chains are identical. Although the dispersions will be more
complicated as will be the matrix elements, $l^i_{ab}(0,0)$, we
do not expect great qualitative differences between a calculation
as done in Chapter 3 and one which accounts for the actual
symmetry when the field is placed along a crystal axis. There
will also be contributions due to the $\bphi$--$\bvec{l}$ correlator;
these are also expected to be small.
There are, however, important manifestations of  having
two inequivalent chains. These will be discussed in the next chapter
when we suggest further experiments.

In conclusion, consideration of the crystal structure
introduces both symmetry breaking terms and two inequivalent
chains per unit cell. The symmetry breaking terms will give small
corrections to the relaxation rate. 

\section{Analysis of the Data}

\begin{figure}
\centerline{
\epsfxsize=13.5 cm
\epsffile{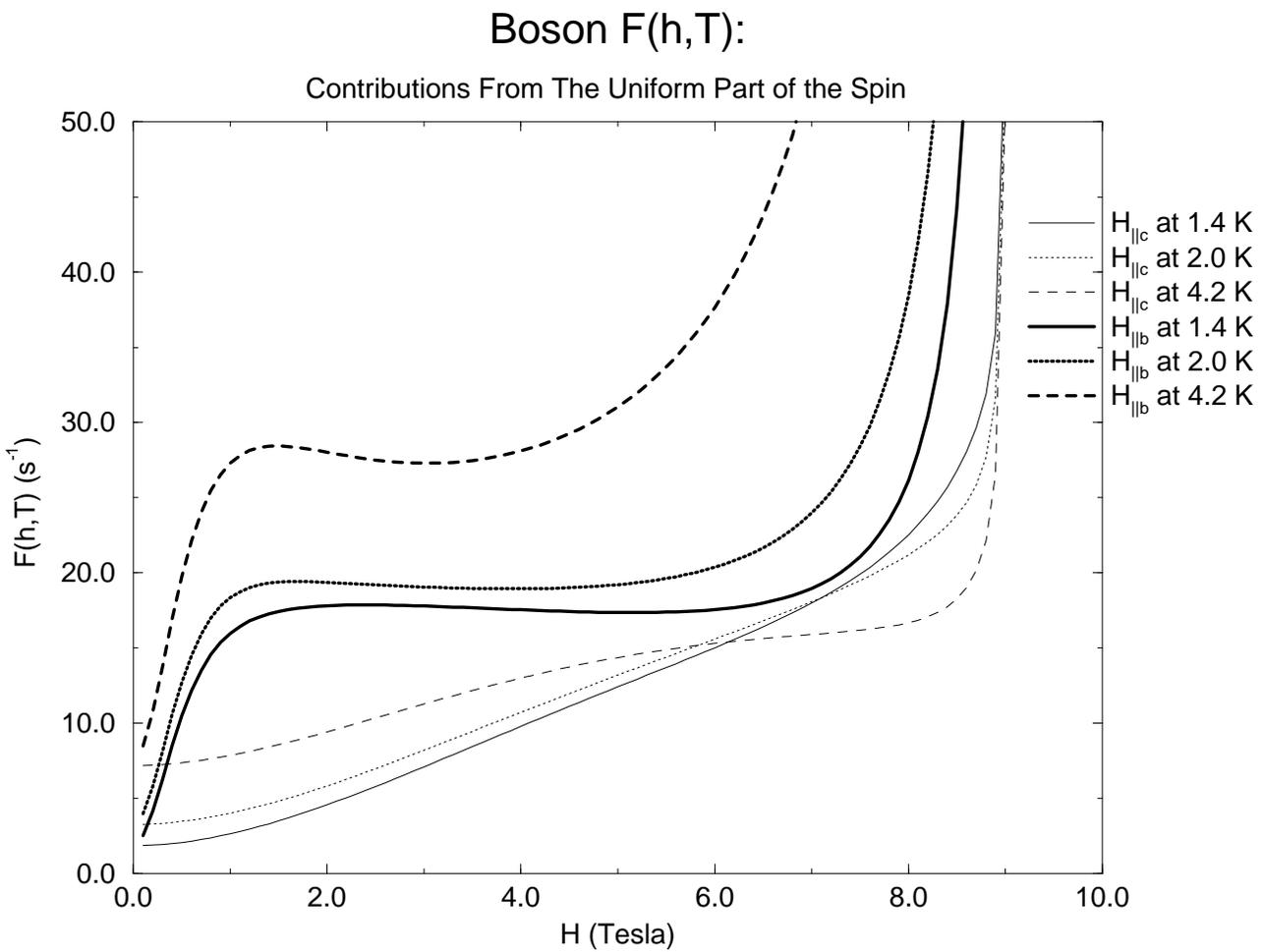}
}
\caption{Boson $F(h,T)$ for fields along the $b$ and $c$ chain directions.}
\label{fig:c}
\end{figure}

By far, the most studied Haldane gap $S=1$ material is NENP.
The most recent measurements of the relaxation rate, $1/T_1$,
on this substance have been made by Fujiwara et. al. \cite{fuji}.
Before directly comparing our results to the data we discuss the
expected results on a pure (infinite) system. The three gaps are
given by neutron scattering: $\Delta_a = 1.17 m$eV,
$\Delta_b = 2.52 m$eV and $\Delta_a = 1.34 m$eV. We use $v=10.9m$eV,
and the generic value of 2.2 for the electronic $g$-factor. Since we
do not have an accurate description for the hyperfine coupling of the
$Ni$ ion to the protons in its surrounding chelate, we assume a uniform
value for all the contributing hyperfine matrix elements in a given
direction of the applied field. Writing 
\begin{eqnarray}
T_1^{-1} = F(h,T) e^{-\Delta_-(h)/T}
\end{eqnarray}
we use the results from Chapter 3 to plot F(h,T) for bosons and fermions
and for fields along the {\em chain} $a$, $b$ and $c$ directions. The results
are shown in Figs. \ref{fig:c} - \ref{fig:a}.
\begin{figure}
\centerline{
\epsfxsize=13.5 cm
\epsffile{fig6}
}
\caption{Fermion $F(h,T)$ for fields along the $b$ and $c$ chain directions.}
\end{figure}

\begin{figure}
\centerline{
\epsfxsize=13.5 cm 
\epsffile{fig3}
}
\caption{Boson $F(h,T)$ for fields along the $b$ and $a$ chain directions.}
\end{figure}

\begin{figure}
\centerline{
\epsfxsize=13.5 cm
\epsffile{fig5}
}
\caption{Fermion $F(h,T)$ for fields along the $b$ and $a$ chain directions.}
\label{fig:a}
\end{figure}

We included multiparticle transitions by
simply replacing the Boltzmann factor by appropriate occupation factors
in Eqn. (\ref{t_rs_final}):
$f_b (1+f_b )= \mbox{cosech}^2(\frac{\beta \omega}{2}) /4$ for bosons,
and $f_f (1-f_f )= \mbox{sech}^2(\frac{\beta \omega}{2}) /4$ for fermions
\cite {mah}.
Within approximations used, multiparticle effects amount to multiplying
the final expressions by $(1 \pm e^{-\beta \omega_s})^{-2}$.
At higher temperatures it is also
necessary to include the $k$-dependence of the integrand past the peak at the
origin. We expect that at
temperatures $T \approx \frac{\Delta}{3}$ and fields $h \approx \frac{2\Delta}
{3}$ the numerically integrated results would differ 
by about 10 percent.

$F(h,T)$ is shown for fields up to 9 Tesla even though
the $(\beta \omega_- \gg 1)$ approximation is no longer valid at such fields.
This is done to contrast the predictions of the boson and fermion models. 
It's easy to see that the boson result for $F(h,T)$ diverges at the critical
field, while no such catastrophe is present in
the fermion result. This divergence is logarithmic and infrared. 
It will persist even after account is made for the staggered part
of the correlation function. Multiparticle scattering
will in fact worsen the effect, since the bose distribution function
diverges as $1/\omega$ with vanishing energy $\omega$.
This again is evidence of the inadequacy
of the free boson model close to criticality.

In NENP, when the field is along the $b$ direction, we expect relevant 
interbranch transitions only for small field. 
In this regime, one must also be careful to include intrabranch
transitions in the second lowest branch. All these processes are of the
same order. Even though the intrabranch rates vanish at low fields, the
interbranch contributions are suppressed by the absence of low momentum
transitions (ie. $Q$ for the interbranch transitions is $O(\sqrt{\Delta^2_1-
\Delta_2^2})$ as opposed to $O(\omega_N)$.)
For this case, only $l^3$ need be calculated.

When the field is along the $c$ direction (corresponding to the middle gap), 
we restrict ourselves to calculating intrabranch transitions along 
the lower branch and interbranch
ones between the lower and $c$ branch. There are no
intrabranch processes along the $c$ axis. Calculating the interbranch 
transitions amounts to calculating $ |l^{1}_{a,c}|^2 $ and
$|l^{2}_{a,c}|^2 $.
When the field is along the $a$ direction, the calculation proceeds as above. 
The crossing of the branches provides for the interesting effect mentioned 
earlier. The peak in $1/T_1$ can be used to locate the true $ac$-axes
for the chain

Notice that $F(h,T)$ for the field parallel to the $b$ axis is nearly
field independent over a large range of the magnetic field. This behaviour
is quite easy to
understand from the universal results valid in the axially symmetric case, 
discussed in Chapter 3. When the field is along $b$, the system is only
slightly anisotropic, and so the axially symmetric results roughly apply.
$F_b$ is roughly independent of field with axial
symmetry since
$l^3_{--}$ is nearly $h$ independent (in fact, $F_b$ exhibits a
logarithmic divergence
as $h \rightarrow 0$). On the other hand, $F_c$ vanishes quadratically as
$h \rightarrow 0$. Including the small breaking of the axial
symmetry corresponding
to $\Delta_c - \Delta_a = 2^{\circ} K$, $F_b$ is essentially constant
down to low fields of order $\Delta_c - \Delta_a \approx 1$T,
before rapidly decreasing as seen in
in the figures.

\begin{figure}
\centerline{
\epsfxsize=13.5 cm
\epsffile{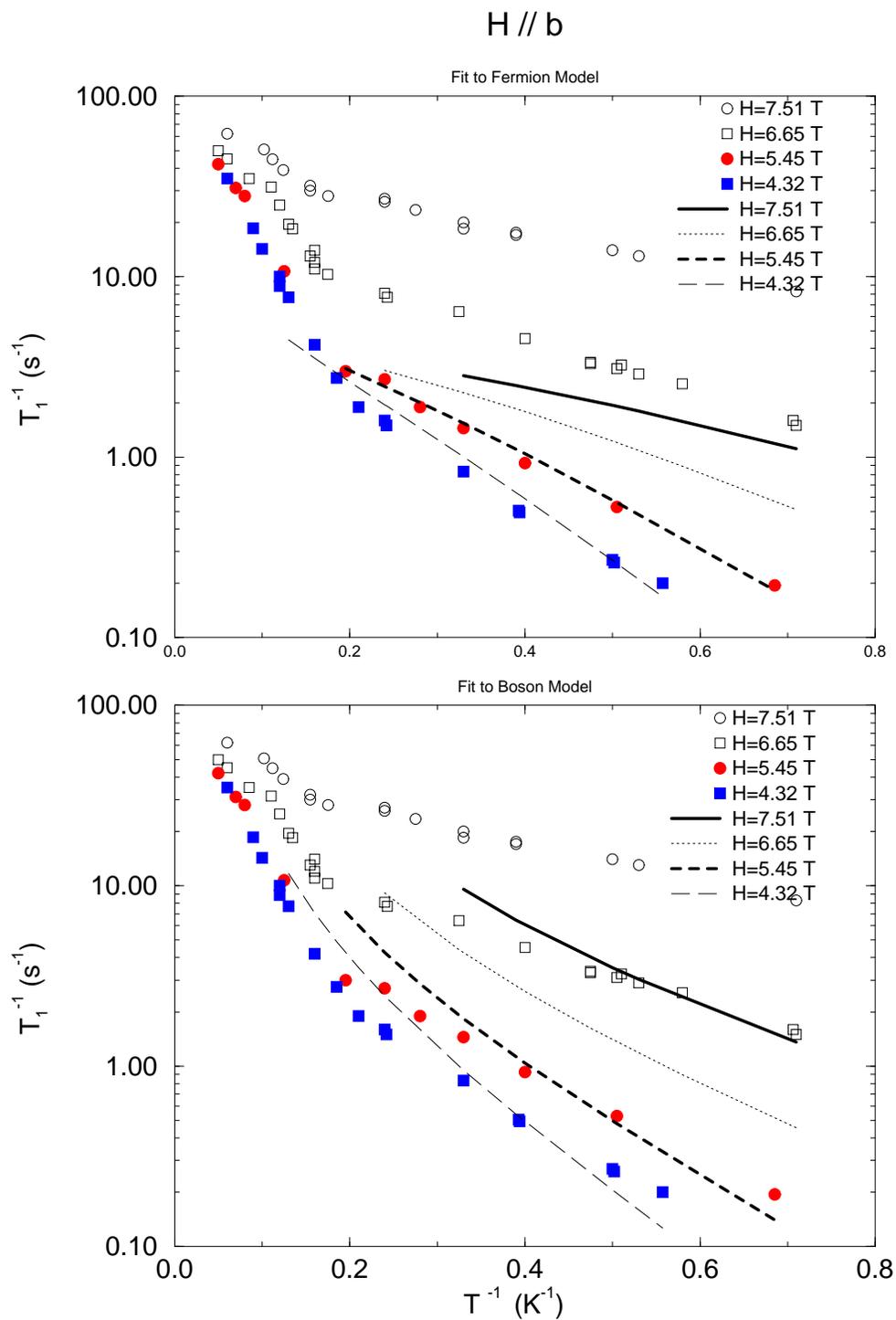}
}
\caption{Theoretical (lines) vs. experimental data (circles and squares)}
\label{fig:fuj_t_b}
\end{figure}

We now proceed to directly compare our results with those of 
Fujiwara et. al. Since the hyperfine coupling is not known, we
find a best fit to it using the experimental data. This is
best done for mid-sized fields: in the low field regime
impurities may dominate, and in the high field regime the staggered
part of the spin is expected to contribute. Figs. \ref{fig:fuj_t_b} and
\ref{fig:fuj_t_c} are such fits to the boson and fermion models.

\begin{figure}
\centerline{
\epsfxsize=13.5 cm
\epsffile{fuj_t_c}
}
\caption{Theoretical (lines) vs. experimental data (circles and squares)}
\label{fig:fuj_t_c}
\end{figure}
In producing these fits we get different values for $A_w$, the hyperfine
coupling for a field placed along the $w$-direction:
\begin{eqnarray}
A_b \sim \left \{ \begin{array}{cc}
8.5 \;\;\;  \mbox{MH}z & \mbox{fermions} \\
7.0 \;\;\;  \mbox{MH}z & \mbox{bosons} \end{array} \right.
\end{eqnarray}
\begin{eqnarray}
A_c \sim \left \{ \begin{array}{cc}
17.7  \;\;\;  \mbox{MH}z & \mbox{fermions} \\
12.0 \;\;\;   \mbox{MH}z & \mbox{bosons} \end{array} \right.
\end{eqnarray}
These values are reasonable for dipolar
hyperfine couplings between
a nuclear spin ($s=1/2$) and the $Ni$ spin at a distance of about 2$\AA$:
\begin{eqnarray}
A \sim \frac{\mu_N \mu_B}{r^3} \sim 2 \mbox{ MH}z
\end{eqnarray}
Also, we can get a similar feeling for the size of the hyperfine couplings
from Knight shift \cite{chiba} and magnetic susceptibility \cite{renard} data
for a field placed along the $b$-axis.
\begin{eqnarray}
\mu_N \delta H \sim A \mu_b \chi_s H 
\end{eqnarray}
At about 4K, the susceptibility is roughly a fourteenth of its maximum
value. Given that $\chi_{\mbox{Max}} \sim 1/J$, and that the Knight shift at 
large fields is about $10^{-3}$, we get $A \sim 8$ MH$z$. It should be
noted, however, that these are order of magnitude estimates; an accurate
evaluation of the hyperfine matrix elements is still unavailable.
Overall, the fermion fit is the better of the two. This is more obvious
at high fields when the anisotropy $\Delta_1-\Delta_2$ is high (ie. when 
the field is along the $c$-axis). For both models, the fit to the $H \| b$
data becomes progressively worse as the field is increased. Fitting
to the lower field data seems to give better overall agreement than
fitting to the higher field results.
This is not the case for the $h \| c$ data (at least with the fermions).
Since the field in the experiment was not actually placed along the 
{\em chain} $c$-axis, we might expect even worse agreement between this set
of data and our calculations! In fact, as mentioned before, we expect a very
{\em weak} field dependence for the $h\|b$ data which would result from 
being close to $U(1)$ symmetry. This was the universal result of Chapter 3.

As is evident from the figures, the slope
of the $h\| b$ data and the calculated results agree. This implies 
that the relaxation is largely mediated through thermal bosons and
that the calculation is off by a $T$-independent multiplicative factor. 
For small
anisotropy, $h \gg E$, this effect cannot come from the matrix elements for the 
transition or the density of states. We believe that we have taken account
of the obvious mechanisms for relaxation. Terms coming from the structural
properties of NENP into the Hamiltonian (as discussed in the last section)
are too small to be responsible for such a large increase in the relaxation
at the mid-field range. Moreover, they would be expected to play a similar
role when the field is placed along the $c$-axis. There are 16 protons in 
the chelate surrounding each $Ni$ ion. Nuclear dipole-dipole interactions among 
them are energetically negligible, and thus could not be the cause
for the increase in relaxation. It is certainly conceivable that
the averaged hyperfine coupling is highly anisotropic, but it's hard
to explain why there would an additional dependence on the {\em magnitude}
of the field. Perhaps the discrepancy is due to reasons intrinsic to
the experiment.

\begin{figure}
\centerline{
\epsfxsize=13.5 cm
\epsffile{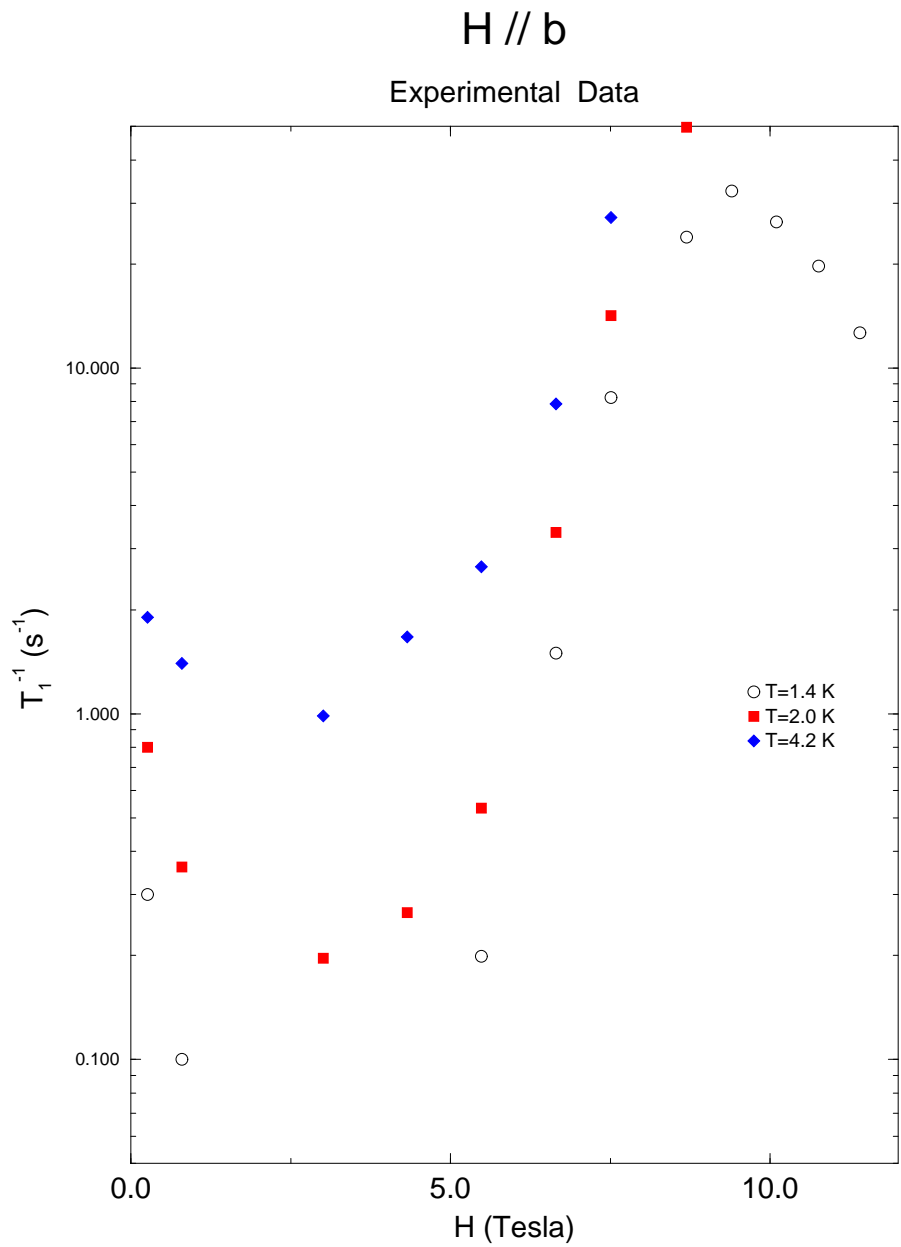}
}
\caption{Relaxation rate 
for field along the $b$-axis. }
\label{fig:fuj_f}
\end{figure}

\begin{figure}
\centerline{
\epsfxsize=7 cm
\epsffile{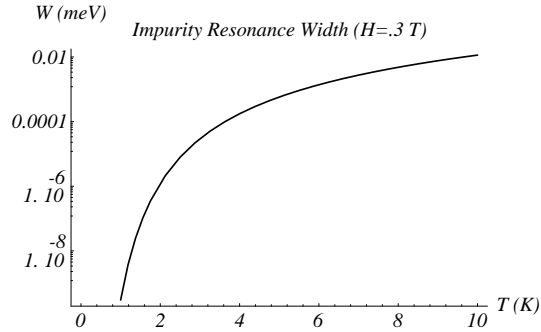}
}
\caption{$\Gamma(T)$---the width of the impurity resonance}
\label{fig:width}
\end{figure}

Next we attempt to fit to the low field measurements taken
for field along the $b$-axis. We find that for fields less than 4
Tesla, it is not sufficient to consider the bulk theory alone. The 
relaxation rate {\em decreases} with increasing field in this regime (see
Fig. \ref{fig:fuj_f}).
We can try to apply the impurity model to explain the data.
Assuming the phase shift constants, $\lambda_{\pm}^i$, in Eqn.
(\ref{phases}) are $O(1)$, the impurity resonance width, $\Gamma$,
derived in the last chapter can be graphed as in Fig. \ref{fig:width}.
As is clear from the plot and Eqn. (\ref{imp1}),
the impurity relaxation rate is essentially
one delta-function peaked at $D'/2=h$ and another peaked at $D'/2 + J' =h$.
This means that we expect {\em two} bumps in the relaxation rate due to impurity
effects. The width of the bumps should correspond to the width of distribution
of impurity couplings. The problem arises when we see that the temperature
dependence of the low field data is roughly exponential: $\sim e^{-J'/T}$,
where the impurity coupling $J'$ is about $4.7$K. 
Furthermore, the sharp
decrease from zero field relaxation suggests $D'=0$.
By analyzing Eqn. (\ref{imp1}) we
see that the second bump should have little temperature dependence. This 
means that assuming the first bump sits near $h=0$, the second must be larger
and separated by about 3.5T. This is clearly not the case. Indeed, we would
need a complicated distribution of couplings, $J'$ and $D'$, to get
a proper fit. Adding an $E$ type anisotropy will not change these conclusions.
We thus do not have a satisfactory explanation for the low field behaviour.
One should take notice, however, that the data was taken for a field along the
$b$-axis, where other problems were present at mid-field.

Finally, we would like to mention some recent NMR data collected on the
1-D $S=1$ spin chain $AgVP_2S_6$ by Takigawa et. al. \cite{taki}. This material
is highly one dimensional with a large gap ($\Delta \sim 320$K)
and very nearly isotropic ($\delta \sim 4$K). These
characteristics make it ideal for analysis using our results. There are,
however, some questions about the properties of the material which would
have to be analyzed before an understanding of the NMR results is possible
within the framework proposed here. The gap deduced from studies on the
Vanadium atom ($\Delta \sim 410$K)
conflicts significantly with those performed on the phosphorus
sites and with neutron scattering data. In addition, the material has
very low symmetry (corresponding to the space group $P2/a$) and very
little is known about the possible small $E$ and $D$ terms in the Hamiltonian
and their corresponding symmetry. There is fair qualitative agreement
between the $ ^{31}P$ NMR data and our theory, and it is possible to explain
some of the discrepancies using a temperature dependent anisotropic gap
structure, but we feel that not enough is yet understood about gross
features of the material to justify such speculation at this time.


\chapter{Suggested Experiments and Curious Predictions}
\resetcounters

We finish by pointing in this final chapter towards further
experimental work which could serve to both better understand
$S=1$ 1DHAF's as well as corroborate and clarify some of the
issues raised in this thesis.

\section{Experimentally Testable Conflicts Between Models}

When discussing the matrix elements, $<k,a|S^i(0)|q,b>$, within
the different models, we noticed that there were some
discrepancies between predictions. We now examine this hoping to
offer experiments that would resolve the issue in favour of one
model or another. 

We start by discussing experiments on isotropic systems. In this
case, the major differences between the predictions of the models
concern large $\theta$ transitions, where we recall from Chapter
2 that
\begin{eqnarray}
<k,a|l^i(0)|q,b> = i\epsilon^{iab} G(\theta) \; \; \; \; \; \;
\cosh(\theta) = -(\omega_k \omega_q - v^2kq)/\Delta^2
\end{eqnarray}
This is especially dramatic in the case of backscattering. The
problem with an experiment which probes large $\theta$
transitions is that contributions from matrix elements of the
staggered part of the spin may be large as well. This can be
cured by looking for a low temperature experiment ($T \ll
\Delta$), where the energy exchanged with the probe is small. As
shown in the analysis of $1/T_1$, the staggered contributions
will be suppressed by a double Boltzmann factor, $e^{-2\Delta/T}$.
A good candidate for such an experiment\footnote{$T^{-1}_1$
relaxation is not an appropriate tool since the transitions are
dominated by {\em small} momentum transitions} is 
{\em elastic} neutron scattering at zero or near zero
magnetic field. 
The cross section is proportional to the spin correlation function;
for elastic scattering, this is
\begin{eqnarray}
{\cal S}(Q,0) \propto \sum_{n,m} |<n|\bvec{S}(0)|m>|^2
\delta(\omega_n - \omega_m) \delta(Q - k_n + k_m) e^{-\omega_m/T}
\end{eqnarray}
At sufficiently low temperatures, this expression is simpler than
the analogous one for the relaxation rate thanks to the momentum
conserving delta function. The energy conserving delta function
ensures that {\em only} backscattering will contribute to the
cross section. Using the results of Chapter 3 we easily integrate
this to give
\begin{eqnarray}
{\cal S}(Q,0) \propto |G(\theta)|^2 \frac{2\omega_{Q/2}}{vQ}
e^{-\omega_{Q/2}/T}
\end{eqnarray}
The NL$\sigma$ model gives
\begin{eqnarray}
|G(\theta)|^2 = \frac{\pi^4}{64}
\frac{|1+(\theta/\pi)^2|}{|1+(\theta/2\pi)^2|} \left |
\frac{\tanh(\theta/2)}{\theta/2} \right |
\end{eqnarray}
At large $Q$ this will behave as $1/\log^2(vQ/\Delta)$. This is
very different from the free boson prediction of $G(\theta)=1$,
and from the free fermion prediction of $G(\theta) \rightarrow
\Delta^2/(vQ)^2$ for large $Q$. We need to qualify what we mean
by `large' $Q$. As discussed in Chapter 1, the field theoretic 
models introduced are expected to be accurate only for $Q$ near 
zero and $\pi$. If we want to explore the two magnon nature of the
structure function, we must be near $Q \sim 0$. What we mean by `large'
momentum elastic scattering is the investigation of the structure
function near the border region where the field theories begin to
diverge from numerical simulations \cite{sorensen};
a region which satisfies all the criterion is $.2\pi \leq
Q \leq .4\pi$. This corresponds to energies three to six times that
of the gap. We expect that the differences between the models should
be discernible in this range. The reason we suggest the experiment be
done at zero or nearly zero magnetic field is to ensure that only
backscattering transitions contribute. For nonzero field,
interbranch transitions can occur at large momentum which will
not necessarily select only backscattering events. This will not
serve to make the interpretation transparent. The condition for
backscattering even in the presence of a magnetic field is
\begin{eqnarray}
Q \gg h/v
\end{eqnarray}

In the case of axial symmetry, we can suggest the same technique
to investigate the difference between the zero field predictions
of the boson and fermion models. Regardless of the size of $D$,
if one only considers the cross section for scattering with $Qv >
J$, then the fermion model predicts a result that vanishes as
$\sim \frac{e^{-Qv/T}}{Q^2}$ while the boson model prediction
only involves the exponential factor. The same comments apply to
the case where axial symmetry is broken as well. This is no
surprise since at large enough momentum, $O(3)$ symmetry is
effectively restored. 

Elastic neutron scattering is a good probe for the matrix
elements involving large momentum and small energy exchange.
Other techniques which explore the opposite regime are
electron spin resonance (ESR) or far infrared absorption 
experiments. In both, one subjects the magnetic system to an
external source of electromagnetic radiation (the microwave frequency
value of the radiation depends on the transitions one is interested in
investigating). The RF field couples to the spins in the same way
that a magnetic field does, assuming that the electric dipole
moment of the electrons on the magnetic ion is much smaller than
the effective spins.\footnote{A rigorous treatment would try to
treat the coupling to the electric dipole moment; this can be
done within the spin manifold using the Wigner-Eckart theorem. We
will not bother with such a treatment here, but we note that it
may be crucial in understanding some experiments on NENP
\cite{halp}} The interaction Hamiltonian is therefore
\begin{eqnarray}
H_I =\bvec{H}_{\mbox{RF}} \bold{G} \cdot \sum_i \bvec{S}_i
\cos(\omega t)
\end{eqnarray}
Since the coupling is to the total spin of the system, the
resonant transitions implied by Fermi's Golden Rule will involve
energy $\omega$ and {\em zero} momentum exchange. At low temperatures,
the power
absorbed when a uniform field is applied to the system will be 
\begin{eqnarray}
I(\omega) \propto \sum_{n,m} |<n|S^i_{q=0}|m>|^2
\delta(\omega_n(h) - \omega_m(h) - \omega) \delta(k_n - k_m) e^{-
\omega_m/T} \nonumber
\\ \approx  |<a;k,\omega_k^a(h)|S^i_{q=0}|b;k,\omega^b_k(h)>|^2
e^{-\omega_k^b/T} \left | \frac{\partial (\omega^a_k(h) -
\omega^b_k(h))}{\partial k} 
\right |^{-1}
\label{esr_abs}
\end{eqnarray}
where $a$ and $b$ denote one magnon states and $k$ satisfies,
$\omega_k^a(h) - \omega_k^b(h) = \omega$. Since the density of
states factor in the above is divergent for $k=0$, it stands that
$I(\omega)$ will have a peak at the value of $h$ for which
$\Delta^a(h) - \Delta^b(h) = \omega$. (The divergence will be
cured by higher dimensional effects as discussed previously.)
In a typical experiment, one judiciously chooses the RF
frequency, $\omega$, to be in the vicinity of desirable
transitions, and the uniform field is then tuned to the peak in
the absorption power. This is much easier to do than to fine tune
the RF field. 

Let us now relate the ESR matrix elements to
$\bvec{l}_{a,b}(0,0)$, calculated in Chapter 2. 
\[ <a;0|S^i_{Q=0}|b;0> = \int dx <a;0|S^i(x)|b;0> =
\int dx <a;0|e^{iPx}S^i(0)e^{-iPx}|b;0> \]
\begin{eqnarray}
= \int dx <a;0|S^i(0)|b;0> \approx L l^i_{a,b}(0,0)
\end{eqnarray}
Interesting conflicts between the models can be seen when there
is some kind of, preferably large, anisotropy. For example,
considering axial symmetry with a large $D$ anisotropy,
\begin{eqnarray}
|l^+_{0,-}(0,0)|^2 = \left \{ \begin{array}{cc}
\frac{1}{2} (\Delta_3/\Delta_{\perp} + \Delta_{\perp}/\Delta_3 +2) &
\mbox{Bosons} \\ 2 & \mbox{Fermions} \end{array} \right.
\end{eqnarray}
The maximal difference corresponds to $\Delta_3/\Delta_{\perp} \sim 2$
which leads to a discrepancy of about 13\% between the models.
The closer the two branches lie, the better the agreement between the
models. This suggests the following experiment on highly
anisotropic materials (NENP being a prime candidate). One chooses
two RF frequencies. The first should correspond to the large
interbranch gap, $D$, and the peak absorption ought to be
measured with a {\em low} field placed along the direction of the
$D$ anisotropy. The second RF frequency should be $O(E)$ if the
material breaks axial symmetry, or $O(h)$ if the material is
axially symmetric. This should then be used to measure the absorbed
ESR power with the positions of the uniform and
RF fields exchanged. This second transition will involve matrix
elements which will be gap independent in both models. The matrix
elements from the first transition can be extracted and
compared to that of the first. If the boson model is a better
description even at these low fields, then the two matrix
elements should be identical. 

One may argue that it is redundant to make both measurements
since, if the gaps are known, Eqn. (\ref{esr_abs}) should give
the correct description. The problem lies in cutting off the
infrared divergence at the absorption peak. This will introduce
an unknown proportionality constant. This divides out when
comparing the two measurements. The ratio of the two measurements
would be
\begin{eqnarray}
\frac{I(\omega_1)}{I(\omega_2)} =
\frac{1-e^{-\omega_1/T}}{1-e^{-\omega_2/T}}
\frac{\Delta_3}{4\Delta_{\perp}} \left| l^+_{0,-}(0,0) \right|^2
\end{eqnarray}
where we assume a small field, $h \ll \Delta_{\perp}$, and small 
$E \ll D$.

To end this section, while on the subject of ESR experiments,
we would like to propose additional
experiments to test the impurity model presented in Chapter 4.
ESR is ideal for such tests. Used in conjunction with $T^{-1}_1$
measurements on a given sample, it would be possible to
characterize the couplings $J'$ and $D'$ of the end spins .

\section{Measuring Small Anisotropies}

Recall that we expect a peak in $T^{-1}_1$ whenever two branches
cross. Experiments on Haldane Gap materials have yet to look for these.
The sharpness of this peak depends on the interchain
couplings which cut off the diverging integral in the calculation
of the relaxation rate. Often, this will be broad because
intrabranch transitions will share the same cutoff (ie. when
$J_I>\omega_N$). However, the bump should be experimentally
observable. We propose that information about the anisotropy
tensor can be extracted from this phenomenon. Essentially, one
looks for the lowest field at which this bump occurs. This would
give the direction of the lowest branch and the size of the
anisotropy. We now explain this further.

We assume the material in question has a well resolved $D$
anisotropy and a seemingly degenerate doublet unresolved by other
experimental techniques, such as neutron scattering. One begins
by placing a uniform magnetic field in the plane perpendicular to
the axial direction (ie. somewhere in the $xy$-plane). 
The magnitude of the field
should be $h^2 > \bar{\delta} (\Delta_3 - \Delta_{perp})$, where
$\bar{\delta}$ is the uncertainty in resolving the doublet. One
then proceeds to measure $T^{-1}_1$ for different angles in the
$xy$-plane spanning a region of at most $180^{\circ}$. If there
is an $E$ type anisotropy, one ought to see some structure to the
data as a function of angle. Moreover, if there is such
structure, we expect a bump at the angle where the branches
cross. Once this angle is found, the experiment is repeated for
somewhat lower field. The angle where the new bump should be seen
would be greater than the old. There is actually enough
information in these two measurements already to determine the
anisotropy tensor. The dispersion relations are a function of the
angle of the field (relative to some axis), the field magnitude
and the gaps. The only unknowns are the absolute angle (or
location of the axes of the anisotropy tensor) and the difference
in gaps, $|\Delta_x-\Delta_y|$. The two measurements could be used
to solve for these two unknowns.

In principle, one could also continue lowering the field and looking
for the bump angle until it's clear that signal is being lost
when the field is reduced further. At this point, one has located
the minimum crossing field which must lie along the direction
associated with the middle gap. This field also gives the
anisotropy: $|\Delta_x - \Delta_y| = h^2/|\Delta_D - \Delta|$.

It would be interesting to perform such an experiment on NENP.
Presumably, one would find two angles corresponding to the two
inequivalent chains in each unit cell. Moreover, one would be
able to verify the claims made in the last chapter regarding the
positions of the local anisotropy tensor in NENP.

\section{ESR for NENP}

\begin{figure}
\centerline{
\epsfxsize=13.5 cm
\epsffile{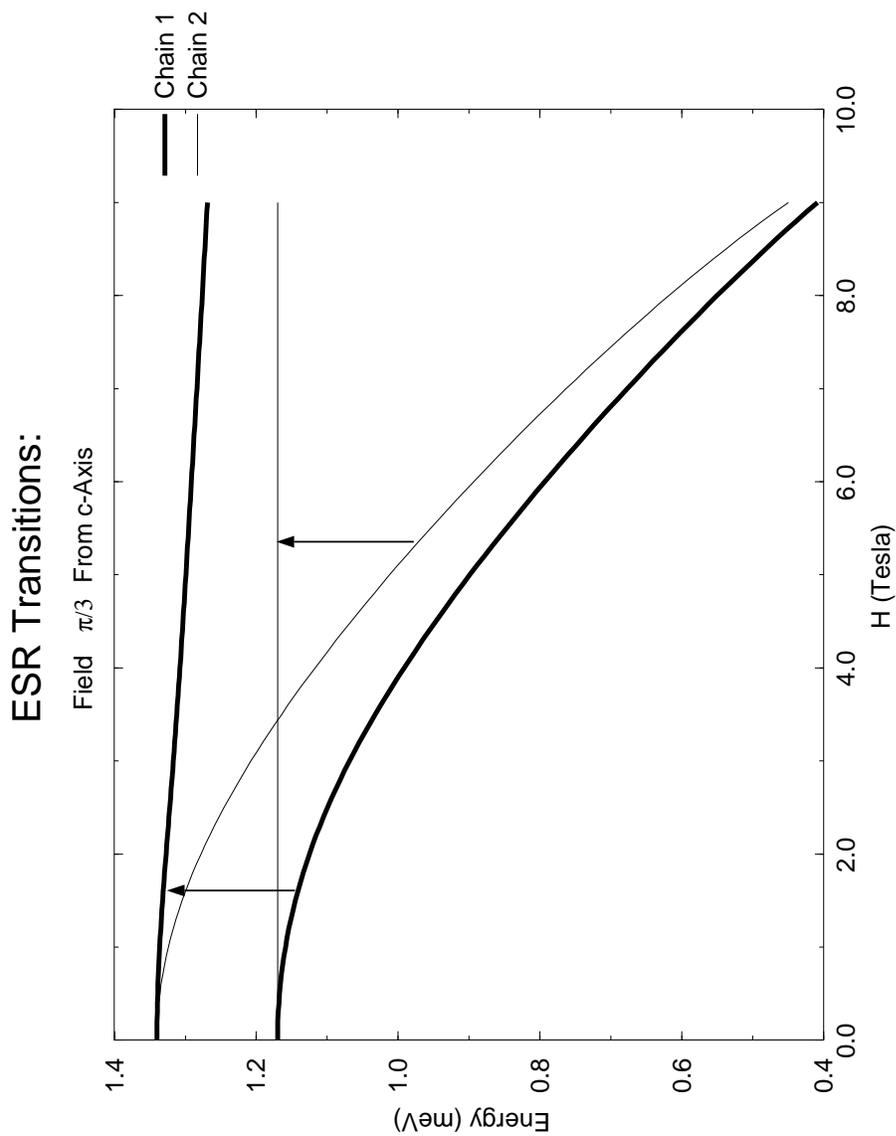}
}
\caption{Dispersions for the two chain conformations and sample 
resonant transitions
for a uniform field placed $60^\circ$ from the crystallographic $c$-axis.}
\label{fig:disp_esr}
\end{figure}

In the last chapter we noted that NENP has two inequivalent
chains per unit cell. Furthermore, their local anisotropy tensor
was argued to have symmetry axes which did not correspond to the
crystal axes. These facts have important ramifications for ESR
experiments on NENP.
Figure \ref{fig:disp_esr}  shows the dispersions for chains 1 and
2 (bold and light
lines, respectively) when the field is $\pi /3$ from the
crystallographic $c$
axis in the $ac$-plane.
This is an example of how transitions at two
field strengths ought to be possible in the ESR experiment.

\begin{figure}
\centerline{
\epsfxsize=13.5 cm
\epsffile{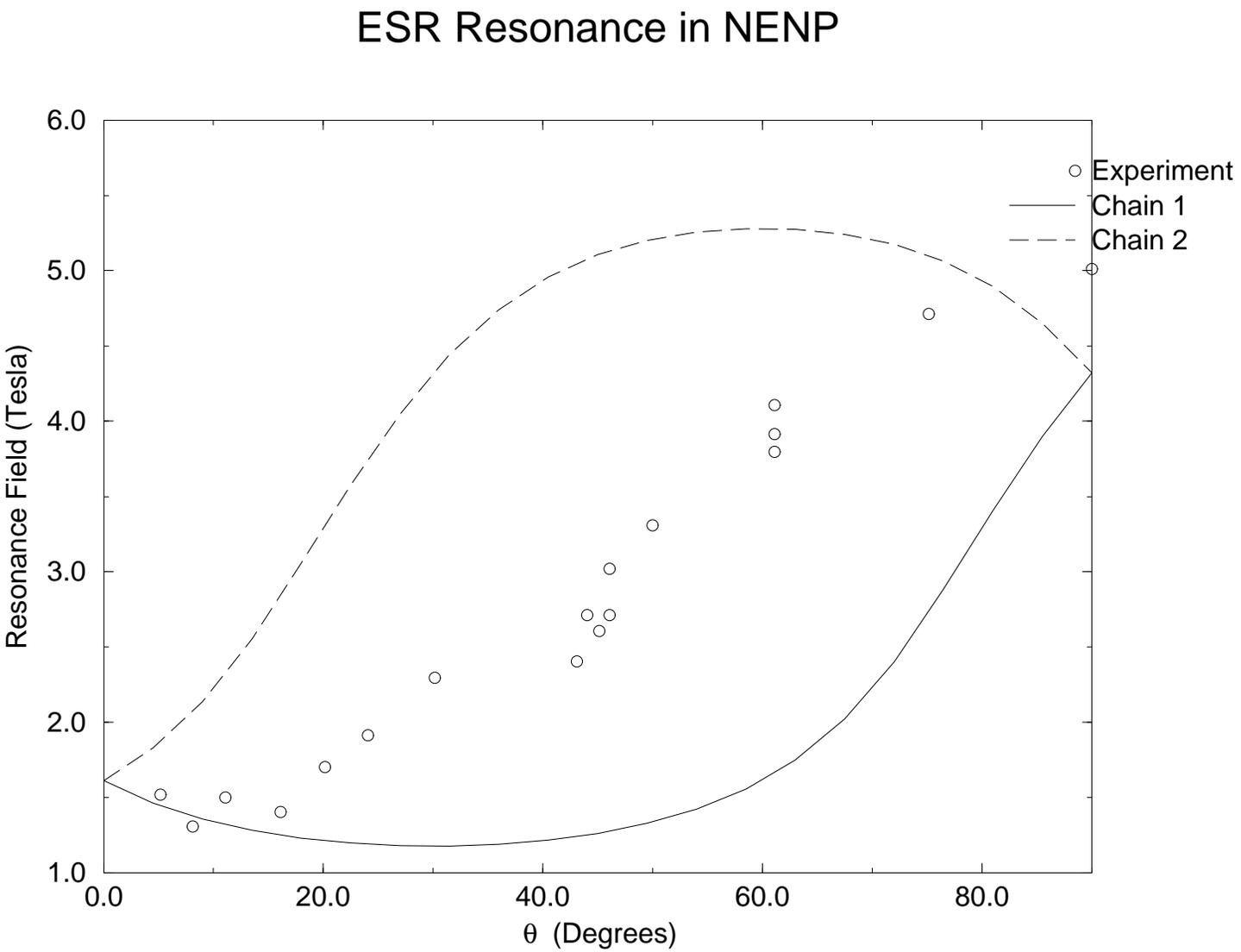}
}
\caption{Resonant field vs. field orientation in the $ac$-plane for
.19 meV transitions.}
\label{fig:esr_res}
\end{figure}

Figure \ref{fig:esr_res} shows the resonance field versus
orientation of field in
the crystallographic $ac$-plane.
The lower branch denotes
transitions in chain 1
while the upper branch corresponds to transitions in chain 2.
The transitions were
calculated at .19 meV.
corresponding to 47 GHz. 
In addition the experimental
results of Date and Kindo \cite{dat} are represented by the
circles. One immediately
sees that the data does not compare well with the predictions
based on the
models we've used so far, for instead of
following one of the branches, the experimental results lie
between them.
Furthermore, it seems unlikely that perturbations will cause such
a significant
shift in the resonance field. One sees that the discrepancy is
$\sim
\pm 1$ Tesla. One possible explanation is that since the ESR
signal
in \cite{dat}
was also $\sim \pm 1$ Tesla in width and symmetric (in conflict
with the
predictions of \cite{aff}), the signal from the
resonances in both
chains was somehow smeared and interpreted as one single peak.
Seen that way
the model predictions are in good agreement except for the large
field
regime. One also has to keep in mind that the high-field boson
dispersions
are not accurate and therefore the predictions at larger angles
could easily be
.5 Tesla (or more) off the mark. We propose that further ESR
experiments be done on NENP which specifically look for the
double resonance predicted here.

To end this discussion, we'd like to elaborate on
a previously made statement regarding the assignment of masses to
the
local Ni symmetry axes. It's easy to see that switching the
masses around
is tantamount to a $\pi/2$ shift in Figure \ref{fig:proj1} 
(the fact that the gyromagnetic
constants are not the same in orthogonal directions will not
change the
ESR resonance graph much since the ratio of the gyromagnetic
constants is 0.98).
Redrawing Figure \ref{fig:esr_res} with this geometry misses the
experimental
results by 4 Tesla at 0 and 90 degrees, 
where the two chain resonances coincide.
This determines the proper labeling of the local symmetry axes.


\chapter{Concluding Remarks}
\resetcounters

With the increasing theoretical interest in low dimensional
systems, there has been a proportionate increase in the number of
both realizable physical systems and experiments. This work offers 
a comprehensive analysis of NMR relaxation in Haldane gap materials,
taking account of anisotropy and other material properties.
As well, our analysis has led to predictions pertaining to other
types of experiments.
It is hoped that our efforts will aid in both extending and
clarifying existing knowledge of the subject.

\end{document}